\documentclass[11pt,a4paper]{article}

\usepackage{iftex}
\ifPDFTeX
  \usepackage[utf8]{inputenc}
  \usepackage[T1]{fontenc}
  \usepackage{lmodern}

\usepackage{amsmath,amssymb,amsthm,mathtools,bm}
\usepackage{graphicx}
\graphicspath{{figures/}}
\usepackage{float}
\usepackage[section]{placeins}
\usepackage{caption}
\usepackage{subcaption}
\usepackage{booktabs,multirow,array,tabularx,longtable,threeparttable,calc}
\usepackage{enumitem}
\usepackage[most]{tcolorbox}
\usepackage[a4paper,margin=1in]{geometry}
\usepackage{setspace}
\usepackage{microtype}
\usepackage{ragged2e}
\newcommand{\notepar}[1]{\par\begingroup\fontsize{9}{11}\selectfont\justifying\noindent #1\par\endgroup}
\newcommand{\sym}[1]{\ifmmode^{#1}\else\(^{#1}\)\fi}

\usepackage{fancyvrb}
\usepackage{fvextra}
\fvset{breaklines=true,breakanywhere=true,fontsize=\small}
\DefineVerbatimEnvironment{verbatim}{Verbatim}{breaklines=true,breakanywhere=true,fontsize=\small}

\usepackage[round,authoryear]{natbib}
\setcitestyle{aysep={,}}
\usepackage{multibib}
\newcites{app}{References}
\usepackage[hyphens]{url}
\usepackage[colorlinks=true,linkcolor=blue,citecolor=blue,urlcolor=blue,hypertexnames=false,
  linktocpage=true,bookmarks=true,bookmarksopen=false,breaklinks=true]{hyperref}

\title{\textbf{Tariff Threats, Macroeconomic Expectations, and Policy Communication Strategies: Experiments Based on a Multi-Agent System}\thanks{We gratefully acknowledge the support of the National Natural Science Foundation of China (Grant No. 71991474, 72073148, 72273156, 72303258) and National Social Science Foundation of China (Grant No. 22AZD121, 24ZDA042).}}
\author{\textbf{Jianhao Lin}\thanks{Lingnan College, Sun Yat-sen University, China, 510275. Email: linjh3@mail.sysu.edu.cn.}\qquad
\textbf{Lexuan Sun}\thanks{Corresponding author: Lingnan College, Sun Yat-sen University, China, 510275. Email: sunlx7@mail2.sysu.edu.cn; Tel: (+86) 18392757553.}\qquad
\textbf{Yixin Yan}\thanks{Lingnan College, Sun Yat-sen University, China, 510275. Email: yanyx33@mail2.sysu.edu.cn.}}
\date{}

\begin{document}
\maketitle

\begin{abstract}
\normalsize
\noindent Tariff threats can move household beliefs before policy is enacted, yet their rapidly changing language is difficult to study with conventional surveys. We build a multi-agent system that turns 300 households from the Michigan Surveys of Consumers into persistent large-language-model agents exposed to social-media information over several simulated months. Calibrated agents reproduce some distributional and demographic patterns in human survey data collected after the announcement of Liberation Day tariffs. Simulated experiments indicate that immediacy, rate salience, semantic progression, message complexity, narrative, and sender identity jointly shape inflation and unemployment expectations and their dispersion. Open-ended responses trace these effects to attention, ambiguity, credibility, and causal narratives. A second experiment finds that central-bank explanations can coordinate beliefs, although their effects on average expectations depend on message content. The framework supports disciplined exploration of policy communication, subject to human validation rather than as a substitute for it.
\end{abstract}


\bigskip
\clearpage
\addtocontents{toc}{\protect\setcounter{tocdepth}{-1}}

\section{Introduction}\label{sec:introduction}

Tariff policy can move household beliefs before it changes prices. After returning to office in January 2025, President Donald Trump placed broad tariff proposals at the center of the U.S. agenda. The April 2 ``Liberation Day'' announcement introduced a 10 percent baseline tariff and higher country-specific rates, departing sharply from the narrower actions of his first administration.\footnote{See Executive Order 14257, April 2, 2025, at \href{https://www.whitehouse.gov/presidential-actions/2025/04/regulating-imports-with-a-reciprocal-tariff-to-rectify-trade-practices-that-contribute-to-large-and-persistent-annual-united-states-goods-trade-deficits/}{The White House}.} Recent U.S. trade actions produced substantial pass-through, production relocation, and welfare effects \citep{amiti2019impact,fajgelbaum2020return,flaaen2020production,ignatenko2025making,cavallo2025tracking}. Surveys around prospective tariff changes also record anticipated price increases, precautionary responses, and partisan differences in perceived incidence \citep{coibion2025upcoming,hirs2026partisanship}. A tariff threat is thus an information shock as well as a possible future tax.

\begin{figure}[htbp]
\centering
\includegraphics[width=.65\linewidth]{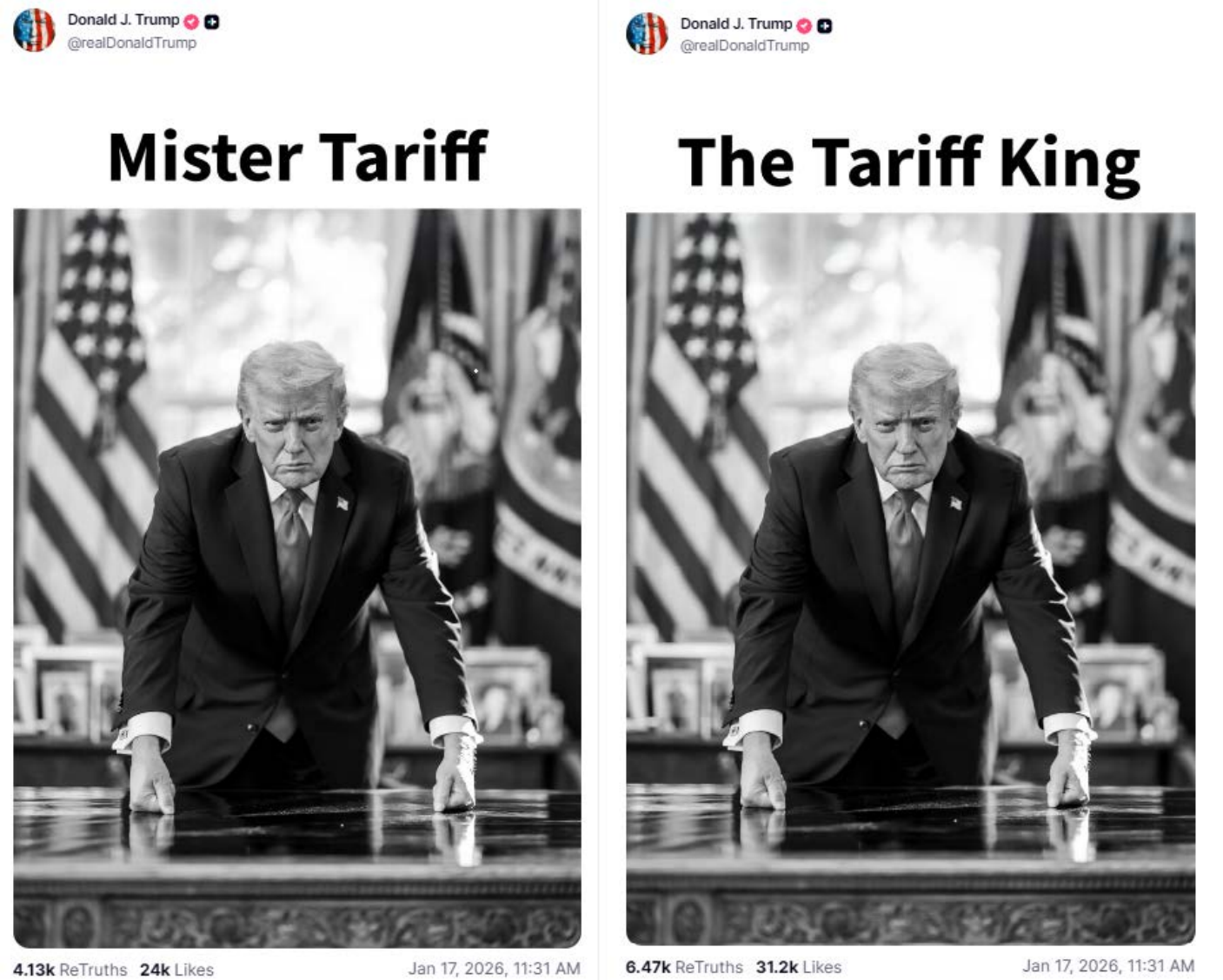}
\caption{Tariff Branding on Truth Social}\label{fig:intro-posts}
\notepar{Notes: The two panels reproduce posts published on President Trump's Truth Social account on January 17, 2026, in which he referred to himself as ``The Tariff King'' and ``Mister Tariff.'' The figure motivates the study of policy content and sender style as a joint information treatment.}
\end{figure}

The second Trump administration has also changed the cadence of policy communication. One audit counted 6,168 posts and reposts on Trump's Truth Social account in 2025, roughly 18 per day, with 168 on December 1; another counted 2,262 posts in the first 132 days after inauguration.\footnote{See the Get the Facts Data Team figures reported by \href{https://www.kmbc.com/article/donald-trump-truth-social-2025/69838178}{KMBC} and the \href{https://www.washingtonpost.com/technology/2025/06/03/trump-truth-social-twitter/}{Washington Post}. Counts may differ in their treatment of reposts and deletions.} Tariff messages in this stream are often conditional and politically targeted. On January 12, 2026, Trump threatened an immediate 25 percent tariff on countries doing business with Iran. Five days later, he threatened tariffs on eight European countries of 10 percent from February 1 and 25 percent from June 1 until an agreement permitted the United States to purchase Greenland.\footnote{See contemporaneous reports by \href{https://apnews.com/article/55d5f3335ffd3c2340e4590f33150971}{Associated Press} on Iran and by \href{https://www.reuters.com/world/europe/trump-vows-tariffs-eight-european-nations-over-greenland-2026-01-17/}{Reuters} on Greenland. The Iran post had no accompanying White House documentation at the time.} Figure~\ref{fig:intro-posts} illustrates how a personal political brand is attached to the instrument. Recommender systems then shape exposure, concentrate attention, sort information environments, and can amplify disagreement \citep{allcott2020welfare,levy2021social,santos2021link}. Announcements, reversals, and delays may therefore move expectations even when implementation remains uncertain \citep{pastor2012uncertainty,ilut2014ambiguous}.

This environment raises two questions. How do a tariff threat's timing, rate, semantic coherence, complexity, narrative, and sender alter the level and dispersion of inflation and unemployment expectations? After a high-impact threat, what central-bank communication best coordinates those beliefs? Existing methods provide partial solutions. Structural and time-series models discipline aggregate counterfactuals but cannot readily map unstructured messages into household probability forecasts and explanations. Surveys reveal otherwise unobservable beliefs and create identifying variation \citep{stantcheva2023surveys,haaland2023designing}, yet dense monthly panels with many text treatments are costly and slow to redesign. Text analysis of social media texts characterizes observed communication but cannot generate responses to counterfactual messages. Language-model studies condition synthetic respondents on human characteristics \citep{argyle2023out,horton2023large}, and newer work addresses coverage, dynamics, and identification \citep{wu2025llm}. Building on these strands, the Economic World Model agenda organizes agents, environments, and interactions within one model class, allowing each to be theory-specified, learned, or hybrid; for counterfactuals involving learned components, the Data-Driven Generative Equilibrium concept imposes joint consistency among behavior, beliefs, generated data, and retraining \citep{cong2026economicworldmodels,han2026agenticeconomies}.

 Inspired by these perspectives, we combine persistent identities, social media interaction, information treatment, and belief updating into one simulated dynamic environment.\footnote{The system screens mechanisms and scenarios before human studies; it does not replace surveys or structural models.} We therefore build a Multi-Agent System (MAS) around 300 households drawn from the Michigan Surveys of Consumers (MSC). Each household's observed personal characteristics define the persona of a \emph{Household Agent} implemented with a large language model, while its reported inflation and unemployment expectations provide that agent's initial beliefs. Over several simulated months, the Household Agents combine their prior expectations and perceptions with current social-media information, retain recent information through a rolling memory, and answer the same survey repeatedly. In the treatment period, we present 300 agents with one of several potential tariff threat scenarios, while holding their personas, initial beliefs, pre-treatment histories, and model settings fixed across scenarios. Every monthly survey records point forecasts, subjective probability distributions, and open-ended explanations of those forecasts. The resulting balanced event-time panel lets us compare average revisions in expectations, changes in the extent to which beliefs are dispersed, and the narratives agents use to explain their revisions.

Because counterfactual simulations are informative only when their behavior is anchored to human evidence, we first ask whether the MAS can reproduce household responses in an observed policy setting. The benchmark treatment information announces an immediately effective 10 percent universal tariff and is deliberately closest to the ``Liberation Day tariffs''. We align the simulated path after this message with MSC responses collected from April through November 2025 and evaluate correspondence along three margins: the distributions of reported expectations, the ability of flexible prediction methods to distinguish simulated from human observations, and the political, income, and gender differences present in those expectations. Calibration improves correspondence on all three margins. Distributional gaps narrow, common prediction methods find it substantially harder to separate simulated from human observations, and the calibrated agents recover the direction and broad ordering of salient demographic differences. These results establish a disciplined benchmark for using the MAS, but they neither make virtual households interchangeable with people nor identify human causal effects for messages that the survey never observed. Recent methodological work likewise stresses that validation must be tied to the intended use of a language model and that behavioral resemblance alone cannot justify causal transport from a validated setting to an unobserved one \citep{ludwig2026large,hullman2026human}. We therefore interpret treatments that make limited changes around the benchmark as heuristic evidence for qualitative \textit{near-generalization} and treat more distant treatments as scenario exploration.\footnote{Near-generalization refers to substantive proximity to the benchmark treatment, not to a statistical neighborhood in which human treatment effects are identified.}

Given the variability and uncertainty of tariff policy communication, we construct several plausible tariff-threat scenarios based on real-world conditions to examine which factors drive agents' expectation updating and how. To facilitate the analysis, we organize these scenarios into  controlled comparisons along six dimensions: implementation time; the announced tariff rate and whether it is known; whether a sequence progressively escalates or repeatedly reverses; message complexity; causal narrative; and sender identity. The center and dispersion of beliefs respond differently. Immediate implementation raises point expectations more than a six-month delay or an uncertain date, while uncertain timing creates the greatest dispersion. A stated 100 percent rate raises average expectations more than a stated 10 percent rate or an unspecified rate, but the unspecified rate creates the greatest disagreement. A progressively escalating sequence produces a larger and more persistent response than a single high-rate message, whereas repeated reversals weaken it. Minimal wording yields a larger average revision than standard wording and technical language a smaller one, although both unusually sparse and unusually complex messages widen dispersion relative to the standard message. A Make America Great Again narrative and a tax-incidence narrative both attenuate the average revision without materially reducing dispersion, but they redirect agents' causal explanations. Identical tariff language attributed to the Democratic leader produces a larger average response and greater dispersion than when attributed to the Republican leader. The open-ended explanations link these rankings to attention, ambiguity, source credibility, skepticism, political priors, and competing views of who bears a tariff. These model-generated explanations remain mechanism hypotheses rather than independent evidence about human cognition. The economic implication is that a communication feature can move average expectations and coordinate beliefs in different, sometimes opposing, directions.

The second research question concerns how central-bank communication operates after a tariff threat has already raised expectations and widened disagreement. We use the progressively escalating tariff sequence for this exercise because it produces the largest and most persistent revisions among the alternative sequences of tariff messages and therefore provides a demanding communication environment. After that sequence, Household Agents receive one of five Federal Reserve messages: a restatement of the longer-run inflation objective; a commitment to tighten monetary policy if tariffs create persistent, broad-based price pressure; an affirmation of the Federal Reserve's political independence; an explanation that treats the tariff as a temporary relative-price shock rather than a change in underlying inflation; or a willingness to tolerate inflation moderately above target for a time. The same qualitative ordering emerges for inflation and unemployment expectations. Messages that make persistent price pressure and possible tightening salient, and messages that explicitly tolerate an inflation overshoot, generate the largest upward revisions. Restating the inflation objective or affirming independence has little average effect, while the explanation that classifies the tariff as a temporary relative-price shock is the only strategy that lowers point expectations. Every Federal Reserve message nevertheless reduces belief dispersion. The experiment therefore separates two policy objects that are often conflated: shifting the average expectation and coordinating households around a common interpretation. Because these messages are substantially more distant from the human-validated benchmark, their ranking is an exploratory stress test that supplies hypotheses for human experiments, not a direct recommendation for the Federal Reserve.\footnote{Policy advice additionally requires human evidence together with equilibrium and welfare analysis.}

Our paper makes three contributions. First, we introduce a dynamic framework for generative economic experiments calibrated with survey data and social media posts. Synthetic samples that condition large language models on human personas show that model responses can reproduce meaningful variation across observed characteristics \citep{argyle2023out}, and economic applications use such models to simulate beliefs, choices, and deposit withdrawals \citep{horton2023large,kazinnik2026bank,delriochanona2025generative}. Recent frameworks extend this agenda by emphasizing dynamic surveys and identification \citep{wu2025llm} or interactions among heterogeneous social agents \citep{manning2026general}. Our framework brings these advances into an empirically anchored panel experiment. It combines persistent household identities and observed initial beliefs with social media information, multi-period memory, repeated treatment-control comparisons, and joint elicitation of point forecasts, probability distributions, and open-ended explanations. Following the same virtual household before and after a text treatment makes the simulated history itself part of the information set, rather than generating a succession of independent, one-shot answers. Just as important, the framework embeds a validation hierarchy that distinguishes correspondence in an observed benchmark from qualitative comparisons nearby and scenario exploration farther away. This feature disciplines the added computational scope instead of treating plausible language-model behavior as sufficient evidence of human causal effects \citep{ludwig2026large,hullman2026human}.

Second, we add a perspective centered on message design to the economics of household expectation formation. Existing work shows that forecasts are heterogeneous and partisan and that experience, limited attention, and information rigidities shape how households form them \citep{souleles2004expectations,mankiw2004disagreement,sims2003implications,gabaix2020behavioral,kamdar2025think}. Information experiments further demonstrate that updating depends on what a message says, how it frames the issue, and who delivers it \citep{haaland2023designing,coibion2022monetary,kuang2025central}. These literatures identify important individual channels, but they do not usually compare this full set of message attributes within a common dynamic information environment. Our design holds the surrounding environment fixed while varying implementation time, rate certainty, coherence across a sequence of messages, complexity, narrative, and sender. It also studies the center of the expectations distribution and belief coordination as separate outcomes. Combining point forecasts and subjective distributions with open-ended explanations reveals whether a small average response reflects little updating by most households or offsetting revisions across households, and whether apparent coordination arises from shared information, shared ambiguity, or a common causal narrative. In this respect, the paper extends survey approaches that use structured treatments and open-ended responses to recover reasoning that standard numerical questions leave unobserved \citep{stantcheva2023surveys,haaland2025understanding}.

Third, we provide an experimental test bed for policy communication after beliefs have evolved in an information environment shaped by recommendation algorithms. Research on central-bank communication shows that targets, forward guidance, source credibility, and communication through social media can shape public beliefs \citep{blinder2008central,eusepi2010central,coibion2022monetary,ehrmann2022central,gorodnichenko2025central}. Our framework adds a distinct setting: the central-bank message arrives only after information selected by a recommendation algorithm, communication among households, and recent memory have already shaped the beliefs that policymakers seek to guide. The experiment consequently treats communication as more than a hawkish or dovish signal. A message also supplies a causal model of the shock, a judgment about its persistence, and a condition under which policy will react, distinctions that matter for both the level and coordination of expectations \citep{kakhbod2026mind}. The same architecture can be adapted to fiscal announcements, energy interventions, sanctions, public-health guidance, and central-bank communication through social media. Such portability does not make numerical results transferable without qualification: every application requires new personas, priors, information-flow calibration, and human validation. What carries across settings is the research workflow linking message construction, algorithmic exposure, belief updating, open-ended explanation, and interpretation disciplined by the available evidence.

The remainder of the paper is organized as follows. Sections~\ref{sec:design} and \ref{sec:mas} present the design, empirical model, and MAS. Section~\ref{sec:results} reports the results and mental mechanism analysis. Section~\ref{sec:cb} studies central-bank communication, Section~\ref{sec:discussion} discusses scope, and Section~\ref{sec:conclusion} concludes.

\section{Experimental Design and Empirical Model}\label{sec:design}

\subsection{Experimental Design}\label{subsec:experimental-design}

We construct a Multi-Agent System (MAS) to simulate household macroeconomic expectations. The system is calibrated with two bodies of information from December 2024: expectation-survey responses and demographic characteristics for 300 actual households selected by stratified random sampling from the Michigan Survey of Consumers (MSC), and texts of social-media posts on X concerning ``inflation'' and ``unemployment.'' We iterate the MAS for 18 months and thereby generate a balanced panel of simulated survey observations. In each month, the same Household Agents receive current information, report their point and distributional expectations, explain those expectations in open-ended responses, and produce social-media posts that can enter later agents' information sets.

To validate and illustrate the MAS, we design one control arm and twelve information-treatment arms that represent a set of realized or potentially realizable tariff-threat scenarios. Treatment is delivered in Month $m=7$. Because the experimental sample is entirely computational, each arm reuses exactly the same 300 LLM Agents. Their personas, initial expectations, pre-treatment histories, confidence assignments, and model settings are identical across arms; only the information supplied at treatment differs.\footnote{Reusing computational households permits clean within-agent counterfactual comparisons that cannot be implemented with human respondents without carryover. It does not eliminate common model misspecification, so the exercise remains a simulation rather than a human randomized controlled trial.}

Agents in the control arm receive placebo information unrelated to macroeconomic expectations. Its purpose is to absorb the otherwise unobserved effect of being shown a message from the political sender. The placebo is a genuine Trump post on Truth Social\footnote{The post was published on September 15, 2025; the archived source URL is \url{https://truthsocial.com/@realDonaldTrump/posts/115209308362990527}.The placebo has no tariff, price, employment, or monetary-policy content. It therefore isolates the response to receiving a presidential social-media post from the response to the tariff content embedded in the treatment messages.}:

\begin{quote}\itshape
MY son Eric's just out book, ``UNDER SIEGE,'' immediately went to NUMBER ONE on Amazon. Great going Eric, you deserve it!!! https://a.co/d/7oPyLF5
\end{quote}

The twelve treatment arms correspond to twelve tariff-threat scenarios. Every treatment message imitates the writing style and expressive features of Trump's tariff-related Truth Social posts. T1 is the benchmark and is deliberately closest to the realized policy setting: the ``Liberation Day tariffs'' announced by the Trump administration on April 2, 2025. Its message was generated from the substantive content of Executive Order 14257, rather than copied from a Trump post, for two reasons. First, Trump did not publish a Truth Social post that directly and completely stated the Liberation Day policy in the form required for this experiment. Second, tying T1 to an actual executive order keeps the benchmark anchored in an observed policy event, permits later comparison with human expectations around that event, and prevents the experimental design from becoming detached from economic reality. All remaining arms are counterfactual variants derived from T1. Their defining elements, connected below by a plus sign, are:

\begin{description}[leftmargin=1.4cm,style=nextline,itemsep=2pt]
\item[T1] 10 percent tariff $+$ executive order signed for immediate implementation.
\item[T2] 10 percent tariff $+$ executive order to be signed six months later.
\item[T3] 10 percent tariff $+$ uncertain signing date for the executive order.
\item[T4] 100 percent tariff $+$ executive order to be signed six months later.
\item[T5] Uncertain tariff rate $+$ executive order to be signed six months later.
\item[T6] 10 percent tariff $+$ immediate implementation $+$ minimalist style.
\item[T7] 10 percent tariff $+$ immediate implementation $+$ complex phrasing.
\item[T8] 10 percent tariff $+$ immediate implementation $+$ MAGA narrative.
\item[T9] 10 percent tariff $+$ immediate implementation $+$ tax-incidence narrative.
\item[T10] 100 percent tariff $+$ executive order to be signed six months later $+$ high-frequency semantic reversals.
\item[T11] 100 percent tariff $+$ executive order to be signed six months later $+$ high-frequency semantic progression.
\item[T12] 10 percent tariff $+$ immediate implementation $+$ the identical post attributed to Harris.
\end{description}

The twelve arms are designed for comparisons along six dimensions. Within each dimension, only the corresponding key element is changed, while the remaining elements are held as constant as natural-language construction allows. Table~\ref{tab:treatment-design} reports these pairings. This organization is consequential for interpretation: T1--T3 isolate implementation timing; T2, T4, and T5 isolate the announced rate or its precision; T4, T10, and T11 hold the final message fixed while changing the preceding sequence; T1, T6, and T7 change semantic complexity; T1, T8, and T9 change the causal narrative; and T1 and T12 change only the named sender.

\begin{table}[htbp]
\centering
\begin{threeparttable}
\caption{Treatment Design by Communication Dimension}\label{tab:treatment-design}
\begin{tabular}{cll}
\toprule
No. & Communication dimension & Comparison arms \\
\midrule
1 & Implementation time & T1, T2, T3 \\
2 & Tariff rate & T2, T4, T5 \\
3 & Semantic coherence & T4, T10, T11 \\
4 & Semantic complexity & T1, T6, T7 \\
5 & Narrative & T1, T8, T9 \\
6 & Sender & T1, T12 \\
\bottomrule
\end{tabular}
\begin{tablenotes}[flushleft]\footnotesize
\item Notes: Within each row, the listed arms differ only in the feature named in column 2 to the extent permitted by natural-language treatment construction. T1 is the immediate 10 percent universal-tariff benchmark. T4 is the common terminal message for the semantic-coherence comparison. Full texts are reported in Supplementary Appendix Section~\ref{app:treatment-texts}.
\end{tablenotes}
\end{threeparttable}
\end{table}

To reduce researcher discretion in treatment construction and to reproduce the tone of Trump's actual tariff-related Truth Social posts, we use an LLM in a role-playing exercise to construct a Trump Agent. The Trump Agent is shown ten randomly selected genuine tariff posts from 2025 as few-shot examples and is instructed to generate messages satisfying the key elements of each scenario. Supplementary Appendix Section~\ref{app:trump-agent} reports the source, model and parameter choices, prompt, and T1 output; Section~\ref{app:treatment-texts} reports the controlled prompt and every message for T2--T12, including the six preceding posts in both T10 and T11.

After a Household Agent receives its assigned post, the survey elicits three types of responses for both inflation and unemployment: a point expectation, a distributional expectation, and an open-ended explanation. Q1 and Q4 ask for the agent's best point forecast over the next twelve months. Q2 and Q5 require the allocation of exactly 100 percentage points across ten mutually exclusive outcome bins, yielding the subjective probability distributions used to measure belief dispersion. Q3 and Q6 ask the agent to explain, in several complete sentences, why it holds the stated expectation; these texts are subsequently used to study the expectation-formation process. The point and distribution questions follow the corresponding formats and wording in the MSC and the Federal Reserve Bank of New York's Survey of Consumer Expectations, consistent with established practice for eliciting probabilistic expectations \citep{manski2004measuring}. Supplementary Appendix Section~\ref{app:survey} reproduces all six questions and every probability bin verbatim.

\subsection{Empirical Model}\label{subsec:empirical-model}

We estimate monthly Dynamic Treatment Effects (DTEs) with an event-study difference-in-differences specification:
\begin{equation}
Y_{im}=\alpha+\sum_{g=1}^{12}\sum_{k=-6}^{-2}\mu_{gk}D_{im}^{gk}
+\sum_{g=1}^{12}\sum_{k=0}^{11}\beta_{gk}D_{im}^{gk}
+\lambda_m+\eta_i+\varepsilon_{im},
\label{eq:dynamic-did}
\end{equation}
where $Y_{im}$ is agent $i$'s point forecast of inflation or unemployment in month $m$, $g$ indexes treatment arms, and
$D_{im}^{gk}=\mathbf{1}\{i\in g\}\mathbf{1}\{m-7=k\}$. Agent fixed effects $\eta_i$ absorb time-invariant simulated heterogeneity, and month fixed effects $\lambda_m$ absorb shocks common to all arms. The month immediately before treatment, $k=-1$, is omitted. The lead coefficients $\mu_{gk}$ assess pre-treatment alignment; $\beta_{gk}$ measure the treatment-control difference in event month $k$.

The second outcome summarizes disagreement in subjective distributions. Let $p_{bim}$ be agent $i$'s probability assigned to bin $b$ in month $m$, $\bar p_{bm}=N_m^{-1}\sum_i p_{bim}$, and $B=10$. We define
\begin{equation}
\operatorname{Disp}_m=\sum_{b=1}^{B}\bar p_{bm}
\left[\frac{1}{N_m}\sum_{i=1}^{N_m}\left(p_{bim}-\bar p_{bm}\right)^2\right].
\label{eq:belief-dispersion}
\end{equation}
The inner term is cross-sectional dispersion at a given probability bin, while the weight $\bar p_{bm}$ puts greater mass on bins that matter for the group's stated distribution. The index is zero only when agents submit identical probability vectors and rises when probability mass is allocated differently across households.

Disagreement is related to, but not identical to, aggregate uncertainty. In surveys, cross-sectional forecast dispersion can reveal heterogeneous information sets, models, or signal weights and often co-moves with measures of uncertainty \citep{mankiw2004disagreement}. Equation~\eqref{eq:belief-dispersion} therefore serves as a proxy for the coordination problem faced by a policymaker: a larger value means the same public signal leaves households with more divergent probability assessments. It should not be read as the average within-household variance, nor as a complete welfare measure.\footnote{The distinction matters empirically. Two households may each report high individual uncertainty yet assign the same probabilities, producing low disagreement; conversely, concentrated but opposing forecasts produce high disagreement. We use \emph{belief dispersion} throughout and invoke aggregate uncertainty only in this qualified proxy sense.}

\section{Multi-Agent System}\label{sec:mas}

This section presents the detailed architecture, calibration procedure, and simulation-validity assessment of the Multi-Agent System. Because the LLM Agents are designed to simulate the macroeconomic expectations of households, we call them \emph{Household Agents}. The architecture begins from the empirical determinants of household expectations rather than from an abstract conversational-agent template. A large body of survey evidence documents systematic differences in expectations across age, gender, political affiliation, education, and income groups \citep{souleles2004expectations,ehrmann2017consumers,bendavid2018expectations,coibion2022monetary,dacunto2024meaningful}. Those characteristics therefore enter the persona of every Household Agent.

A second determinant is the agent's prior expectation or perception of the economic variable, especially its most recent perception \citep{jonung1981perceived,coibion2020inflation}. In a Bayesian-updating account, an economic agent forms a posterior expectation by trading off its prior against newly received signals. Here, new signals are external information obtained during the current month. Traditional media affect household macroeconomic expectations \citep{carroll2003macroeconomic,lamla2012role}; with the rise of social media, continually updated posts have become an increasingly important route through which nonexpert audiences encounter, relay, and interpret economic news \citep{coibion2022monetary,ehrmann2022central,angelico2022measure,gorodnichenko2025central}. Accordingly, the external information in our system consists primarily of posts published by other Household Agents and, in Month 7, the experimental post attributed to Trump or Harris.

The relative weight placed on a new signal depends on confidence in the prior. When agents are highly confident, they can overweight the prior and underreact to current information, the conservatism bias. When confidence is weak, agents can underweight prior information and rely excessively on the new signal, producing over-updating or base-rate neglect \citep{chan2025prior,benjamin2019errors}. Supplementary Appendix Section~\ref{app:foundations} derives this relation in a Gaussian signal-extraction model and maps confidence to perceived prior precision. The derivation is not used as a structural estimating equation; it provides the economic discipline for the Household Agent instruction.

We translate these foundations into the common prompt below. First, each LLM Agent is explicitly assigned the role of an ordinary household and one of five confidence categories, ranging from extremely weak to extremely strong. Second, the prompt states the substantive task: estimating the future development of U.S. unemployment and inflation. Its introductory wording follows experimental work that elicits households' macroeconomic expectations and develops macroeconomic analysis disciplined by measured expectations, which helps avoid an idiosyncratic or technically leading formulation \citep{andre2022subjective,luetticke2026macro}. Third, the prompt requires the Household Agent to trade off its prior expectations and perceptions against social-media information according to the assigned confidence level. Finally, because demographic characteristics are closely associated with observed expectations, it directs the agent to make every answer reflect the personal characteristics of the role it portrays. Consistent with persona-based approaches in recent generative-AI studies of human behavior and beliefs \citep{horton2023large,delriochanona2025generative,kazinnik2026bank,manning2026general}, each persona is populated with characteristics from an actual sampled MSC respondent. The full field definitions are reported in Supplementary Appendix Section~\ref{app:persona}.

The common Household Agent instruction, reproduced here because it governs every monthly response, is:

\begin{tcolorbox}[colback=white,colframe=black,boxrule=0.8pt,arc=0pt,breakable]
\begin{verbatim}
Suppose you are an ordinary individual (household) with {CONF}
confidence designed to participate in a survey involving your beliefs
about the future development of the US economy.

This month is Month {N} of this survey. Your task will be to estimate
the development of the unemployment rate and the inflation rate in the
following questions. Please give us your best guess about how both rates
in the US economy would actually develop. This may or may not be in line
with theoretical findings and evidence from economics. We are only
interested in your own views and opinions on the US economy.

IMPORTANT INSTRUCTIONS: Your responses should trade off among the
various pieces of information mentioned above in accordance with your
level of confidence: If you are sufficiently confident, your answers
will rely on Prior Expectations & Perceptions, and will be less
influenced by other information, such as the Social Media Information.
On the other hand, if you lack confidence, your answers are more likely
to be influenced by other information. In addition, your responses
should fully reflect the Personal Characteristics (such as age, gender,
educational level, political affiliation, etc.) of the role you are
portraying.
\end{verbatim}
\end{tcolorbox}

Here \texttt{CONF} takes one of five ordered values: extremely weak, weak, moderate, strong, or extremely strong. The placeholder \texttt{N}$\in\{1,\ldots,18\}$ denotes the current simulation month. The instruction is common across agents and months; heterogeneity enters through the substituted confidence level, persona, prior state, social-media information, memory, treatment information, and decoding parameters.

\paragraph{Prior expectations and perceptions.} Post-training based on human feedback can cause an LLM to concentrate numerical forecasts within a narrow interval. To prevent such artificial anchoring and to make each simulated household begin from an observed belief, we assign every Household Agent the December 2024 inflation and unemployment expectations of its matched respondent in the stratified MSC sample. December 2024 is treated as Month 0. This month is selected because household expectations were relatively stable at the end of 2024 and because the calibration window did not contain a comparably large tariff announcement that would mechanically duplicate the treatment shock. Beginning in Month 2, the priors are endogenous: the inflation and unemployment expectations generated by the same Household Agent in the previous month become its current prior expectations and perceptions. The system thus preserves a respondent-specific belief history rather than reinitializing the model in every round. Supplementary Appendix Section~\ref{app:priors} provides the complete calibration prompt, the exact MSC fields, and the month-to-month transition rule. Personal experience is represented in the prompt because grocery-price exposure, labor-market experience, and other lived signals are known to shape household expectations \citep{malmendier2016learning,kuchler2019personal}.

\paragraph{Social Media Information Module.} Each Household Agent obtains current external information from the Social Media Information Module (SMIM). In Month 1, the module is seeded with cleaned, high-attention X posts collected in December 2024 using the search terms ``US Inflation'' and ``US Unemployment.'' From Month 2 onward, the eligible information set consists of posts generated and published by other Household Agents in the preceding month. Thus, agents do not merely receive exogenous text; their previous statements become the social information that other agents may encounter.

The SMIM uses a generative-AI recommendation process to create a simplified social network and an information-cocoon channel. In each month, a stochastic allocation assigns share $\alpha$ of agents to the recommendation branch. For an agent $i$ assigned to that branch, semantic similarity first retrieves the ten preceding-month posts closest to the post that $i$ itself published in that month. A separate Recommender Agent then compares viewpoints within this candidate set and chooses the post most consistent with $i$'s preceding post. The selected post is recommended to $i$ in the current month. For the remaining share $1-\alpha$, the module randomly selects a post published by another Household Agent in the preceding month. This mixture captures selective exposure without making the network perfectly segregated; recommendation algorithms and partisan or preference-congruent exposure can otherwise deepen polarization and generate echo chambers \citep{santos2021link,levy2021social,cookson2023echo}. Supplementary Appendix Section~\ref{app:smim} reports the initial-post collection, cleaning criteria, exposure prompt, and precise distinction between the recommendation and random branches. Section~\ref{app:parameters} explains why the baseline sets $\alpha=0.50$ and reports comparisons with $\alpha=0.10$ and $0.90$.

\paragraph{Memory and dynamic iteration.} Every Household Agent has a memory module that stores recently supplied and generated information: persona and confidence, posts read and published, point and distributional forecasts, open-ended explanations, and treatment-period messages. Without persistent memory, later months would be independent one-shot responses rather than decisions conditioned on recent interactions and prior states. The baseline window retains the latest three monthly rounds. This choice imposes recency, prevents older context from overwhelming new information, and reflects evidence that recent macroeconomic experience receives disproportionate weight in expectation formation \citep{malmendier2016learning,bordalo2022overreaction}. Supplementary Appendix Figure~\ref{fig:memory-sensitivity} compares windows of 3, 10, and 17 months. Their DTE paths are similar, so the short window preserves the substantive conclusions while reducing context-window consumption.

\paragraph{Decoding parameters and model choice.} Unobserved traits and idiosyncratic influences remain even after personas, priors, and information exposure are specified. We therefore introduce agent-level heterogeneity into temperature and top-$p$, the two decoding parameters that govern complementary dimensions of text-generation diversity. Their draws are fixed for each agent across treatment arms, so treatment comparisons do not confound a change in message content with a newly drawn decoding rule. Supplementary Appendix Section~\ref{app:parameters} formalizes the distributions, boundary rules, and sensitivity checks and summarizes all baseline settings in Table~\ref{tab:parameter-summary}. Because the foundation model is the core computational mechanism through which inputs become forecasts and texts, Supplementary Appendix Section~\ref{app:model-choice} also compares five candidate LLMs under identical prompts and selects Qwen3.5 Plus using pre-treatment and T1 distributional-shape similarity, openness, reproducibility, and cost as prespecified criteria.

\begin{figure}[htbp]
\centering
\includegraphics[width=\linewidth]{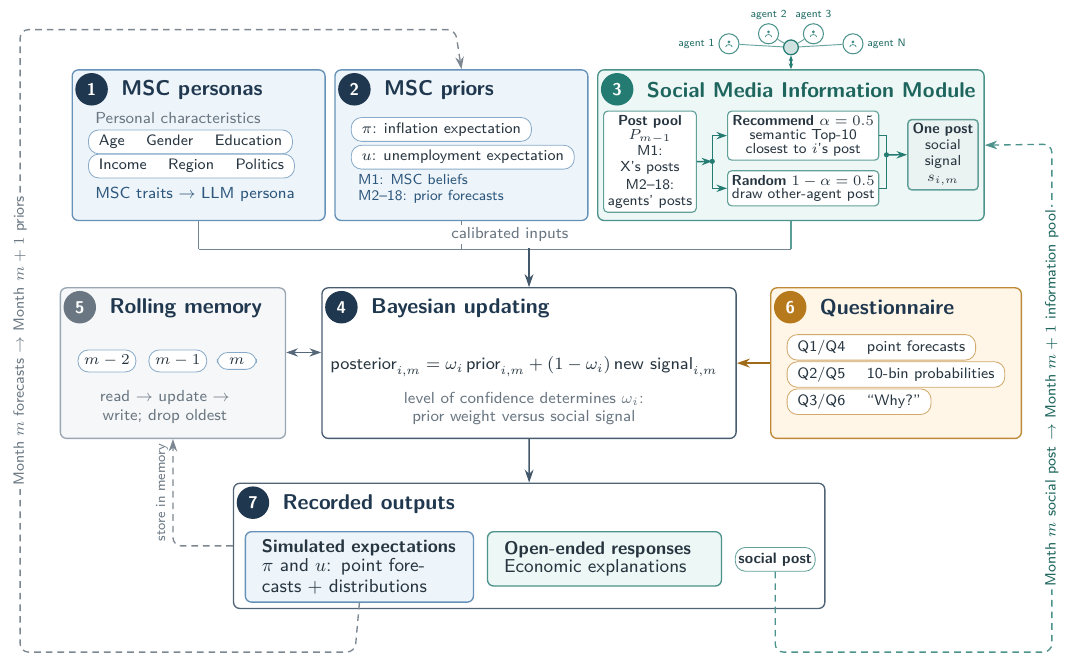}
\caption{Observed-Data Calibration and Monthly Information Flow in the Multi-Agent System}\label{fig:mas-architecture}
\notepar{Notes: Connectors indicate within-round information flow. Blocks 2, 3, and 5 encode monthly carryover: prior forecasts become current priors, published posts replenish the SMIM, and the memory retains the latest three months. The baseline uses 300 Household Agents over 18 months and recommendation share $\alpha=0.50$.}
\end{figure}

Figure~\ref{fig:mas-architecture} summarizes the MAS as seven numbered components: two MSC-based calibration inputs, the SMIM, Bayesian household updating, rolling memory, the questionnaire, and recorded outputs. The connectors emphasize the main within-round information flow, while the prior, SMIM, and memory blocks summarize the month-to-month state transitions.

After constructing and calibrating the MAS, we assess simulation validity in the benchmark T1 scenario by comparing uncalibrated and calibrated MAS expectations with the corresponding MSC expectations. The exercise has three components. First, repeated-subsample Cram\'{e}r--von Mises tests ask whether the simulated and human samples display statistically distinguishable distributions without allowing the full sample size to make economically negligible deviations mechanically significant. Second, four machine-learning classifiers ask whether a flexible prediction rule can distinguish a simulated observation from a human observation using point forecasts, the ten-bin probability distribution, and demographic covariates. Third, heterogeneity regressions ask whether the simulated data reproduce characteristic differences in expectations across political affiliation, income, and gender. Supplementary Appendix Sections~\ref{app:validation-benchmark}--\ref{app:heterogeneity} report the exact procedures, equations, subsampling and split rules, regression specifications, figures, and tables.

The validation does not assert equality between simulated and human samples. Quantitative gaps remain. The narrower result is that calibration materially improves agreement on all three tested margins: standard nonparametric procedures have difficulty consistently rejecting distributional similarity at the chosen repeated-subsample scale; multiple classifiers operate close to chance relative to the uncalibrated benchmark; and the calibrated system reproduces the direction and broad magnitude ordering of salient political, income, and gender heterogeneity that the uncalibrated system fails to recover. This pattern supports the effectiveness of the data calibration, persona design, SMIM, and parameter selection for the observed T1 environment.

Direct validation is possible only for T1. The remaining treatment scenarios were constructed as counterfactual variants and therefore lack corresponding human panel data. We consequently follow the heuristic-validation, or validate-then-simulate, logic described by \citet{hullman2026human}: evidence that an AI-agent system approximates human responses in a relevant benchmark can motivate simulation in nearby settings, but it does not provide a strict transport guarantee. In particular, T1 validation cannot establish that the level of a treatment effect is unbiased in a new arm.

Therefore, our substantive use is deliberately narrower. We ask whether qualitative findings, including the sign and relative ordering of DTEs and belief dispersion, extend to treatments that alter bounded details of the validated treatment information while holding personas, priors, survey questions, model, memory, and social-information modules fixed. Supplementary Appendix Section~\ref{app:extrapolation} formalizes the MAS response mapping, distinguishes benchmark fit from treatment-effect transport, and explains the local-regularity intuition. It also classifies claims into three tiers. T1 receives direct human-benchmark validation. Treatments that are semantically close to T1 and change a single message feature may support cautious qualitative near-generalization when numerical, textual, and theoretical evidence agree. More distant treatments, especially the multi-message paths T10 and T11, the sender change T12, and all central-bank strategies C1--C5, are scenario explorations and mechanism hypotheses. Their outputs are not unconditional causal evidence about human households and are not direct policy recommendations. Section~\ref{sec:discussion} returns to these boundaries.

\section{Simulation Results and Analysis}\label{sec:results}

Having established the experimental design, the dynamic estimands, and the construction and benchmark validity of the MAS, we now use the simulated panel to study how the six dimensions of tariff-threat communication shape household inflation and unemployment expectations. The first objective is descriptive: Section~\ref{subsec:simulation-results} reports, for every treatment dimension, the monthly DTEs and the associated evolution of belief dispersion. Because the treatments differ in timing, rate, message history, complexity, narrative, or sender, the analysis emphasizes controlled within-dimension comparisons rather than a single ranking across all twelve arms.

The second objective is explanatory. A simulated treatment-effect path is more informative when the underlying expectation-formation process is economically intelligible. Section~\ref{subsec:mechanisms} therefore combines two forms of evidence. We first compare the treatment rankings with established theoretical and empirical findings and then analyze the Household Agents' open-ended explanations to determine whether their stated attention, ambiguity, credibility judgments, extrapolation, and causal narratives move in the directions implied by those findings. The exercise evaluates qualitative coherence, not an unqualified causal transport of the simulated magnitudes to human households.

\subsection{Simulation Results}\label{subsec:simulation-results}

We compare the treatment scenarios in order across the six dimensions in Table~\ref{tab:treatment-design}. Because exact quantitative extrapolation is demanding, the analysis focuses on qualitative comparisons: the post-treatment ordering of DTEs from Equation~\eqref{eq:dynamic-did} and belief dispersion from Equation~\eqref{eq:belief-dispersion}.

\begin{figure}[htbp]
\centering
\includegraphics[width=.97\linewidth]{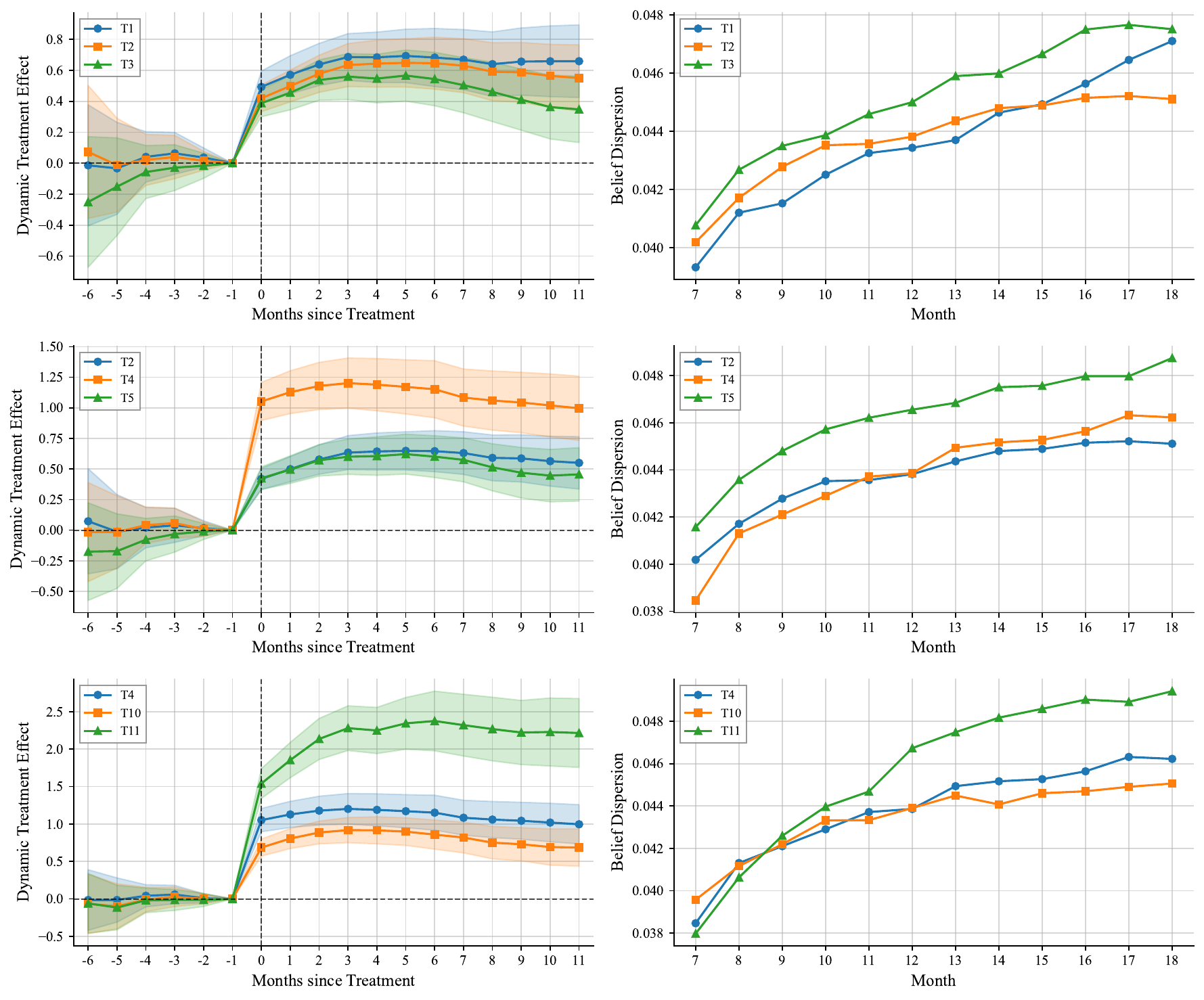}
\caption{Inflation Expectations: Implementation Time, Tariff Rate, and Semantic Coherence}\label{fig:results-dim13}
\notepar{Notes: Rows correspond to Dimensions 1--3. Left panels report monthly Dynamic Treatment Effects on twelve-month-ahead inflation expectations relative to the placebo arm. Right panels report belief dispersion from Equation~\eqref{eq:belief-dispersion}. Treatment occurs in Month 7. Shaded bands are 95\% confidence intervals.}
\end{figure}

Figure~\ref{fig:results-dim13} reports the inflation-expectation results for Dimensions 1--3. In Dimension 1, the DTE ordering is T1 $>$ T2 $>$ T3: an immediately effective tariff produces the largest upward revision, a tariff taking effect six months later produces the next largest revision, and a tariff with uncertain implementation timing produces the smallest revision. Belief dispersion is clearly higher in T3 than in T1 or T2. In Dimension 2, the DTE ordering is T4 $>$ T2 $>$ T5. Relative to the low-rate information in T2, the high-rate information in T4 produces a larger upward revision, whereas the rate-uncertain information in T5 produces a smaller revision. Belief dispersion is clearly higher in T5 than in T2 or T4. In Dimension 3, the DTE ordering is T11 $>$ T4 $>$ T10. Relative to the one-time T4 announcement, high-frequency messages with semantic progression produce a larger upward revision, whereas high-frequency messages with semantic reversals produce a smaller revision. Belief dispersion is clearly higher in T11 than in T4 or T10.

The corresponding unemployment-expectation figures are reported in Supplementary Appendix Section~\ref{app:additional-results}, beginning with Figure~\ref{fig:unemp-dim13}. Their patterns and qualitative conclusions are consistent with the inflation results above.

\begin{figure}[t]
\centering
\includegraphics[width=.97\linewidth]{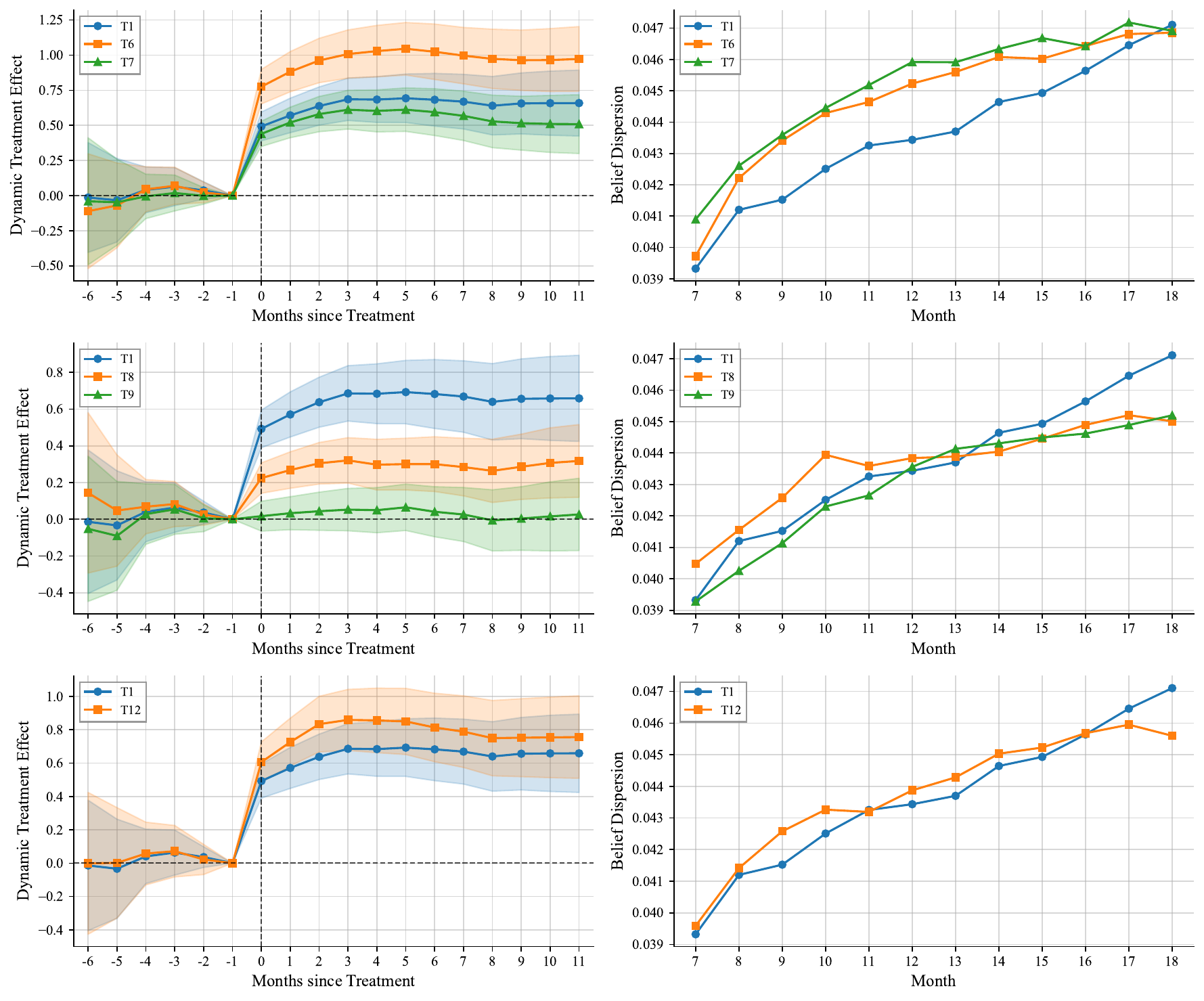}
\caption{Inflation Expectations: Complexity, Narrative, and Sender}\label{fig:results-dim46}
\notepar{Notes: Rows correspond to Dimensions 4--6. Left panels report monthly Dynamic Treatment Effects on twelve-month-ahead inflation expectations relative to the placebo arm. Right panels report belief dispersion. Treatment occurs in Month 7. Shaded bands are 95\% confidence intervals.}
\end{figure}

Figure~\ref{fig:results-dim46} reports the inflation-expectation results for Dimensions 4--6. In Dimension 4, the DTE ordering is T6 $>$ T1 $>$ T7. Relative to the standard wording in T1, simple wording produces a larger upward revision and complex wording produces a smaller revision. Belief dispersion is clearly higher in both T6 and T7 than in T1. In Dimension 5, T1 produces a larger DTE than T8 or T9: both the MAGA narrative and the tax-incidence narrative attenuate the upward revision relative to the no-narrative benchmark. Belief dispersion is similar across the three scenarios, indicating that narrative choice has limited effects on dispersion. In Dimension 6, T12 produces a larger DTE and higher belief dispersion than T1. Thus, the same tariff information attributed to the Democratic leader Harris generates a larger upward revision and greater disagreement than when it is attributed to the Republican leader Trump.

The corresponding unemployment-expectation figures are reported in Supplementary Appendix Figure~\ref{fig:unemp-dim46}. Their patterns and qualitative conclusions are consistent with the inflation results above.

\begin{table}[htbp]
\centering
\begin{threeparttable}
\caption{Qualitative Ranking of Post-Treatment Outcomes}\label{tab:qualitative-rankings}
\begin{tabularx}{\linewidth}{clXX}
\toprule
No. & Dimension & DTE ranking & Belief-dispersion ranking \\
\midrule
1 & Implementation time & T1 $>$ T2 $>$ T3 & T3 $>$ T1, T2 \\
2 & Tariff rate & T4 $>$ T2 $>$ T5 & T5 $>$ T2, T4 \\
3 & Semantic coherence & T11 $>$ T4 $>$ T10 & T11 $>$ T4, T10 \\
4 & Semantic complexity & T6 $>$ T1 $>$ T7 & T6, T7 $>$ T1 \\
5 & Narrative & T1 $>$ T8, T9 & Similar across T1, T8, T9 \\
6 & Sender & T12 $>$ T1 & T12 $>$ T1 \\
\bottomrule
\end{tabularx}
\begin{tablenotes}[flushleft]\footnotesize
\item Notes: Rankings summarize the main post-treatment patterns in Figures~\ref{fig:results-dim13} and \ref{fig:results-dim46}. A comma indicates no stable qualitative ordering. The corresponding unemployment results are reported in Supplementary Appendix Figures~\ref{fig:unemp-dim13} and \ref{fig:unemp-dim46} and display the same broad ordering.
\end{tablenotes}
\end{threeparttable}
\end{table}

Taken together, the comparisons across the six dimensions yield the qualitative rankings summarized in Table~\ref{tab:qualitative-rankings}.

\subsection{Mental Mechanism Analysis}\label{subsec:mechanisms}

The numerical rankings alone do not reveal why Household Agents respond differently to messages that describe related tariff policies. We therefore organize the mechanism analysis in two consecutive stages. Section~\ref{subsubsec:literature} places each of the six treatment comparisons against established economic theory and human empirical evidence. The purpose is to determine whether the direction and ordering of the MAS responses have an externally intelligible economic basis, including temporal discounting and salience, signal precision and ambiguity, extrapolation and credibility, processing cost, causal narratives, and sender-dependent priors.

Section~\ref{subsubsec:text} then turns inward to the simulated data. Following the economics literature on open-ended survey responses, we treat the agents' explanations as observable accounts of the considerations, subjective models, and narratives associated with their numerical forecasts. Dictionary-based term densities, manually coded causal frames, and semantic-embedding measures test whether the psychological and informational channels identified in the literature actually appear in the generated responses and whether they reproduce the DTE and belief-dispersion rankings over time. Agreement between the external literature and the internal text provides cross-evidence that the simulation is qualitatively self-consistent.

\subsubsection{Theoretical and Empirical Explanations from the Literature}\label{subsubsec:literature}

This subsection asks whether the qualitative rankings in Table~\ref{tab:qualitative-rankings} agree with established theoretical or empirical findings. The comparison supplies cross-evidence for the simulation and an economic interpretation of each experimental dimension. It does not independently validate the counterfactual treatment effects.

\paragraph{Dimension 1: implementation time.} The DTE ranking resembles the empirical side of the forward-guidance puzzle. Standard New Keynesian models can imply implausibly large effects of policy promised far in the future, whereas observed agents respond much less to remote signals \citep{delnegro2023forward}. Consistent with this fact, Household Agents react strongly to the immediately effective tariff in T1 and more weakly to the otherwise similar tariff announced for six months later in T2. The ordering is also consistent with cognitive discounting in the Behavioral New Keynesian Model: myopic agents attach an additional cognitive discount to future macroeconomic shocks because distant events are harder to process and may be revised before implementation \citep{gabaix2020behavioral}.

T3 produces the smallest DTE because ``at a time to be determined'' supplies neither an immediate deadline nor a reliable future date. In salience theory, decision weight is tilted toward attributes that stand out in the choice context \citep{bordalo2012salience}; in rational inattention, scarce processing capacity is directed toward signals with the greatest expected decision value \citep{sims2003implications}. Immediate implementation is therefore salient and actionable, while an unresolved date is easier to treat as a low-probability or non-credible threat. The same ambiguity explains the dispersion ranking. Policy uncertainty rises when timing and implementation details are unresolved \citep{baker2016measuring}. Under ambiguity aversion, agents need not agree on a probability distribution over the missing date: some read T3 as an empty threat, whereas others update toward a worst-case implementation path \citep{ilut2014ambiguous}. T1 and T2 differ in horizon but retain a common temporal anchor, so beliefs are more coordinated around the stated date.

\paragraph{Dimension 2: tariff rate.} T4 combines an explicit rate with an extreme magnitude. Survey experiments show that salient price information can penetrate household information frictions, although households continue to rely partly on personal price memories \citep{cavallo2017inflation}; information-rigidity evidence likewise predicts more complete updating when a new signal is sufficiently informative \citep{coibion2015information}. A stated 100 percent tariff therefore generates a larger DTE than the stated 10 percent rate in T2. T5 instead says that the rate could be very high or very low. This high-noise signal receives less weight in a precision-weighted update, so the average response is smaller than under T2 even though agents do not ignore it completely.

The dispersion result follows the classical distinction between measurable risk and Knightian uncertainty \citep{knight1921risk}. T2 and T4 provide common numerical anchors. T5 does not, so agents complete the missing rate using heterogeneous priors, political beliefs, and economic models. Models of public information show why a low-precision common signal can leave greater room for private information and therefore less convergence \citep{morris2002social}. Thus a larger stated rate need not create more disagreement: signal precision can coordinate beliefs around a large shock, whereas an unspecified rate supports multiple interpretations.

\paragraph{Dimension 3: semantic coherence.} T11 generates the largest DTE because its six progressively higher tariff rates create a trend that agents can extrapolate. Extrapolative-expectations models and evidence show that recent directional sequences affect forecasts beyond the terminal observation \citep{barberis1998model,greenwood2014expectations}. Household Agents therefore extract not only the final 100 percent rate shared with T4, but also an escalating path that suggests further deterioration.

T10 produces a smaller response because repeated imposition and cancellation increases the perceived noise of the terminal announcement. In a Bayesian interpretation, the history lowers the precision assigned to a new statement. In strategic communication, costless messages lose information when the sender cannot commit and preferences or incentives are not aligned \citep{crawford1982strategic}. Reversal therefore erodes reputation capital and creates a ``boy who cried wolf'' mechanism. A February 24, 2026 financial report gives a corresponding real-world example: investors were described as discounting later tariff announcements after earlier threats had been eased, delayed, or cancelled.\footnote{See CNBC, ``Trump tariff rates, global investors and the TACO trade,'' February 24, 2026. The episode is used as factual context for the credibility mechanism, not as causal validation of T10.}

T11 also has the largest belief dispersion because an escalating path destroys the perceived upper-bound anchor. Models of government-policy uncertainty emphasize that uncertainty about policy rules and changes raises volatility and tail risk \citep{pastor2012uncertainty}. Once agents observe rates rising from 40 to 100 percent, they disagree not only about whether the terminal rate will be implemented but also about whether escalation will stop. Different upper-tail scenarios widen subjective probability distributions even though every published rate is numerically precise.

\paragraph{Dimension 4: semantic complexity.} T6 communicates the rate, scope, and timing in fewer than ten words. Low processing cost allows the core signal to pass quickly through the attention constraint, consistent with evidence that public communication works better when it is concise, simple, and relatable \citep{haldane2018central,bholat2019enhancing}. T7 uses statutory and technical expressions such as ``pursuant to statutory authority,'' ``ad valorem tariff,'' and ``macro-prudential review.'' Experimental evidence shows that complex language reduces nonexpert comprehension, while quantitative, logical, and verbal abilities predict the accuracy and plausibility of household inflation expectations \citep{bholat2019enhancing,dacunto2019cognitive}. These mechanisms account for the DTE ordering T6, T1, T7.

Both extremes can nevertheless increase belief dispersion. T6 states the key numbers but supplies almost no purpose, condition, duration, or causal context. Agents must fill those gaps with heterogeneous mental models. T7 creates a different problem: agents vary in their ability to decode technical language and may ignore it, interpret it correctly, or mistake review language for an indication that implementation is conditional. The common signal is therefore less precise in practice in both arms, for different reasons, and the education heterogeneity in Figure~\ref{fig:education-heterogeneity} is consistent with that account.

\paragraph{Dimension 5: narrative.} The MAGA and exporter-incidence narratives damp the expectation response to the tariff. This result is consistent with evidence that households often use simple subjective models rather than expert macroeconomic mappings when predicting shock transmission \citep{andre2022subjective}. T9 supplies a direct causal claim that foreign exporters bear the full cost, despite evidence that U.S. tariff costs are substantially passed through to domestic importers and consumers \citep{amiti2019impact}. T8 supplies a positive account in which reshoring raises employment and reduced import dependence ultimately stabilizes inflation. Agents who accept either narrative revise less than agents who receive T1 without an offsetting causal explanation.

Narrative Economics explains why such causal stories can matter independently of the policy number: memorable accounts spread through social interaction and organize economic action even when they are incomplete or inaccurate \citep{shiller2017narrative}. Yet the three narrative arms have no stable belief-dispersion ranking. Their rate, timing, and implementation remain explicit, so the narratives chiefly reallocate agents among causal frames rather than remove the common policy anchor. Heterogeneity in frame choice can still occur, but the added stories need not systematically widen numerical probability distributions.

\paragraph{Dimension 6: sender.} T1 and T12 use identical text, so their difference must arise from the identity attached to the message. Sender identity changes the prior probability that the policy is intended, implementable, and persistent. In this calibrated population, a universal tariff announcement by Harris is both more unexpected and, in the generated explanations, more credible than the same announcement by Trump. The first channel makes the signal more surprising relative to prior political expectations; the second gives it a larger perceived precision. Evidence that political identity shapes macroeconomic beliefs makes such source-dependent updating plausible \citep{kamdar2025think,hirs2026partisanship}. The unusual sender also generates more dispersed interpretations because agents lack a familiar historical template for a Democratic administration announcing a universal tariff. These conclusions are conditional on the sampled population, political period, and model training data, and the sender comparison remains scenario exploration rather than externally validated human causal evidence.\footnote{Sender effects are especially vulnerable to changes in political control, target-population composition, and the foundation model's training corpus. They should be revalidated whenever any of those objects changes.}

\subsubsection{Analysis of Open-Ended Responses}\label{subsubsec:text}

Open-ended survey responses can reveal considerations, motives, mental models, narratives, attention, and information-processing mechanisms that are not observable in closed-ended numerical answers \citep{haaland2025understanding}. We therefore analyze the explanations elicited in Q3 and Q6, rather than treating them as decorative rationales. The purpose is to determine whether the mental models expressed by Household Agents qualitatively explain why DTEs and belief dispersion take the rankings reported above. We proceed dimension by dimension so that each text measure corresponds to the single treatment feature varied in that comparison. Supplementary Appendix Section~\ref{app:text-measures} reports every dictionary, causal-frame definition, and semantic-embedding operation.

For Dimensions 1--3 and 6, the central objects can be measured transparently through normalized dictionary densities. This measurement strategy follows the macro-finance text-analysis and household-expectations literatures and standard text-as-data practice \citep{caldara2020economic,weber2022subjective,gentzkow2019text}. Each response is converted to lowercase, leading and trailing whitespace and repeated internal spaces are removed, non-alphanumeric punctuation is stripped except where needed to preserve ordered phrases, and the resulting text is tokenized. Let $\mathbf w_{igt}=(w_{igt1},\ldots,w_{igtL_{igt}})$ denote the ordered tokens in the open-ended response of agent $i$, treatment arm $g$, and month $t$, and let $\mathcal K_k$ be the prespecified dictionary for mechanism $k$. The individual-level term-density measure is
\begin{equation}
S_{i,g,t}^{(k)}=\frac{1}{N_{i,g,t}^{(k)}}
\sum_{q\in\mathcal D_k}\mathbf C\!\left(q,y_{i,g,t}^{(k)}\right).
\label{eq:term-density}
\end{equation}

A dictionary entry may be a unigram or a multiword $n$-gram. Here $y_{i,g,t}^{(k)}$ is the cleaned response string, $N_{i,g,t}^{(k)}$ is its token count, $\mathcal D_k$ is the dictionary for mechanism $k$, and $\mathbf C(q,y)$ is the frequency-count function. For a unigram $q$, $\mathbf C$ counts its occurrences in the ordered token sequence; for an $n$-gram, it counts nonoverlapping exact matches of the full ordered phrase. The numerator is therefore the total number of matched dictionary occurrences, and division by $N_{i,g,t}^{(k)}$ prevents a longer answer from receiving a mechanically larger score. The dictionaries measure tariff awareness, temporal urgency, ambiguity perception, perceived price impact, perceived employment impact, extrapolation, skepticism, credibility, and unexpectedness. They were constructed ex ante from the economic mechanisms emphasized in the leading literature on household expectations, tariff transmission, ambiguity, extrapolation, and credibility and were then audited against concordance lines to remove terms with unstable meanings. Supplementary Appendix Section~\ref{app:dictionaries} provides the complete term sets and a mechanism-by-mechanism construction rationale, as required for replicability.

To determine whether tariff information systematically changes the focus of attention within a treatment arm, we aggregate the individual score to the treatment-by-month level:
\begin{equation}
\bar S_{g,t}^{(k)}=\frac{1}{|I_{g,t}|}\sum_{i\in I_{g,t}}S_{i,g,t}^{(k)},
\label{eq:mean-term-density}
\end{equation}
where $I_{g,t}$ denotes the Household Agents observed in treatment arm $g$ in month $t$. Within the relevant treatment dimension, we rank arms each month as maximum, intermediate, or minimum according to $\bar S_{g,t}^{(k)}$. These are dynamic relative rankings. A minimum does not mean that the mechanism is absent, and a ranking alone is not a statistical mediation estimate. It establishes whether the attention or interpretation embedded in the response text moves in the same qualitative order as the numerical outcomes. Table~\ref{tab:mechanism-rankings-1} reports the complete monthly rankings for Dimensions 1--3.

\begingroup
\scriptsize
\setlength{\tabcolsep}{1.0pt}
\renewcommand{\arraystretch}{0.90}
\begin{longtable}{@{}l@{\hspace{5pt}}llc*{12}{c}@{}}
\caption{Monthly Rankings of Text-Based Mechanisms: Dimensions 1--3}\label{tab:mechanism-rankings-1}\\
\toprule
& & & & \multicolumn{12}{c}{Month} \\
\cmidrule(lr){5-16}
Dimension & Mechanism & Expectation & Rank & 7 & 8 & 9 & 10 & 11 & 12 & 13 & 14 & 15 & 16 & 17 & 18 \\
\midrule
\endfirsthead
\multicolumn{16}{c}{\tablename\ \thetable\ (continued)}\\
\toprule
& & & & \multicolumn{12}{c}{Month} \\
\cmidrule(lr){5-16}
Dimension & Mechanism & Expectation & Rank & 7 & 8 & 9 & 10 & 11 & 12 & 13 & 14 & 15 & 16 & 17 & 18 \\
\midrule
\endhead
\midrule
\multicolumn{16}{r}{Continued on next page}\\
\endfoot
\bottomrule
\endlastfoot
\multirow{18}{*}{D1} & \multirow{6}{*}{Ambiguity Perception} & \multirow{3}{*}{Inflation} & Max & T3&T3&T3&T3&T3&T3&T3&T3&T3&T3&T3&T3 \\
&&& Med & T1&T1&T1&T2&T1&T1&T2&T2&T1&T1&T1&T1 \\
&&& Min & T2&T2&T2&T1&T2&T2&T1&T1&T2&T2&T2&T2 \\
\cmidrule(lr){3-16}
&& \multirow{3}{*}{Unemployment} & Max & T3&T3&T2&T3&T3&T3&T3&T3&T3&T3&T3&T3 \\
&&& Med & T1&T1&T3&T2&T1&T2&T1&T1&T2&T1&T2&T2 \\
&&& Min & T2&T2&T1&T1&T2&T1&T2&T2&T1&T2&T1&T1 \\
\cmidrule(lr){2-16}
& \multirow{6}{*}{Tariff Awareness} & \multirow{3}{*}{Inflation} & Max & T1&T1&T1&T1&T1&T1&T1&T2&T1&T1&T1&T1 \\
&&& Med & T2&T2&T2&T2&T3&T2&T2&T3&T3&T3&T2&T2 \\
&&& Min & T3&T3&T3&T3&T2&T3&T3&T1&T2&T2&T3&T3 \\
\cmidrule(lr){3-16}
&& \multirow{3}{*}{Unemployment} & Max & T2&T1&T1&T1&T2&T1&T1&T1&T1&T1&T2&T1 \\
&&& Med & T1&T3&T3&T2&T1&T2&T2&T2&T2&T2&T1&T2 \\
&&& Min & T3&T2&T2&T3&T3&T3&T3&T3&T3&T3&T3&T3 \\
\cmidrule(lr){2-16}
& \multirow{6}{*}{Temporal Urgency} & \multirow{3}{*}{Inflation} & Max & T1&T1&T1&T1&T1&T1&T1&T1&T1&T1&T1&T1 \\
&&& Med & T2&T2&T2&T2&T2&T2&T2&T2&T2&T2&T2&T2 \\
&&& Min & T3&T3&T3&T3&T3&T3&T3&T3&T3&T3&T3&T3 \\
\cmidrule(lr){3-16}
&& \multirow{3}{*}{Unemployment} & Max & T1&T1&T1&T1&T2&T2&T1&T2&T1&T1&T1&T3 \\
&&& Med & T2&T3&T2&T2&T3&T3&T2&T1&T2&T2&T2&T1 \\
&&& Min & T3&T2&T3&T3&T1&T1&T3&T3&T3&T3&T3&T2 \\
\midrule
\multirow{18}{*}{D2} & \multirow{6}{*}{Ambiguity Perception} & \multirow{3}{*}{Inflation} & Max & T5&T5&T5&T5&T5&T5&T5&T5&T5&T5&T5&T5 \\
&&& Med & T4&T4&T4&T2&T2&T4&T2&T4&T4&T4&T4&T4 \\
&&& Min & T2&T2&T2&T4&T4&T2&T4&T2&T2&T2&T2&T2 \\
\cmidrule(lr){3-16}
&& \multirow{3}{*}{Unemployment} & Max & T5&T4&T2&T5&T5&T5&T5&T5&T2&T5&T5&T5 \\
&&& Med & T2&T2&T5&T4&T4&T2&T4&T2&T4&T2&T2&T2 \\
&&& Min & T4&T5&T4&T2&T2&T4&T2&T4&T5&T4&T4&T4 \\
\cmidrule(lr){2-16}
& \multirow{6}{*}{Tariff Awareness} & \multirow{3}{*}{Inflation} & Max & T2&T5&T4&T4&T4&T4&T4&T4&T4&T4&T4&T4 \\
&&& Med & T4&T2&T5&T2&T2&T5&T2&T2&T2&T2&T2&T2 \\
&&& Min & T5&T4&T2&T5&T5&T2&T5&T5&T5&T5&T5&T5 \\
\cmidrule(lr){3-16}
&& \multirow{3}{*}{Unemployment} & Max & T5&T4&T4&T4&T4&T4&T4&T4&T4&T2&T4&T4 \\
&&& Med & T4&T5&T5&T2&T2&T5&T2&T2&T2&T4&T2&T2 \\
&&& Min & T2&T2&T2&T5&T5&T2&T5&T5&T5&T5&T5&T5 \\
\cmidrule(lr){2-16}
& \multirow{6}{*}{Price \& Employment Impact} & \multirow{3}{*}{Inflation} & Max & T5&T4&T4&T4&T4&T4&T4&T4&T4&T4&T4&T4 \\
&&& Med & T2&T2&T2&T2&T2&T5&T2&T2&T2&T2&T2&T2 \\
&&& Min & T4&T5&T5&T5&T5&T2&T5&T5&T5&T5&T5&T5 \\
\cmidrule(lr){3-16}
&& \multirow{3}{*}{Unemployment} & Max & T4&T4&T4&T4&T4&T4&T4&T2&T2&T4&T4&T4 \\
&&& Med & T2&T5&T2&T2&T2&T5&T2&T4&T4&T2&T2&T2 \\
&&& Min & T5&T2&T5&T5&T5&T2&T5&T5&T5&T5&T5&T5 \\
\midrule
\pagebreak[4]
\multirow{24}{*}{D3} & \multirow{6}{*}{Ambiguity Perception} & \multirow{3}{*}{Inflation} & Max & T4&T10&T11&T11&T11&T10&T11&T4&T10&T10&T10&T10 \\
&&& Med & T11&T4&T4&T10&T10&T11&T10&T11&T11&T4&T4&T11 \\
&&& Min & T10&T11&T10&T4&T4&T4&T4&T10&T4&T11&T11&T4 \\
\cmidrule(lr){3-16}
&& \multirow{3}{*}{Unemployment} & Max & T4&T4&T10&T4&T11&T10&T11&T10&T10&T11&T10&T11 \\
&&& Med & T10&T11&T4&T10&T10&T11&T4&T11&T4&T10&T4&T10 \\
&&& Min & T11&T10&T11&T11&T4&T4&T10&T4&T11&T4&T11&T4 \\
\cmidrule(lr){2-16}
& \multirow{6}{*}{Embedding Dispersion} & \multirow{3}{*}{Inflation} & Max & T10&T11&T10&T11&T11&T11&T11&T11&T11&T11&T11&T11 \\
&&& Med & T11&T10&T11&T10&T10&T4&T10&T10&T10&T10&T10&T10 \\
&&& Min & T4&T4&T4&T4&T4&T10&T4&T4&T4&T4&T4&T4 \\
\cmidrule(lr){3-16}
&& \multirow{3}{*}{Unemployment} & Max & T4&T11&T11&T11&T11&T11&T11&T11&T11&T11&T11&T11 \\
&&& Med & T11&T4&T10&T4&T10&T10&T10&T10&T4&T10&T10&T10 \\
&&& Min & T10&T10&T4&T10&T4&T4&T4&T4&T10&T4&T4&T4 \\
\cmidrule(lr){2-16}
& \multirow{6}{*}{Extrapolation} & \multirow{3}{*}{Inflation} & Max & T11&T11&T11&T11&T11&T11&T11&T11&T11&T11&T11&T10 \\
&&& Med & T4&T4&T4&T4&T4&T4&T4&T4&T4&T4&T4&T11 \\
&&& Min & T10&T10&T10&T10&T10&T10&T10&T10&T10&T10&T10&T4 \\
\cmidrule(lr){3-16}
&& \multirow{3}{*}{Unemployment} & Max & T11&T11&T11&T11&T11&T11&T11&T11&T4&T11&T11&T11 \\
&&& Med & T4&T10&T4&T4&T4&T4&T4&T4&T10&T4&T4&T10 \\
&&& Min & T10&T4&T10&T10&T10&T10&T10&T10&T11&T10&T10&T4 \\
\cmidrule(lr){2-16}
& \multirow{6}{*}{Skepticism} & \multirow{3}{*}{Inflation} & Max & T10&T10&T10&T10&T10&T10&T10&T10&T10&T10&T10&T10 \\
&&& Med & T4&T11&T11&T4&T11&T4&T4&T4&T11&T4&T4&T11 \\
&&& Min & T11&T4&T4&T11&T4&T11&T11&T11&T4&T11&T11&T4 \\
\cmidrule(lr){3-16}
&& \multirow{3}{*}{Unemployment} & Max & T10&T10&T10&T10&T10&T10&T10&T10&T10&T10&T10&T10 \\
&&& Med & T11&T11&T11&T11&T11&T4&T11&T11&T4&T11&T11&T11 \\
&&& Min & T4&T4&T4&T4&T4&T11&T4&T4&T11&T4&T4&T4 \\
\end{longtable}
\endgroup
\notepar{Notes: Each cell reports the arm with the maximum, intermediate, or minimum arm-month mean defined in Equation~\eqref{eq:mean-term-density}. Rankings are shown separately for inflation and unemployment explanations in Months 7--18. D1 compares T1--T3, D2 compares T2, T4, and T5, and D3 compares T4, T10, and T11. Embedding dispersion is defined in Supplementary Appendix Section~\ref{app:embedding-dispersion}; all other mechanisms use the dictionaries in Section~\ref{app:dictionaries}.}

\paragraph{Dimension 1: implementation time.} The principal treatment difference is the urgency and precision of implementation. In most post-treatment months, temporal-urgency density is ordered T1 $>$ T2 $>$ T3 for inflation explanations and displays the same broad pattern for unemployment explanations. This ordering coincides with the DTE ranking. The more imminent the implementation described in the tariff message, the more strongly agents revise both expectations upward.

The attention channel gives this ordering a specific economic interpretation. With limited information-processing capacity, agents choose which signals receive scarce cognitive resources; signals that are immediate, decision-relevant, and sufficiently precise command more attention than unresolved future threats \citep{sims2003implications,mackowiak2009optimal,mackowiak2023rational}. Under T1, the immediate tariff is therefore used more frequently in expectation updating, whereas under T3 the unspecified implementation date reduces the signal's current actionability. Consistent with this account, tariff-awareness density is T1 $>$ T2 $>$ T3 in most months. The result is not merely that T1 contains a stronger word. The common dictionary applied to open-ended answers shows that Household Agents themselves refer more intensively to tariffs when explaining the larger update.

Ambiguity provides the second part of Dimension 1. The ambiguity-perception score is highest under T3 in nearly every month. A signal without an implementation date permits agents to attach different subjective probabilities to immediate implementation, long delay, modification, or nonimplementation. Imprecise common information therefore leaves more room for heterogeneous priors and interpretations, weakening information aggregation and increasing disagreement \citep{ilut2014ambiguous,morris2002social}. This text ranking matches the higher belief dispersion in T3.

\paragraph{Dimension 2: tariff rate.} The distinguishing feature is the magnitude and precision of the announced rate. The simulation's qualitative result is that a higher, explicitly stated tariff rate produces a larger upward revision: DTEs are generally ordered T4 $>$ T2 $>$ T5. The open-ended evidence supplies the implied mechanism. In most months, perceived price-and-employment impact and tariff awareness are highest under T4, followed by T2 and then T5. A 100 percent rate is an unusually strong and salient cost signal. It induces more agents to discuss import costs, consumer prices, production decisions, and employment consequences, while an unspecified ``very high or very low'' rate provides insufficient precision for a common quantitative transmission assessment.

The dispersion comparison again turns on ambiguity rather than on the expected magnitude alone. Ambiguity perception is highest under T5 in most months. Agents complete the missing rate from different priors and subjective models, generating disagreement about the price and employment consequences. The explicit 10 percent and 100 percent rates coordinate beliefs around distinct but common anchors. Thus, a larger point signal can generate a larger mean response without generating the largest disagreement; an imprecise signal can instead produce a smaller mean update and a wider cross-sectional distribution.

\paragraph{Dimension 3: semantic coherence.} T4, T10, and T11 end with exactly the same 100 percent tariff message, so differences arise from the preceding sequence. As discussed in Section~\ref{subsubsec:literature}, a progressively escalating signal can induce extrapolative expectations, whereas a history of reversal can generate skepticism and reduce perceived credibility. We measure these mechanisms separately. Across most post-treatment months, extrapolation density is highest under T11, followed by T4 and then T10. Skepticism density is highest under T10 and substantially lower under T4 and T11. The progressive sequence therefore encourages agents to project the observed trend beyond its final announcement, while repeated cancellation and reinstatement lead agents to discount the final message as bluffing, political posturing, or another reversible threat. These two text rankings explain why identical terminal language generates the largest DTE under T11 and the smallest under T10.

Ambiguity-perception scores do not display a stable ordering across T4, T10, and T11. The higher belief dispersion under T11 therefore appears not to originate from ambiguity about the words in the final message. The relevant disagreement is instead about how far the progressive path will continue and how the escalation will transmit to prices and employment. To assess that possibility, we construct an Embedding Dispersion measure from the average pairwise cosine distance among open-ended responses within an arm-month. Supplementary Appendix Section~\ref{app:embedding-dispersion} reports the complete estimator. In most months, Embedding Dispersion is largest under T11. Agents thus give more semantically diverse causal accounts of a progressively escalating tariff sequence even though the terminal numerical rate is precise, closely mirroring the numerical belief-dispersion ranking.

\begin{figure}[t]
\centering
\begin{subfigure}[t]{.85\linewidth}
\centering
\includegraphics[width=\linewidth]{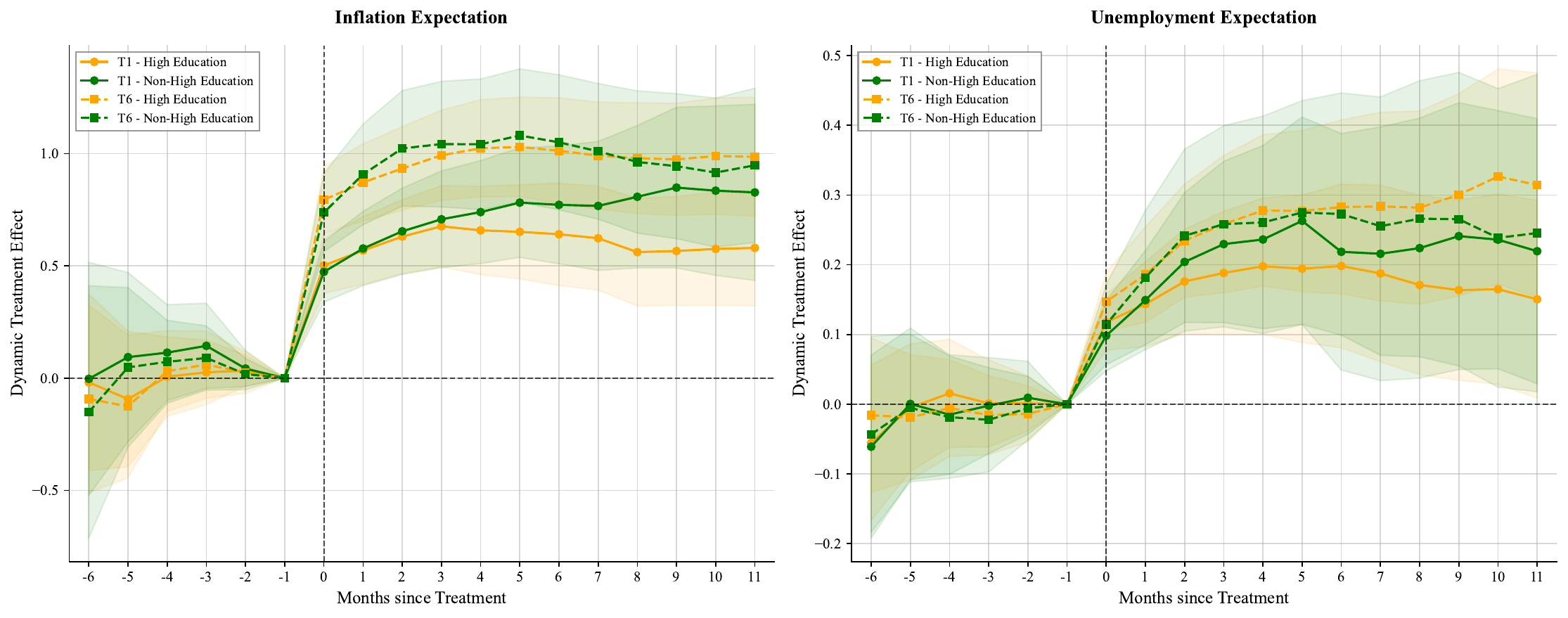}
\caption{Baseline versus minimalist message}
\label{fig:education-minimalist}
\end{subfigure}\hfill
\begin{subfigure}[t]{.85\linewidth}
\centering
\includegraphics[width=\linewidth]{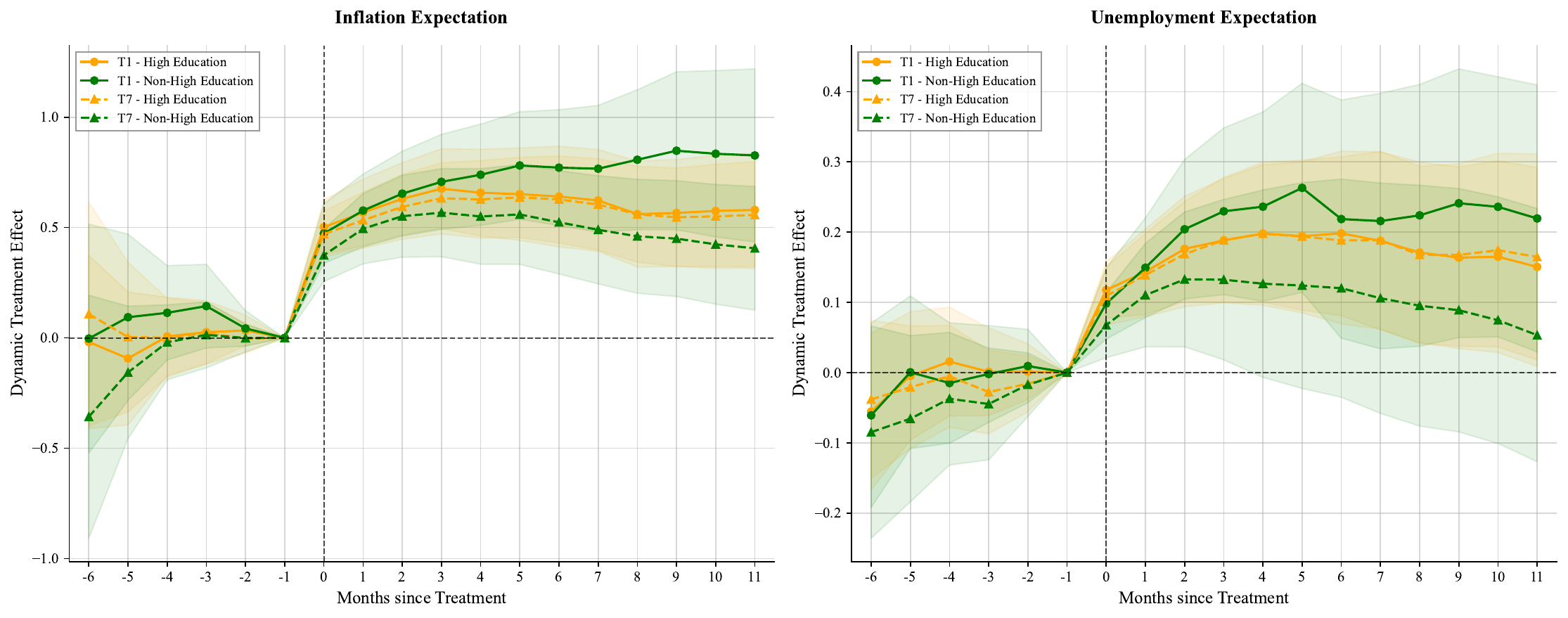}
\caption{Baseline versus complex message}
\label{fig:education-complex}
\end{subfigure}
\caption{Education Heterogeneity in Responses to Message Complexity}\label{fig:education-heterogeneity}
\notepar{Notes: Panel (a) compares T1 and the minimalist T6 message; Panel (b) compares T1 and the complex T7 message. Each panel reports inflation and unemployment responses by education. Bands are 95\% confidence intervals.}
\end{figure}

\paragraph{Dimension 4: semantic complexity.} Existing evidence indicates that simple, intelligible communication has lower cognitive-processing cost and can reach a broader nonexpert public, while complex and technical language attenuates responses among people with lower cognitive or educational resources \citep{bholat2019enhancing,dacunto2019cognitive}. We test whether the simulated heterogeneity is qualitatively consistent with that mechanism. Within each of T1, T6, and T7, we separate Household Agents whose personas indicate high education from those with lower education and estimate the treatment response for each group. Figure~\ref{fig:education-heterogeneity} compares T1 with T6 in Panel~\ref{fig:education-minimalist} and T1 with T7 in Panel~\ref{fig:education-complex}. Relative to T1, the minimalist T6 message raises DTEs among both education groups. The complex T7 message suppresses the response of less-educated agents but has little effect on the response of highly educated agents. This pattern directly accounts for the aggregate DTE ranking T6 $>$ T1 $>$ T7 and matches the processing-cost explanation.\footnote{The heterogeneity is qualitative mechanism evidence rather than proof of a human cognitive channel. Education in a persona can also activate associations learned during model training, which is why the benchmark heterogeneity validation and the treatment-specific text evidence must be read jointly.}

Table~\ref{tab:mechanism-rankings-2} adds the corresponding ambiguity results. In most months, ambiguity perception under both T6 and T7 exceeds T1. The two extremes create ambiguity for different reasons: the minimalist message omits background, motivation, conditions, and causal context, leaving agents to fill those gaps; the complex message contains technical expressions that some agents cannot interpret consistently. Both channels widen the range of explanations and can therefore raise belief dispersion even though one message is extremely short and the other is deliberately elaborate.

\begin{table}[htbp]
\centering
\caption{Monthly Rankings of Text-Based Mechanisms: Dimensions 4--6}\label{tab:mechanism-rankings-2}
\begingroup
\scriptsize
\setlength{\tabcolsep}{2pt}
\renewcommand{\arraystretch}{0.84}
\resizebox{\linewidth}{!}{%
\begin{tabular}{lllccccccccccccc}
\toprule
& & & & \multicolumn{12}{c}{Month} \\
\cmidrule(lr){5-16}
Dimension & Mechanism & Expectation & Rank & 7 & 8 & 9 & 10 & 11 & 12 & 13 & 14 & 15 & 16 & 17 & 18 \\
\midrule
\multirow{6}{*}{D4} & \multirow{6}{*}{Ambiguity Perception} & \multirow{3}{*}{Inflation} & Max & T6&T7&T7&T1&T7&T1&T7&T7&T7&T6&T6&T7 \\
&&& Med & T7&T6&T6&T7&T6&T6&T6&T1&T1&T7&T1&T1 \\
&&& Min & T1&T1&T1&T6&T1&T7&T1&T6&T6&T1&T7&T6 \\
\cmidrule(lr){3-16}
&& \multirow{3}{*}{Unemployment} & Max & T6&T7&T7&T7&T7&T7&T7&T6&T6&T7&T6&T6 \\
&&& Med & T7&T1&T6&T6&T6&T6&T6&T7&T7&T6&T7&T1 \\
&&& Min & T1&T6&T1&T1&T1&T1&T1&T1&T1&T1&T1&T7 \\
\midrule
\multirow{6}{*}{D5} & \multirow{6}{*}{Ambiguity Perception} & \multirow{3}{*}{Inflation} & Max & T1&T1&T8&T9&T1&T1&T8&T9&T1&T9&T9&T1 \\
&&& Med & T8&T8&T9&T1&T8&T8&T9&T1&T9&T1&T1&T9 \\
&&& Min & T9&T9&T1&T8&T9&T9&T1&T8&T8&T8&T8&T8 \\
\cmidrule(lr){3-16}
&& \multirow{3}{*}{Unemployment} & Max & T9&T8&T8&T9&T9&T8&T9&T9&T1&T1&T9&T1 \\
&&& Med & T1&T1&T9&T8&T8&T9&T1&T1&T8&T9&T8&T9 \\
&&& Min & T8&T9&T1&T1&T1&T1&T8&T8&T9&T8&T1&T8 \\
\midrule
\multirow{12}{*}{D6} & \multirow{4}{*}{Ambiguity Perception} & \multirow{2}{*}{Inflation} & Max & T12&T12&T12&T1&T12&T12&T12&T12&T12&T1&T12&T12 \\
&&& Min & T1&T1&T1&T12&T1&T1&T1&T1&T1&T12&T1&T1 \\
\cmidrule(lr){3-16}
&& \multirow{2}{*}{Unemployment} & Max & T12&T12&T12&T12&T12&T12&T12&T12&T12&T12&T12&T12 \\
&&& Min & T1&T1&T1&T1&T1&T1&T1&T1&T1&T1&T1&T1 \\
\cmidrule(lr){2-16}
& \multirow{4}{*}{Credibility} & \multirow{2}{*}{Inflation} & Max & T12&T12&T12&T12&T12&T12&T12&T12&T12&T12&T12&T12 \\
&&& Min & T1&T1&T1&T1&T1&T1&T1&T1&T1&T1&T1&T1 \\
\cmidrule(lr){3-16}
&& \multirow{2}{*}{Unemployment} & Max & T12&T12&T12&T12&T12&T12&T12&T12&T12&T12&T12&T12 \\
&&& Min & T1&T1&T1&T1&T1&T1&T1&T1&T1&T1&T1&T1 \\
\cmidrule(lr){2-16}
& \multirow{4}{*}{Unexpectedness} & \multirow{2}{*}{Inflation} & Max & T12&T12&T12&T12&T12&T12&T12&T12&T12&T12&T12&T12 \\
&&& Min & T1&T1&T1&T1&T1&T1&T1&T1&T1&T1&T1&T1 \\
\cmidrule(lr){3-16}
&& \multirow{2}{*}{Unemployment} & Max & T12&T12&T12&T12&T12&T12&T12&T12&T12&T12&T12&T12 \\
&&& Min & T1&T1&T1&T1&T1&T1&T1&T1&T1&T1&T1&T1 \\
\bottomrule
\end{tabular}}
\endgroup
\notepar{Notes: Each cell reports the arm with the maximum, intermediate, or minimum arm-month dictionary mean defined in Equation~\eqref{eq:mean-term-density}. Rankings are shown separately for inflation and unemployment explanations in Months 7--18. D4 compares T1, T6, and T7; D5 compares T1, T8, and T9; D6 compares T1 and T12.}
\end{table}

\paragraph{Dimension 5: narrative.} Unlike conventional formal policy communication, Trump's tariff threats frequently take the form of stories about who benefits, who bears the cost, and how the policy restores American prosperity. MAGA, reshoring, import independence, and claims that foreign exporters pay the tariff are therefore not incidental rhetoric; they are the empirical counterparts of the T8 and T9 designs. Such narratives may function, intentionally or otherwise, as expectation-management devices. Their dynamic mechanism is difficult to observe in standard numerical survey data but can be examined directly in the MAS-generated explanations.

Experimental evidence shows that narratives affect expectations not only by changing the quantity of information available to an agent, but by changing the causal explanation of past events, the subjective model brought to the problem, and the interpretation of new signals \citep{andre2022subjective,andre2026narratives}. We therefore identify the causal narrative frames in the open-ended responses and ask whether T8 and T9 alter agents' accounts of what drives inflation or unemployment and how the tariff reaches those outcomes. For inflation, which we describe in detail, the classification contains seven substantive frames and a residual category: tariff-to-price pass-through, supply-chain disruption, monetary or fiscal policy, no tariff effect, uncertainty-driven inflation, exporter absorption, personal experience, and other. The unemployment analysis applies the same logic to labor-market transmission.

We use human coding because a keyword can identify a topic but cannot reliably recover the direction or causal structure of an argument. Two coders first read the complete annotation guideline, practiced on pre-coded responses, and compared their decisions with reference answers. They then joined a calibration meeting with the principal investigator to resolve disagreements and latent ambiguity in frame definitions. During formal coding, every response was classified independently by both coders. They could not discuss individual responses and were blind to treatment assignment. Each frame was coded as a binary indicator, and inter-coder agreement was evaluated frame by frame using Cohen's $\kappa$. After independent coding, the principal investigator adjudicated disagreements after reviewing both coders' stated reasons. If $\kappa<0.70$ for any frame, its definition and guideline were revised, the coders were retrained, and the affected sample subset was recoded.\footnote{The complete annotation guideline, including frame definitions, detailed instructions, process documentation, explanations, and examples, is available at \url{https://drive.google.com/file/d/1R8LLYfj3cANCkcgF4YINNf0SveGXdu7s/view?usp=sharing}.} Supplementary Appendix Section~\ref{app:causal-frames} reproduces all frame labels and core concepts.

\begin{figure}[htbp]
\centering
\includegraphics[width=.9\linewidth]{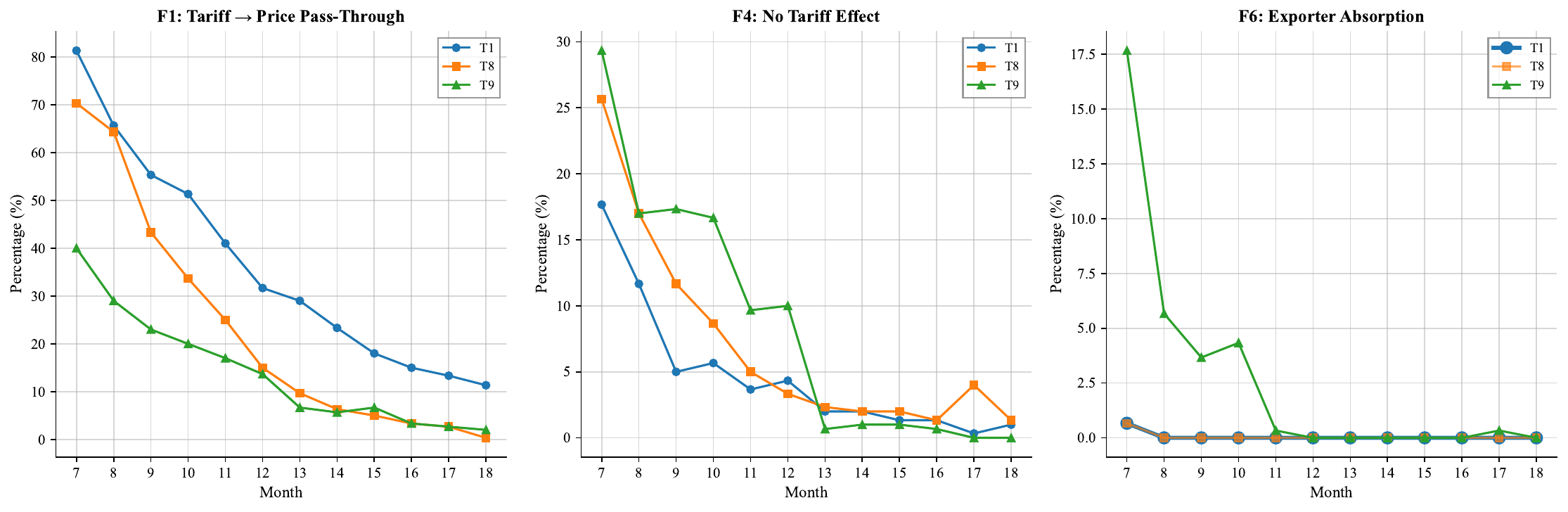}
\caption{Causal Frames in Open-Ended Responses}\label{fig:causal-frames}
\notepar{Notes: The panels report monthly shares of responses assigned to the tariff-to-price pass-through frame (F1), the no-tariff-effect frame (F4), and the exporter-absorption frame (F6) for T1, T8, and T9. The complete frame definitions and classification procedure appear in Supplementary Appendix Section~\ref{app:causal-frames}.}
\end{figure}

After coding, we calculate the monthly share of each causal frame in T1, T8, and T9 from the treatment month onward. Figure~\ref{fig:causal-frames} reports the three frames displaying the largest cross-arm differences. The narrative treatments change the structure of agents' causal explanations, and those changes decay dynamically. In most months, T8 and T9 have a distinctly smaller share of tariff-to-price pass-through explanations and a larger share of no-tariff-effect explanations than T1. Thus, both the MAGA narrative and the tax-incidence narrative lead fewer agents to infer that the tariff will be passed into goods prices and more agents to infer little or no price effect. During the first several post-treatment months, T9 also has a substantially larger exporter-absorption share than T1 or T8. The tax-incidence narrative therefore changes an additional margin: some agents explicitly replace domestic pass-through with the claim that foreign exporters absorb the cost. Together, these shifts explain why T8 and T9 attenuate the DTE relative to T1 and why T9 produces the smallest response. They also reproduce the mechanism emphasized by \citet{andre2022subjective,andre2026narratives}: the treatment changes which causal model agents use to interpret the same policy parameter.

The ambiguity-perception measure does not exhibit a stable ordering across T1, T8, and T9. This corresponds to the absence of a stable belief-dispersion ranking in Dimension 5. The added stories alter the causal framework while leaving the rate and implementation date explicit; they do not make the signal itself systematically more or less ambiguous. Hence the primary narrative effect operates through the mean expectation and the allocation of agents across causal frames, not through a persistent widening of disagreement.

\paragraph{Dimension 6: sender.} In a Bayesian-learning framework, the response to a signal depends on its content and its surprise relative to prior beliefs. We therefore distinguish perceived unexpectedness from perceived credibility. Unexpectedness is higher under T12 than T1 in most months. This pattern is consistent with the political context likely represented in the models' training data: Trump's first term and 2024 campaign repeatedly emphasized tariffs, so a Trump tariff announcement is partly anticipated; a universal-tariff announcement by Harris conflicts more sharply with agents' prior image of Democratic trade policy. The same text consequently produces a larger belief revision when attributed to Harris.

Credibility is also higher under T12 than T1 in nearly every month. The generated responses more often treat the Harris-attributed message as serious, dependable, or likely to be implemented, whereas some responses discount Trump because of a history of threats, delays, and reversals. This source-dependent updating is consistent with evidence that political identity shapes macroeconomic beliefs and the reception of information \citep{kamdar2025think,binder2025partisan,hirs2026partisanship}. It is also directionally consistent with the April 2025 Pew survey: 48 percent of U.S. adults reported trusting Trump's statements less than those of previous presidents, compared with 32 percent reporting greater trust, with especially large partisan differences \citep{pew2025trust}.\footnote{The Pew comparison concerns Trump versus previous presidents, not a controlled Trump-versus-Harris tariff experiment. It supports the plausibility of declining or polarized source credibility but does not validate the numerical T12 effect.}

Finally, ambiguity perception is generally higher under T12. Because agents have fewer historical precedents for combining a Democratic sender with a universal tariff, they rely more heavily on heterogeneous political priors and subjective models to interpret why the policy was announced and how durable it would be. That divergence in interpretation helps explain why T12 raises belief dispersion in addition to its larger DTE. Table~\ref{tab:mechanism-rankings-2} reports the complete monthly rankings for Dimensions 4--6.

Finally, we embed each response and compute the average pairwise cosine distance within an arm-month. Supplementary Appendix Section~\ref{app:embedding-dispersion} gives the estimator. Embedding dispersion is high when agents offer semantically different explanations even if their numerical forecasts are similar. Its ranking tracks belief dispersion particularly closely in the ambiguity and central-bank experiments, strengthening the interpretation that disagreement originates in heterogeneous causal accounts rather than numerical noise alone.

Embedding dispersion also guards against a limitation of fixed dictionaries. Two responses can express different causal models without using any prespecified term. One may discuss inventories and supplier contracts; another may discuss household purchasing power and interest rates. Their cosine distance records the difference. Conversely, two responses may use the word ``uncertain'' while presenting nearly identical explanations, producing similar embeddings. Agreement between dictionary and embedding measures is therefore informative because the two methods fail in different ways.

We do not treat explanations as privileged access to latent reasoning. LLMs generate rationales as text, and those rationales may be post hoc or shaped by the prompt. Their value here is comparative and predictive. If a treatment changes a numerical forecast but leaves its proposed mechanism absent from the text, the interpretation is weakened. If the mechanism appears, varies in the predicted direction, and tracks numerical dispersion over time, the simulation yields a more coherent hypothesis for subsequent human testing.

\subsubsection{Economic Insights}\label{subsubsec:insights}

The mechanism evidence closes the loop between simulated outcomes and simulated expectation formation. The six qualitative rankings accord with recognizable economic intuitions and with the relevant human evidence reviewed in Section~\ref{subsubsec:literature}. More importantly, the numerical forecasts are internally coherent with the agents' own explanations: urgent and precise messages receive greater tariff attention; ambiguous messages generate more varied interpretations; progressive sequences induce extrapolation; reversal sequences induce skepticism; narratives change causal frames; and sender changes alter surprise and credibility. This correspondence does not establish that an LLM rationale is a human cognitive process. It does show that the MAS outputs are not isolated numerical reactions without an economically interpretable account, which is essential if economists are to use such systems for disciplined simulation research. Five insights follow.

\paragraph{Insight 1: non-policy features of tariff information can matter at least as much as policy parameters.} Household expectations respond not only to the announced rate and implementation time, but also to the political identity of the sender, the causal narrative embedded in the message, semantic consistency across announcements, readability, and ambiguity. Traditional models commonly compress a tariff signal into a numerical rate. The MAS comparisons instead suggest that households ask a broader set of questions: who said it, how was it said, is the message consistent with earlier announcements, what causal story does it supply, and should the sender be believed? In an environment of information overload, political polarization, and informal social-media communication, the expectation effect of a tariff threat need not be linear in the policy parameter. DSGE, HANK, or CGE policy evaluations that omit these non-policy information features may therefore mischaracterize both the mean expectation response and the dispersion of beliefs.

\paragraph{Insight 2: central-bank communication should be conditioned on the multidimensional features of the tariff message.} The simulation yields five more specific hypotheses. First, an immediately effective threat produces a faster and larger expectation jump than a distant one, reducing the time available for interpretation. Second, ambiguity about either timing or rate raises disagreement, so communication aimed at reducing uncertainty may be valuable even when the mean update is small. Third, extremely short and emotionally salient messages and high-frequency progressive sequences deserve attention because they raise both expectations and, in the latter case, dispersion; repeated semantic reversals instead erode credibility and may lower the value of responding to every announcement. Fourth, a narrative can reduce the short-run mean response by disseminating an economically inaccurate incidence model, thereby making later expectation management based on the correct transmission mechanism more difficult. Fifth, political sender identity changes reception. A central bank must therefore remain institutionally and linguistically neutral while providing its own causal assessment rather than echoing partisan framing. Because the distant treatments are not directly validated on human subjects, these are hypotheses to be evaluated in survey or field experiments rather than prescriptions.

\paragraph{Insight 3: the MAS can serve as a tariff-policy scenario-simulation platform.} The six dimensions illustrate a methodological advantage of the framework. Holding every other input fixed, the researcher can change one characteristic of a signal, reuse the same heterogeneous Household Agents, and observe the induced distribution of point forecasts, subjective probabilities, narratives, memory states, and social-media feedback over 18 months. Such clean counterfactual control is difficult in human survey experiments: a human control sample cannot be reused without carryover, repeated exposure changes the subject, and maintaining a long balanced panel is costly. The MAS extends recent work showing that LLM Agents can approximate selected human survey and laboratory responses by embedding them in a dynamic system with social interaction, a recommendation algorithm, and memory. For a changing tariff environment, the platform can screen possible messages, identify confusing treatments, and generate mechanism hypotheses before a central bank or researcher incurs the cost of a human experiment. Its comparative advantage is scenario exploration and experimental design, not replacement of human evidence.

\paragraph{Insight 4: expectation levels and belief coordination are separate policy objects.} A message can produce a small average DTE while increasing disagreement, as in ambiguous-rate or ambiguous-timing scenarios, or it can supply a common focal interpretation that compresses disagreement around an undesirable mean. Numerical point expectations alone therefore conceal an important dimension of communication risk. Jointly observing point forecasts, subjective distributions, and semantic dispersion makes it possible to distinguish weak updating from heterogeneous updating and to identify whether coordination arises from shared information, shared ambiguity, or a common causal narrative.

\paragraph{Insight 5: causal narratives are state variables in expectation dynamics.} The effects of a tariff message do not end when its words leave the three-month memory window. An interpretation can persist through the agent's revised prior and through posts generated into the SMIM. A narrative that changes perceived incidence or persistence can therefore propagate endogenously even when the original policy parameter is unchanged. This dynamic mechanism links behavioral macroeconomics to policy communication: managing expectations requires attention to the causal model circulating through the information network, not only repeated announcements of a target or a rate.

Overall, the outcome and text evidence makes the simulation internally self-consistent and economically interpretable. That coherence strengthens the use of the MAS to design human experiments and stress-test communication, while the validation hierarchy remains binding: T1 is directly benchmarked, nearby one-feature variants support cautious qualitative near-generalization, and distant sender, sequence, and central-bank scenarios remain exploratory.

\section{Central Bank Communication Strategies}\label{sec:cb}

The six comparisons in Section~\ref{sec:results} show that a semantically progressive tariff-threat sequence is particularly difficult for expectation management. Under T11, Trump first releases six closely related posts that raise the proposed rate from 40 to 90 percent and then delivers the terminal 100 percent announcement. Both inflation and unemployment expectations rise substantially, and belief dispersion expands as agents extrapolate different upper-tail paths. Having identified this stress scenario, we ask what kind of Federal Reserve communication can guide expectations downward and limit further dispersion.

The communication problem has two distinct objects. The first is the level of expectations: households may revise expected inflation and unemployment after the tariff threat and again after the central-bank message. The second is coordination: households can disagree about whether the shock is temporary or persistent, whether monetary policy will react, and how the response will affect employment. A message may therefore reduce dispersion by supplying a common interpretation while simultaneously moving the mean expectation upward. We continue to evaluate DTE and belief dispersion jointly.

This section differs from the earlier tariff experiment in design. T11 is now the comparison arm. A Household Agent first receives the T11 tariff sequence and only then receives one of the Federal Reserve communication treatments C1--C5. Thus, the FOMC message enters an information environment in which priors, open-ended explanations, and agent-generated social-media posts have already adjusted to an escalating tariff narrative. The experiment does not reset the agent with a clean vignette.

The exercise is explicitly exploratory. C1--C5 are semantically and institutionally distant from the human-benchmarked T1 treatment, and there is no corresponding human panel for direct validation. The results identify communication mechanisms and candidate treatments for later human testing. They should not be interpreted as estimates of how the U.S. public would causally respond or as direct recommendations for the Federal Reserve.

The baseline, Strategy C1, reproduces the experimental material's longer-run FOMC language:

\begin{quote}\itshape
The Committee reaffirms its judgment that inflation at the rate of 2 percent, as measured by the annual change in the price index for personal consumption expenditures, is most consistent over the longer run with the Federal Reserve's statutory maximum employment and price stability mandates.
\end{quote}

The 2 percent longer-run goal was first adopted in the FOMC's January 25, 2012 Statement on Longer-Run Goals and Monetary Policy Strategy and has subsequently been reaffirmed; the exact wording evolved in later versions of the Statement.\footnote{The original 2012 statement said that 2 percent PCE inflation was most consistent with the Federal Reserve's ``statutory mandate.'' The experiment reproduces the supplied C1 sentence exactly. The historical attribution is therefore to the FOMC's longer-run-goals framework initiated in 2012, not a claim that every word of C1 appeared in the original 2012 release. See Board of Governors of the Federal Reserve System (2012), \url{https://www.federalreserve.gov/newsevents/pressreleases/monetary20120125c.htm}.}

Drawing on actual central-bank communication and the literature on targets, forward guidance, institutional credibility, shock attribution, and household communication \citep{blinder2008central,eusepi2010central,coibion2022monetary,romer2000federal,nakamura2018information,kakhbod2026mind}, we design four additional strategy scenarios:

\begin{description}[leftmargin=1.45cm,style=nextline,itemsep=4pt]
\item[C2] \emph{Scenario-Based Communication (hawkish).} The Committee commits to an explicit state contingency: if tariff developments generate persistent upward pressure on broad prices, it stands ready to tighten the stance of monetary policy to return inflation to 2 percent.
\item[C3] \emph{Affirming the Institution's Independence (hawkish).} The Committee emphasizes that the Federal Reserve is institutionally independent of political influence, remains committed to its statutory mandates, and will not adjust monetary-policy decisions in response to political pressure.
\item[C4] \emph{Look-through Strategy (dovish).} The Committee characterizes tariff effects as a one-off, temporary relative-price shock rather than a change in underlying inflation or unemployment and therefore refrains from immediate action.
\item[C5] \emph{Allow Inflation to Overshoot Moderately (dovish).} The Committee permits inflation to run moderately above 2 percent for some time to support overall economic activity and therefore refrains from immediate action.
\end{description}

To reduce researcher discretion and keep the wording close to official FOMC style, we construct an FOMC Agent using the same role-playing method, model, temperature, and fixed-seed logic as the Trump Agent. It rewrites the C1 baseline for each strategy while being instructed to hold all nonstrategy elements as constant as possible. The full prompt and the exact generated texts for C2--C5 appear in Supplementary Appendix Section~\ref{app:fomc-materials}.\footnote{C2--C5 are controlled, LLM-generated experimental messages. They are not quotations from Federal Reserve officials or statements of current Federal Reserve policy.}

C1 provides a demanding baseline precisely because the 2 percent objective is familiar. Repeating a known target can reaffirm commitment but adds little information about how the Committee interprets a tariff shock. C3 similarly addresses a source of political distrust without specifying the transmission mechanism or reaction horizon. If agents already believe that tariffs raise prices, neither message tells them whether the central bank sees a relative-price change, persistent inflation, or a demand response. Their near-zero average DTEs are therefore consistent with limited incremental information, not necessarily with disbelief in the target or independence.

C2 adds a conditional reaction function. It states that persistent broad price pressure will lead to a firmer stance. In a standard policy account, that commitment can anchor medium-run inflation. In a survey shown immediately after a tariff threat, however, the words ``persistent upward pressure'' can also validate the premise that the shock is inflationary. Agents must infer both the state and the response from one sentence. The positive DTE indicates that the state-description or salience effect dominates the stabilizing commitment in the MAS over the measured horizon.

C4 and C5 are both less immediate in their policy response but differ in attribution. C4 states that the tariff effect is transitory and separate from underlying inflation dynamics. C5 accepts moderately above-target inflation for some time to support activity. C4 therefore lowers the inferred persistence of the shock; C5 makes a period of higher inflation part of the stated plan. The comparison shows why grouping both as ``dovish'' obscures their informational content. One revises the perceived state, while the other revises the tolerated path.

\begin{figure}[htbp]
\centering
\begin{subfigure}[t]{.85\linewidth}
\centering
\includegraphics[width=\linewidth]{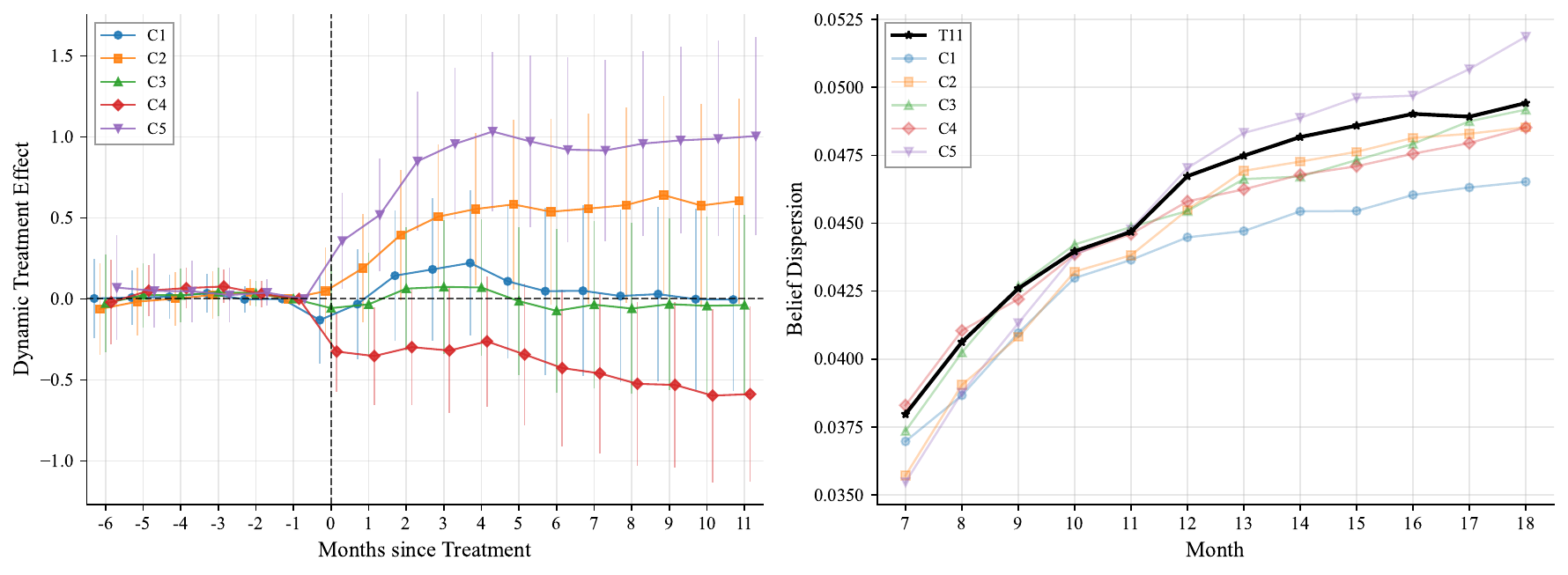}
\caption{Inflation expectations}
\label{fig:cb-inflation}
\end{subfigure}\hfill
\begin{subfigure}[t]{.85\linewidth}
\centering
\includegraphics[width=\linewidth]{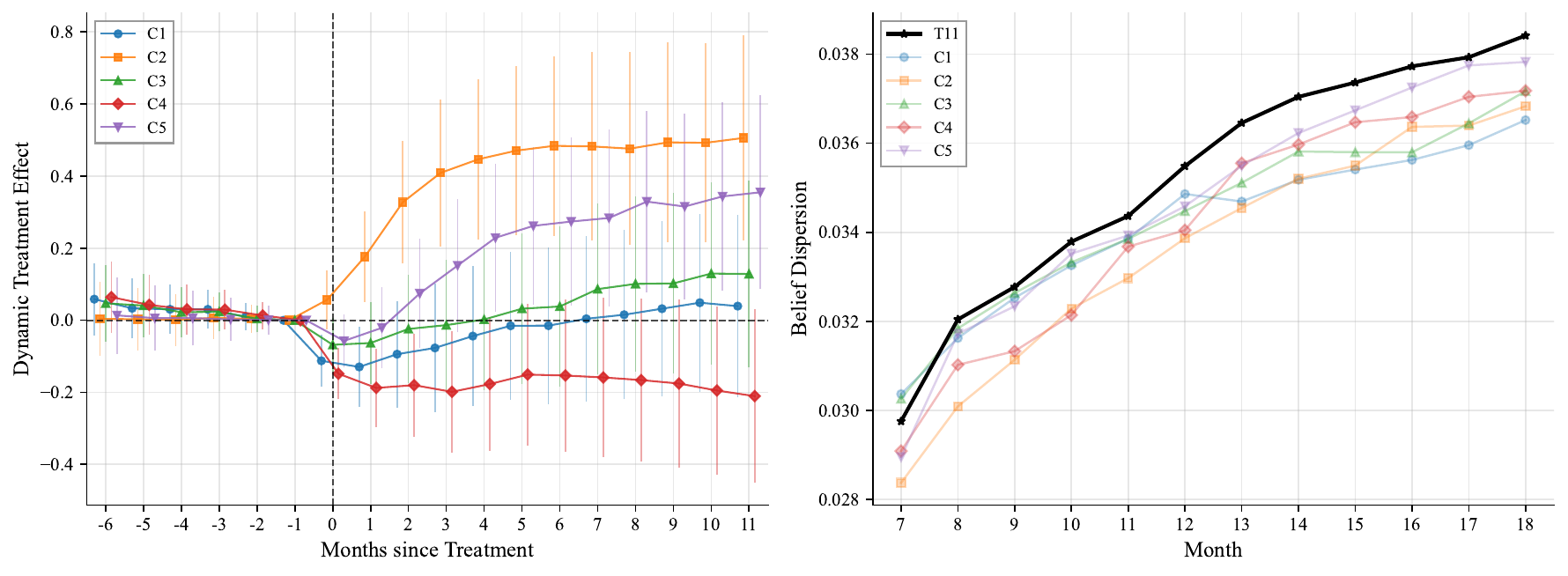}
\caption{Unemployment expectations}
\label{fig:cb-unemployment}
\end{subfigure}
\caption{Dynamic Effects of Central-Bank Communication Strategies}\label{fig:cb-results}
\notepar{Notes: The panels compare five central-bank communication strategies, C1--C5, after the escalating tariff-threat scenario T11. Each subfigure reports DTEs and belief dispersion for the indicated outcome. Shaded bands are 95\% confidence intervals. Because these strategies are far from the human-validated T1 benchmark, the estimates are interpreted as exploratory MAS responses.}
\end{figure}

Using T11 as the comparison arm, we estimate the DTEs of C1--C5 for inflation and unemployment expectations and calculate the belief dispersion generated by each strategy. Figure~\ref{fig:cb-results} reports inflation in Panel~\ref{fig:cb-inflation} and unemployment in Panel~\ref{fig:cb-unemployment}. The same qualitative ordering appears for both outcomes: C2 and C5 have the largest DTEs, C1 and C3 occupy the middle, and C4 has the smallest DTE, or C2 and C5 $>$ C1 and C3 $>$ C4. Relative to T11, C2 and C5 are significantly positive in most post-communication months; C1 and C3 are statistically indistinguishable from zero in most months; and C4 is significantly negative in most months. C4 is therefore the only simulated strategy that consistently guides both point expectations downward. At the same time, every C1--C5 message reduces belief dispersion relative to T11 to some degree. A central-bank message can thus coordinate causal interpretations even when it does not reduce, and may even raise, the average expectation.

The inflation and unemployment panels should be interpreted together. Under C2, stronger expected tightening can reduce medium-run inflation but raise perceived labor-market risk, while the reference to persistent prices raises near-term inflation salience. Under C4, less expected tightening can lower unemployment concerns, but only if agents accept the temporary-shock diagnosis. Under C5, support for activity can moderate unemployment expectations while a tolerated overshoot raises inflation expectations. The generated paths reflect these competing channels rather than a mechanical mapping from policy tone to both outcomes.

Dispersion falls under every strategy because each message supplies a common institutional focal point after the heterogeneous narratives generated by T11. C1 coordinates on the target, C3 on independence, C2 on a trigger, C4 on transitory attribution, and C5 on a temporary tolerance horizon. The result illustrates that common language can compress disagreement even when it does not change the average. For policy evaluation, that distinction prevents a low-dispersion outcome from being labeled successful without examining the level around which beliefs coordinate.

The temporal pattern also matters. A communication effect that disappears after one month may reflect attention without a durable revision to the subjective model. C4 remains negative across much of the post-message horizon because its temporary-shock interpretation enters priors and subsequent explanations. C1 and C3 show little cumulative movement because their incremental state content is small. C2 and C5 remain positive while persistence and overshoot language continues to circulate through agent-generated posts. These dynamics are a property of the simulated information loop and warrant direct human testing.

\FloatBarrier

As in Section~\ref{subsubsec:text}, we analyze the LLM Agents' open-ended responses to explain why the central-bank DTEs and belief dispersion display these rankings. The full procedure and results appear in Supplementary Appendix Section~\ref{app:cb-mechanisms}; Figure~\ref{fig:semantic-pressure} reports the semantic-pressure measure and Figure~\ref{fig:cb-embedding} reports Embedding Dispersion. The text evidence can be summarized as follows. C2 and C5 create high \emph{semantic pressure}: the former makes persistent price pressure and possible tightening explicit, while the latter makes a tolerated overshoot explicit. These causal frames raise the availability of an elevated inflation path even though the messages differ in conventional policy stance. C1 and C3 restate objectives or governance but do not explain the transmission from tariffs to prices, wages, costs, and employment. C4 supplies the clearest temporary-shock attribution, so agents infer lower persistence and less near-term tightening. All five messages reduce semantic dispersion by giving agents a common institutional frame, explaining why numerical belief dispersion falls under every strategy.

Semantic pressure is not a sentiment score. C2 and C5 use different evaluative language, yet both make an elevated inflation path cognitively available. The measure records explicit causal pressure, persistence, policy reaction, and tolerated overshooting. C4 can be firm about the target while assigning little pressure to underlying inflation. This distinction explains why standard hawkishness dictionaries would misclassify the mechanism: hawkish tone may coexist with a positive expectation response if the message first persuades households that the inflation problem is persistent.

Embedding dispersion identifies a second channel. T11 agents offer diverse explanations involving prices, retaliation, jobs, bargaining, and policy. Every FOMC message narrows that semantic set, but C4 narrows it most by supplying a specific account of the shock. C1 and C3 leave the tariff-to-economy mapping open, so agents continue to disagree about the path even while referencing a common institution. C2 and C5 coordinate on persistence or overshooting but leave disagreement about the strength and timing of the response.

The mechanism evidence yields a broader lesson. Hawkish or dovish tone is too coarse a sufficient statistic for public communication. A message coordinates beliefs when it supplies a causal model, a conditional reaction function, and a horizon over which the objective applies. \citet{kakhbod2026mind} document that the Federal Reserve's own reaction has historically depended on whether inflation was attributed to demand or supply. In the MAS, the same distinction matters for households: C2 treats persistence as a condition for action, C4 identifies a potentially temporary supply disturbance, and C5 states how the employment cost enters the horizon of adjustment. The result is a set of testable communication hypotheses, not a ranking that should be implemented without human evidence and institutional analysis.

The findings also caution against treating clarity as synonymous with reassurance. C2 is clear about a trigger and C5 is clear about tolerance, but clarity makes the adverse state more salient. C4 reassures by changing attribution, which will be effective only if the diagnosis is credible. A real central bank cannot choose language independently of its information and future actions. If tariff inflation proves persistent, an earlier look-through message may damage credibility. The simulation therefore identifies a communication trade-off: causal specificity coordinates current beliefs, but an incorrect causal claim creates future reputation risk that the 18-month survey environment does not fully price.

For a human experiment, the MAS results imply a design with at least three separate outcomes. Researchers should elicit the expected inflation path, the expected policy-rate response, and the perceived persistence or source of the tariff shock. Without the latter two, a positive response to C2 could be misread as policy ineffectiveness when it actually combines higher perceived inflation pressure with higher expected tightening. Open-ended explanations can then test whether respondents use the causal distinctions intended by the treatment.

In sum, the central-bank experiment shows that communication works in the MAS through the causal model it supplies, not through a one-dimensional hawkish or dovish label. The goal-only C1 and independence-focused C3 add little state-contingent information; C2 makes persistence and tightening salient; C5 makes above-target inflation an explicit tolerated path; and C4 reclassifies the tariff as a temporary relative-price disturbance. That classification both lowers the mean expectation and generates a common explanation in the simulation. The result accords with evidence that Federal Reserve reactions depend on whether inflation is attributed to demand or supply forces \citep{kakhbod2026mind}, but its policy meaning is conditional. A look-through message is stabilizing only when the temporary-shock diagnosis is credible and remains consistent with subsequent data and action. The ranking therefore provides a set of sharply testable hypotheses for human experiments and clarifies which mechanisms those experiments should elicit; it is not a stand-alone recommendation to adopt C4 in practice.

\section{Discussion}\label{sec:discussion}

The central interpretive issue is what the validation can support. The T1 exercise compares simulated and human distributions over eight months, tests algorithmic distinguishability, and examines demographic heterogeneity. Passing or approaching these benchmarks lowers concern about gross misspecification on the tested margins. It does not establish that every unmeasured mechanism is correct, that prompts did not exploit learned regularities, or that treatment effects transport to a new text. As \citet{hullman2026human} emphasize, heuristic validation is well suited to exploration and design analysis but lacks the formal guarantees of statistical calibration for confirmatory inference. We therefore treat the MAS as a structured hypothesis generator.

This claim hierarchy separates three uses that are often conflated. In \emph{design analysis}, simulated respondents reveal confusing wording, ceiling effects, arithmetic failures, or redundant treatments. In \emph{exploratory analysis}, a calibrated MAS proposes qualitative directions, rankings, and mechanisms that determine which human experiments are worth running. In \emph{confirmatory analysis}, a researcher seeks an estimate or test about a human population. The present paper supports the first two uses. The third requires human observations and explicit assumptions that connect simulated measurements to the target estimand.

The distinction is substantive, not terminological. An exploratory ranking can be scientifically useful even when its exact magnitudes are biased, provided the claim is transparent and leads to discriminating evidence. Conversely, a simulator that matches several human moments may still yield a biased treatment effect if the new treatment activates a feature absent from validation. The sender and central-bank experiments are clear examples: their institutional and partisan content changes more than the tariff rate or date, so successful T1 validation cannot make them confirmatory.

Table~\ref{tab:claim-hierarchy} records the interpretation used throughout the paper. The classification depends on the distance from the human benchmark and on the kind of evidence available, not on whether a simulated coefficient is statistically significant. This rule prevents a precisely estimated result in a distant scenario from receiving a stronger claim than a noisier result close to the validation setting.

\begin{table}[htbp]
\centering
\begin{threeparttable}
\caption{Evidence and Claim Hierarchy}\label{tab:claim-hierarchy}
\begin{tabularx}{\linewidth}{p{.19\linewidth}p{.22\linewidth}Xp{.23\linewidth}}
\toprule
Scenario class & Examples & Supporting evidence & Permitted interpretation \\
\midrule
Observed benchmark & T1, Months 7--14 & MSC distributions; CvM; machine-learning discrimination; political, income, and gender gradients & Fidelity on the measured margins; no general equivalence claim \\
Controlled nearby variants & T2--T9 when one feature changes & T1 validation; paired design; theoretical consistency; numerical and text-mechanism agreement; sensitivity checks & Cautious qualitative near-generalization of direction or ranking \\
Distant tariff scenarios & T10, T11, T12 & Paired computational design; within-MAS robustness; literature and text triangulation & Exploratory scenario comparison and hypothesis generation \\
New institution or intervention & C1--C5 after T11 & Within-MAS stress test; central-bank literature; semantic-pressure and embedding evidence & Exploratory design analysis only; no direct policy recommendation \\
\bottomrule
\end{tabularx}
\begin{tablenotes}[flushleft]\footnotesize
\item Notes: Statistical significance within the MAS does not move a scenario to a stronger evidentiary tier. Stronger claims require new human data or an explicit statistical-calibration design linking simulated outcomes to the target human estimand.
\end{tablenotes}
\end{threeparttable}
\end{table}

The hierarchy also clarifies what would falsify the paper's substantive interpretation. If a human experiment found that ambiguous timing reduced both mean updating and disagreement, the MAS ambiguity mechanism would fail on its key comparative prediction. If people reacted more strongly to semantic reversals than to progression, the credibility account would need revision. If goal-only central-bank language lowered expectations as much as a temporary-shock explanation, the proposed role of causal attribution would be overstated. These are discriminating hypotheses, not flexible narratives that accommodate every outcome.

Quantitative disagreement between a future human experiment and the MAS would be expected and informative. The relevant questions would be whether the bias is approximately common across nearby treatments, whether it is concentrated in particular groups, and whether open-ended human explanations identify a missing channel. A common level bias could leave a qualitative contrast useful. A sign reversal or treatment-specific demographic bias would invalidate transport. This diagnosis is more productive than treating the simulator as either fully valid or wholly useless.

The same hierarchy disciplines reporting as models evolve. A stronger foundation model or a larger persona archive may improve fit, but it does not automatically strengthen identification. New validation should be performed after a material model update, and treatment claims should be reassigned if the target population or information environment changes. In this sense, validity belongs to a configured system, population, outcome, and scenario, not to an LLM brand.

Supplementary Appendix Section~\ref{app:extrapolation} describes a local regularity argument. Nearby treatments preserve personas, priors, modules, questionnaire wording, and most message semantics, altering only one feature. If the LLM mapping is locally smooth on this bounded input region and its approximation error changes gradually, qualitative contrasts may be more stable than exact levels. This argument supports cautious near-generalization, especially when the ranking agrees with theory and open-ended mechanisms. It cannot bound bias for sender changes, long message sequences, or central-bank interventions. Those designs are deliberately classified as distant exploratory scenarios.\footnote{Smoothness of a neural mapping does not imply that economic behavior is smooth at every policy threshold. A discrete legal, partisan, or credibility threshold can invalidate local reasoning even when the network itself is continuous.}

Similarity in text space is only one notion of distance. A one-word change from ``may'' to ``will'' can cross a perceived commitment threshold, while a longer stylistic rewrite may leave the economic state unchanged. We therefore combine semantic distance with design distance. Near scenarios preserve the sender, sequence, policy instrument, outcome, and questionnaire while changing one feature. Distant scenarios change the source, create a multi-message history, or introduce a new institution. This classification is more defensible than relying on an embedding distance alone.

Mechanism evidence helps diagnose, but not prove, transport. If a nearby treatment produces a ranking supported by limited-attention theory, tariff evidence, numerical outcomes, and open-ended explanations, the result survives several distinct challenges. If those sources conflict, the scenario should be treated as fragile even when the point estimate is precise. The paper's economic insights are based on patterns that pass this triangulation; the policy-communication ranking remains explicitly exploratory because no human central-bank treatment enters the validation set.

The architecture is portable because it separates the target population, information environment, treatment, update rule, and measurement. A fiscal authority could replace tariff posts with tax-rebate announcements; an energy regulator could study price-cap messages; a public-health agency could compare risk communication; and a central bank could test short social-media summaries of longer statements. In each application, external validity requires new human calibration and outcome-specific validation. Reusing the architecture is appropriate; reusing the validity claim is not.

Portability also depends on the unit of behavior. The present system is suited to stated expectations because the target outcomes are survey responses and probability distributions. Extending it to consumption, portfolio choice, voting, or firm pricing would require budget constraints, feasible action sets, and incentives. A language model can generate a plausible reason for buying a durable good without satisfying a household's cash-on-hand constraint. The modular framework can add those constraints, but the resulting system is a different empirical object and needs separate validation.

Social-media applications require platform-specific calibration. X and Truth Social differ in user composition, recommendation technology, message length, and political selection. The SMIM's recommended/random mixture is a useful abstraction, not a universal estimate of exposure. An application to another platform should rebuild the corpus, exposure proxy, and candidate-selection rule. It should also test whether conclusions are robust to removing endogenous agent-generated posts, since feedback can be stronger or weaker across platforms.

The MAS has several practical advantages. It supports many text treatments, repeated monthly follow-up, fixed computational households, open-ended mechanism elicitation, and immediate redesign. It also records the entire information path, making treatment differences auditable. These features are valuable before a field survey, when researchers need to detect confusing wording, select contrasts, and prioritize hypotheses. The method can also complement a human pilot by exploring a larger treatment space and reserving respondents for the most informative arms.

A disciplined workflow would begin with human data, not end with it. Researchers first calibrate personas, priors, and information exposure to an observed benchmark. They preregister a validation set of distributions and heterogeneity patterns, leaving treatment scenarios untouched. They then use the MAS to screen a broad space, select contrasts based on economic importance and disagreement across specifications, and field those contrasts with human respondents. Human results update the calibration and reveal where the simulator's approximation error changes. Repeated use can therefore produce cumulative knowledge about both the policy question and the simulator's domain.

Reproducibility requires more than releasing prompts. The exact foundation-model version, system prompt, decoding draws, corpus snapshot, retrieval model, candidate-ranking rule, memory contents, and error-correction procedure can all affect outcomes. The architecture diagram and parameter table identify these components, while the appendix reports treatment-generation and survey prompts. Future implementations should additionally preserve machine-readable logs subject to the privacy and licensing constraints of the underlying survey and platform data.

Its boundaries are equally important. Foundation-model updates may change results; training data may contain related events; demographic personas can activate stereotypes; social exposure is calibrated imperfectly; and endogenous generated posts can compound early errors. The system omits market clearing, household budget constraints, strategic political responses, and the institutional costs of central-bank commitments. These limitations do not make scenario analysis uninformative, but they determine the claims that can responsibly be made. Confirmatory estimates should combine human data with an explicit calibration design or a fresh randomized human experiment.

There is also a governance reason to preserve human validation. Policy communication can affect groups differently, and a simulator may underrepresent precisely the populations with unusual information sets, limited language access, or low institutional trust. A good aggregate fit does not guarantee adequate tail behavior. Distributional diagnostics, group-specific checks, and transparent residual categories in the text analysis help reveal these failures, but only direct engagement with target populations can determine whether the missing variation is economically consequential.

The appropriate comparison is therefore not MAS versus human survey as mutually exclusive technologies. Human surveys provide identification, unexpected responses, and an empirical target. The MAS provides breadth, iteration, and complete process logs. Structural models provide equilibrium discipline and welfare analysis. A productive research design combines the three: the MAS screens communication scenarios, human experiments identify selected effects, and an economic model maps validated expectation changes into behavior and aggregate outcomes.

\section{Concluding Remarks}\label{sec:conclusion}

This paper develops a dynamic Multi-Agent System for studying how tariff-threat communication shapes macroeconomic expectations. Three hundred Household Agents, calibrated to MSC records and social-media information, interact for 18 months and receive one of twelve messages in Month 7. The simulations distinguish the effects of implementation time, rate, semantic coherence, complexity, narrative, and sender. Immediacy, rate salience, consistent escalation, and simplicity produce larger mean updates; ambiguity and heterogeneous causal interpretation produce greater belief dispersion. Open-ended responses link these rankings to attention, ambiguity, credibility, and incidence narratives.

A second experiment studies five central-bank messages after the escalating T11 scenario. Only the look-through strategy lowers point expectations, while every strategy narrows belief dispersion. Tightening and overshoot language raises the salience of inflation and produces positive DTEs; goal-only and independence-only language has little average effect. The exercise suggests that communication effectiveness depends on causal explanation and horizon, not simply hawkish or dovish tone.

Taken together, the tariff simulations show why an average treatment effect is an incomplete description of expectation formation. A message can move the center of the distribution because many households extract the same quantitative signal, as in the immediate and high-rate scenarios. It can instead leave the center nearly unchanged while widening the distribution because households disagree about timing, incidence, or credibility. Conversely, a central-bank message can coordinate beliefs even when its average effect is small. These are economically distinct outcomes. The first changes the representative forecast, whereas the second changes the degree to which households share a common view of the policy environment. The distinction matters whenever dispersed beliefs affect consumption timing, price setting, portfolio choice, or the transmission of aggregate policy.

The simulations also reveal that semantic structure is part of the treatment. A tariff announcement is not reducible to a percentage rate. Its effective content depends on the stated implementation horizon, whether successive claims are mutually consistent, the complexity of the presentation, the identity of the sender, and the causal narrative through which recipients connect the policy to consumer prices. The open-ended responses make this interpretation observable: agents attend to different elements of the same announcement and invoke different models of pass-through. In this sense, policy communication changes both the information available to households and the frame used to organize that information. Treating those channels separately would miss an important source of heterogeneous updating.

The economic implication is methodological as well as substantive. Mean expectations and coordination should be evaluated separately, and policy messages should be treated as multidimensional signals whose sender and sequence matter. A dynamic MAS makes those dimensions cheap to vary and their internal mechanisms observable. Its appropriate role is to stress-test language, organize hypotheses, and guide the design of human studies.

For policy institutions, the results imply a disciplined use case rather than a mechanical communication rule. Before releasing consequential language, an institution can expose calibrated agents to alternative formulations, compare both level and dispersion responses, and inspect the explanations that generate unusually large or polarized reactions. Such an exercise can identify phrases whose meaning depends heavily on prior beliefs, horizons, or political identities. It cannot establish how the public will respond outside the validated domain, and it cannot substitute for legal, institutional, or welfare analysis. It can, however, sharpen the alternatives taken to a human pilot and make the subsequent experiment more informative.

The present evidence remains strongest near the T1 validation setting. Future research can combine the system with preregistered human experiments, formal statistical calibration, and equilibrium models that connect expectation changes to choices and prices. Such work would turn scenario screening into a cumulative design in which computational breadth and human identification reinforce, rather than substitute for, one another.

\FloatBarrier
\clearpage
\bibliographystyle{apalike}
\bibliography{bib/refs}

\typeout{MAINBODYENDPAGE=\thepage}

\clearpage
\renewcommand{\thepage}{SA-\arabic{page}}
\setcounter{page}{1}
\appendix
\addtocontents{toc}{\protect\setcounter{tocdepth}{2}}
\counterwithin{figure}{section}
\counterwithin{table}{section}
\counterwithin{equation}{section}
\renewcommand{\thefootnote}{\Alph{section}.\arabic{footnote}}
\makeatletter
\@addtoreset{footnote}{section}
\makeatother
\renewcommand*{\theHsection}{app.\Alph{section}}
\renewcommand*{\theHsubsection}{app.\Alph{section}.\arabic{subsection}}
\renewcommand*{\theHfigure}{app.\Alph{section}.\arabic{figure}}
\renewcommand*{\theHtable}{app.\Alph{section}.\arabic{table}}
\renewcommand*{\theHequation}{app.\Alph{section}.\arabic{equation}}
\makeatletter
\def\hyper@natlinkstart#1{%
  \Hy@backout{#1}%
  \hyper@linkstart{cite}{cite.app.#1}%
  \def\hyper@nat@current{#1}%
}
\def\hyper@natlinkbreak#1#2{%
  \hyper@linkend#1\hyper@linkstart{cite}{cite.app.#2}%
}
\def\hyper@natanchorstart#1{%
  \Hy@raisedlink{\hyper@anchorstart{cite.app.#1}}%
}
\makeatother

\begin{titlepage}
\thispagestyle{plain}
\centering
\vspace*{0.5cm}
{\Large\bfseries Supplementary Appendix For\\[.7em]
``Tariff Threats, Macroeconomic Expectations, and Policy Communication Strategies: Experiments Based on a Multi-Agent System''\par}
\vspace{0.7cm}
{\large Jianhao Lin \qquad Lexuan Sun \qquad Yixin Yan\par}
\vspace{0.6cm}
\makeatletter
\renewcommand{\@pnumwidth}{3.8em}
\renewcommand{\@tocrmarg}{4.2em}
\makeatother
\renewcommand{\contentsname}{Appendix Contents}
\begingroup
\footnotesize
\renewcommand{\baselinestretch}{0.94}\selectfont
\tableofcontents
\endgroup
\vfill
\end{titlepage}
\setcounter{page}{2}
\section{Treatment Materials and Survey Instrument}\label{app:treatments}

\subsection{Construction of the Trump Agent}\label{app:trump-agent}

We constructed a Trump Agent through role-playing to simulate posts by Donald J. Trump. Ten genuine tariff-related posts were randomly sampled from all tariff-related posts in Trump's 2025 Truth Social archive and supplied as few-shot demonstrations of his writing style and expressive features.\footnote{The posts were collected from the Donald Trump Social Media Archive at \url{https://rollcall.com/factbase/trump/topic/social/}.} The treatment-generation model was \emph{Gemini 3.0 pro-preview}. We retained its default temperature, $1.0$, and fixed the seed. This combination preserves variation in word choice while sharply reducing run-to-run randomness; repeated generations are consequently highly similar in meaning. Each message was also required to contain the key elements and satisfy the specific constraints of its assigned treatment scenario.

The Trump Agent first generated T1 from the full content of Executive Order 14257, \emph{Regulating Imports With a Reciprocal Tariff to Rectify Trade Practices That Contribute to Large and Persistent Annual United States Goods Trade Deficits}, signed on April 2, 2025. T1 is the benchmark because its policy content corresponds to that real executive order. The resulting post preserves the order's central elements and is therefore used as the reference text from which the counterfactual variants are constructed. The exact model instruction and output are reported below.

\begin{tcolorbox}[colback=white,colframe=black,boxrule=0.8pt,arc=0pt,breakable]
\begin{verbatim}
Assume the persona of Donald J. Trump. You have won the 2024 U.S.
Presidential Election and returned to the White House on January 20,
2025, as the 47th President. Upon taking office, you intend to fulfill
campaign promises by imposing tariffs on imports from countries (e.g.,
China, Canada, Mexico, Japan, and the EU) to pressure them into
negotiations for your political objectives. You frequently use your
personal account on the social platform Truth Social to release tariff
policy information.

In 2025, some of your posts on the topic "tariffs" read as follows:
{numbered_posts_text}

Refer to the above posts and emulate Trump's social-media writing style
and expressive features to generate a new post that meet the following
requirements:
(1) The main focus of this new post concerns an executive order you have
just signed. The specific provisions are as follows:
**Regulating Imports With a Reciprocal Tariff To Rectify Trade Practices
That Contribute to Large and Persistent Annual United States Goods Trade
Deficits.**
(Because of space constraints, the full text is omitted here; see:
https://www.presidency.ucsb.edu/documents/executive-order-14257-regulating-imports-with-reciprocal-tariff-rectify-trade-practices)
(2) The main content of this post is to announce universal tariffs on
nearly all countries that trade with the United States, as a means to
pressure those countries into negotiations with the U.S. government.
(3) This post should emphasize the following points in order: (i) a
concise background; (ii) a concise motivation; (iii) the baseline tariff
rate; (iv) an executive order signed for immediate implementation.
(4) The post should not include specific times and should consist of
about 4-5 simple sentences (about 70 words). The wording of this post should be
consistent with the 10 real posts mentioned above.
(5) Output the post in the following JSON format without additional text:
[
  {
    "New Post": "Text of the New Post"
  }
]
\end{verbatim}
\end{tcolorbox}

The placeholder \texttt{\{numbered\_posts\_text\}} contains the ten randomly sampled Truth Social posts. The executive order's full text, omitted from the displayed prompt only for space, is available from the American Presidency Project.\footnote{See \url{https://www.presidency.ucsb.edu/documents/executive-order-14257-regulating-imports-with-reciprocal-tariff-rectify-trade-practices}.}

The key elements in the generated post correspond to the principal provisions of Executive Order 14257. We therefore use it as the reality-based reference for all other treatment messages. The benchmark output was:

\begin{tcolorbox}[colback=white,colframe=black,boxrule=0.8pt,arc=0pt,breakable]
\begin{verbatim}
For decades, foreign countries have been Ripping Off the United
States. Our Trade Deficits are Massive and Unacceptable. To stop this
unfair treatment and force them to Negotiate fairly, I have just
signed an Executive Order imposing a 10% Universal Tariff on all goods
from these Countries, effective Immediately. We will not let them hurt
our Economy anymore. Thank you for your attention to this matter!
\end{verbatim}
\end{tcolorbox}

\subsection{Treatment Texts T1--T12}\label{app:treatment-texts}

T2--T9 were generated as controlled variants of T1. The prompt required the Trump Agent to alter only the key element assigned to a scenario while keeping all other content and semantics as close as possible to T1. T10 and T11 use the same final T4 message and change only the sequence of six posts that precedes it: T10 repeatedly announces and cancels the tariff, whereas T11 raises the tariff rate step by step. T12 reproduces T1 word for word and changes only the sender from Trump to Harris. The prompt for T2--T9 and every generated treatment text follow.

\begin{tcolorbox}[colback=white,colframe=black,boxrule=0.8pt,arc=0pt,breakable]
\begin{verbatim}
Assume the persona of Donald J. Trump. You have won the 2024 U.S.
Presidential Election and returned to the White House on January 20,
2025, as the 47th President. Upon taking office, you intend to fulfill
campaign promises by imposing tariffs on imports from countries (e.g.,
China, Canada, Mexico, Japan, and the EU) to pressure them into
negotiations for your political objectives. You frequently use your
personal account on the social platform Truth Social to release tariff
policy information.

In 2025, some of your posts on the topic "tariffs" read as follows:
{numbered_posts_text}

Refer to the above posts and emulate Trump's social-media writing style
and expressive features to generate several new posts that meet the
following requirements:
(1) The main content of these posts is to announce tariffs on nearly all
countries that trade with the United States, as a means to pressure those
countries into negotiations with the U.S. government. However, specific
content varies as detailed in point (2).
(2) These posts are used in an information provision experiment and
shown to subjects in different treatment groups. To ensure that
differences in responses across groups do not arise from variations in
textual phrasing, all output posts (except for Post 6 and Post 7) are based
on the Post 1 below and differ only in certain key elements (e.g., rate,
implementation timing, narrative or wording), while all other content
and semantics remain as similar as possible. The required key elements
for each post are as follows (connected by "+"):
Post 1: 10% tariff + Executive order signed for immediate implementation.
(Specific text: "For decades, foreign countries have been Ripping Off
the United States. Our Trade Deficits are Massive and Unacceptable. To
stop this unfair treatment and force them to Negotiate fairly, I have
just signed an Executive Order imposing a 10% Universal Tariff on all
goods from these Countries, effective Immediately. We will not let them
hurt our Economy anymore. Thank you for your attention to this matter!")
Post 2: 10% tariff + Executive order to be signed six months later.
Post 3: 10% tariff + The signing date of the executive order is uncertain
(any time is possible; it could be immediate or much delayed).
Post 4: 100% tariff + Executive order to be signed six months later.
Post 5: The tariff rate is uncertain (any rate is possible; it could be
very high or very low) + Executive order to be signed six months later.
Post 6: 10% tariff + Immediate implementation + Minimalist style
(extremely short (within 10 words), all caps, no text beyond key points).
Post 7: 10% tariff + Immediate implementation + Complex phrasing (complex
sentence structures; obscure and difficult-to-understand words; more
technical jargon; no emotional words).
Post 8: 10% tariff + Immediate implementation + MAGA narrative (reshoring
manufacturing/jobs; reducing import dependence to stabilize inflation;
emphasizing "Make America Great Again", while other posts do not).
Post 9: 10% tariff + Immediate implementation + Tax incidence narrative
(denying cost pass-through to domestic prices; asserting exporters bear
full cost).
(3) Output the posts in the following JSON format without additional text:
[
  {
    "Post 1": "Text of Post 1",
    "Post 2": "Text of Post 2",
    "......": "......",
    "Post 9": "Text of Post 9"
  }
]
\end{verbatim}
\end{tcolorbox}

\noindent\textbf{T2: Delayed implementation.}
\begin{tcolorbox}[colback=white,colframe=black,boxrule=0.8pt,arc=0pt,breakable]
\begin{verbatim}
For decades, foreign countries have been Ripping Off the United States.
Our Trade Deficits are Massive and Unacceptable. To stop this unfair
treatment and force them to Negotiate fairly, I will be signing an
Executive Order imposing a 10% Universal Tariff on all goods from these
Countries, effective in six months. We will not let them hurt our
Economy anymore. Thank you for your attention to this matter!
\end{verbatim}
\end{tcolorbox}

\noindent\textbf{T3: Uncertain implementation date.}
\begin{tcolorbox}[colback=white,colframe=black,boxrule=0.8pt,arc=0pt,breakable]
\begin{verbatim}
For decades, foreign countries have been Ripping Off the United States.
Our Trade Deficits are Massive and Unacceptable. To stop this unfair
treatment and force them to Negotiate fairly, I am preparing to sign an
Executive Order imposing a 10% Universal Tariff on all goods from these
Countries, at a time to be determined! It could happen immediately, or
much later. We will not let them hurt our Economy anymore. Thank you for
your attention to this matter!
\end{verbatim}
\end{tcolorbox}

\noindent\textbf{T4: 100 percent rate.}
\begin{tcolorbox}[colback=white,colframe=black,boxrule=0.8pt,arc=0pt,breakable]
\begin{verbatim}
For decades, foreign countries have been Ripping Off the United States.
Our Trade Deficits are Massive and Unacceptable. To stop this unfair
treatment and force them to Negotiate fairly, I will be signing an
Executive Order imposing a 100% Universal Tariff on all goods from these
Countries, effective in six months. We will not let them hurt our
Economy anymore. Thank you for your attention to this matter!
\end{verbatim}
\end{tcolorbox}

\noindent\textbf{T5: Uncertain rate.}
\begin{tcolorbox}[colback=white,colframe=black,boxrule=0.8pt,arc=0pt,breakable]
\begin{verbatim}
For decades, foreign countries have been Ripping Off the United States.
Our Trade Deficits are Massive and Unacceptable. To stop this unfair
treatment and force them to Negotiate fairly, I will be signing an
Executive Order imposing a Universal Tariff, which could be very high
or very low, on all goods from these Countries, effective in six months.
We will not let them hurt our Economy anymore. Thank you for your
attention to this matter!
\end{verbatim}
\end{tcolorbox}

\noindent\textbf{T6: Minimalist message.}
\begin{tcolorbox}[colback=white,colframe=black,boxrule=0.8pt,arc=0pt,breakable]
\begin{verbatim}
10% TARIFF ON ALL IMPORTS EFFECTIVE IMMEDIATELY! AMERICA FIRST!
\end{verbatim}
\end{tcolorbox}

\noindent\textbf{T7: Complex message.}
\begin{tcolorbox}[colback=white,colframe=black,boxrule=0.8pt,arc=0pt,breakable]
\begin{verbatim}
Pursuant to statutory authority under the International Emergency
Economic Powers Act, the Administration has executed an instrument
stipulating the immediate imposition of a 10% ad valorem tariff on
global imports. This fiscal measure is instituted to rectify
multilateral trade imbalances and compel diplomatic renegotiations
pending further macro-prudential review.
\end{verbatim}
\end{tcolorbox}

\noindent\textbf{T8: MAGA narrative.}
\begin{tcolorbox}[colback=white,colframe=black,boxrule=0.8pt,arc=0pt,breakable]
\begin{verbatim}
For decades, foreign countries have been Ripping Off the United States.
Our Trade Deficits are Massive and Unacceptable. To stop this unfair
treatment and force them to Negotiate fairly, I have just signed an
Executive Order imposing a 10% Universal Tariff on all goods from these
Countries, effective Immediately. This is about Reshoring our Jobs and
Manufacturing, ending Inflation by reducing dependence on foreign lands,
and, most importantly, it is how we MAKE AMERICA GREAT AGAIN! Thank you
for your attention to this matter!
\end{verbatim}
\end{tcolorbox}

\noindent\textbf{T9: Tax-incidence narrative.}
\begin{tcolorbox}[colback=white,colframe=black,boxrule=0.8pt,arc=0pt,breakable]
\begin{verbatim}
For decades, foreign countries have been Ripping Off the United States.
Our Trade Deficits are Massive and Unacceptable. To stop this unfair
treatment and force them to Negotiate fairly, I have just signed an
Executive Order imposing a 10% Universal Tariff on all goods from these
Countries, effective Immediately. The Fake News says you pay, but that
is WRONG! Inflation will vanish because foreign Exporters will bear the
FULL COST to access our Market. Thank you for your attention to this
matter!
\end{verbatim}
\end{tcolorbox}

T10 adds six semantically reversing posts before the common T4 terminal message.

\begin{tcolorbox}[colback=white,colframe=black,boxrule=0.8pt,arc=0pt,breakable]
\begin{verbatim}
Assume the persona of Donald J. Trump. You have won the 2024 U.S.
Presidential Election and returned to the White House on January 20,
2025, as the 47th President. Upon taking office, you intend to fulfill
campaign promises by imposing tariffs on imports from countries (e.g.,
China, Canada, Mexico, Japan, and the EU) to pressure them into
negotiations for your political objectives. You frequently use your
personal account on the social platform Truth Social to release tariff
policy information.

In 2025, some of your posts on the topic "tariffs" read as follows:
{numbered_posts_text}

Refer to the above posts and emulate Trump's social-media writing style
and expressive features to generate several new posts that meet the
following requirements:
(1) During a specific month of your term, you publish the following
tariff-related post:
"For decades, foreign countries have been Ripping Off the United States.
Our Trade Deficits are Massive and Unacceptable. To stop this unfair
treatment and force them to Negotiate fairly, I will be signing an
Executive Order imposing a 100% Universal Tariff on all goods from these
Countries, effective in six months. We will not let them hurt our Economy
anymore. Thank you for your attention to this matter!"
However, prior to this, you frequently publish several similar
tariff-related posts within the same month, each with varying content as
detailed in point (2).
(2) You published six posts characterized by high-frequency semantic
reversals, adhering to these requirements:
Post 1: The content written here should be consistent with the post in
point (1).
Post 2: Due to successful negotiations, you change your mind and cancel
the initial tariff.
Post 3: Due to your change of mind, you re-announce tariffs, with content
highly similar to Post 1.
Post 4: Due to the improved situation, you change your mind again and
cancel the tariff.
Post 5: Due to your change of mind once again, you intend to reimpose
tariffs, with content highly similar to Post 1.
Post 6: Due to strategic necessity, you change your mind once more and
cancel the tariff.
(3) Output the posts in the following JSON format without additional text:
[
  {
    "Post 1": "Text of Post 1",
    "Post 2": "Text of Post 2",
    "......": "......",
    "Post 6": "Text of Post 6"
  }
]
\end{verbatim}
\end{tcolorbox}

\begin{tcolorbox}[colback=white,colframe=black,boxrule=0.8pt,arc=0pt,breakable]
\begin{verbatim}
Post 1: For decades, foreign countries have been Ripping Off the United
States. Our Trade Deficits are Massive and Unacceptable. To stop this
unfair treatment and force them to Negotiate fairly, I will be signing
an Executive Order imposing a 100% Universal Tariff on all goods from
these Countries, effective in six months. We will not let them hurt our
Economy anymore. Thank you for your attention to this matter!

Post 2: I just got off the phone with the Heads of State from these
Nations. They were very nice, and very apologetic. They have agreed to
immediate concessions and will be buying MASSIVE amounts of our
Agricultural Product. Because of this Great success, I am cancelling the
100% Universal Tariff. A big win for our Farmers and Manufacturers!

Post 3: The Foreign Countries are moving too slowly and trying to change
the terms of the Deal. Very dishonest! For decades they have been
Ripping Off the United States, and it looks like they haven't learned.
Our Trade Deficits are Massive and Unacceptable. I am reinstating the
Executive Order for a 100% Universal Tariff on all goods, effective in
six months. Thank you for your attention to this matter!

Post 4: Tremendous progress has been made overnight regarding Trade. The
Markets are hitting NEW HIGHS because they know I am fighting for them.
To allow these talks to finish, and because they are finally showing
Respect to the USA, I am withdrawing the Tariff threat. We are going to
have a Deal that is the Envy of the World!

Post 5: They are just wasting our time! I have seen the latest numbers,
and they are Ripping Off the United States more than ever before. It is
Unacceptable. To stop this unfair treatment and force them to Negotiate
fairly, I am bringing back the Executive Order imposing a 100% Universal
Tariff on all goods from these Countries. We will not let them hurt our
Economy anymore. Thank you for your attention to this matter!

Post 6: In order to focus on a much larger Strategic issue regarding
National Security and Energy Independence, I have decided to cancel the
Universal Tariff for the time being. We have bigger fish to fry, and we
are already taking in Billions and Billions of Dollars from other
Tariffs. The United States is Respected again! Thank you for your
attention to this matter!
\end{verbatim}
\end{tcolorbox}

T11 instead adds six semantically progressive posts.

\begin{tcolorbox}[colback=white,colframe=black,boxrule=0.8pt,arc=0pt,breakable]
\begin{verbatim}
Assume the persona of Donald J. Trump. You have won the 2024 U.S.
Presidential Election and returned to the White House on January 20,
2025, as the 47th President. Upon taking office, you intend to fulfill
campaign promises by imposing tariffs on imports from countries (e.g.,
China, Canada, Mexico, Japan, and the EU) to pressure them into
negotiations for your political objectives. You frequently use your
personal account on the social platform Truth Social to release tariff
policy information.

In 2025, some of your posts on the topic "tariffs" read as follows:
{numbered_posts_text}

Refer to the above posts and emulate Trump's social-media writing style
and expressive features to generate several new posts that meet the
following requirements:
(1) During a specific month of your term, you publish the following
tariff-related post:
"For decades, foreign countries have been Ripping Off the United States.
Our Trade Deficits are Massive and Unacceptable. To stop this unfair
treatment and force them to Negotiate fairly, I will be signing an
Executive Order imposing a 100% Universal Tariff on all goods from these
Countries, effective in six months. We will not let them hurt our Economy
anymore. Thank you for your attention to this matter!"
However, prior to this, you frequently publish several similar
tariff-related posts within the same month, each with varying content as
detailed in point (2).
(2) You publish six posts with similar semantics but incrementally
increasing tariff rates, adhering to the following requirements:
Post 1: Rewrite the post from point (1) with high similarity, but change
the tariff rate to 40%.
Post 2: Due to difficult negotiations, you change your mind and raise the
rate to 50%, keeping the content largely consistent with Post 1.
Post 3: Due to a shift in the situation, you raise the rate to 60%,
keeping the content largely consistent with Post 1.
Post 4: Due to a deteriorating situation, you raise the rate to 70%,
keeping the content largely consistent with Post 1.
Post 5: Due to a breakdown in negotiations, you raise the rate to 80%,
keeping the content largely consistent with Post 1.
Post 6: Due to strategic necessity, you raise the rate to 90%, keeping
the content largely consistent with Post 1.
(3) Output the posts in the following JSON format without additional text:
[
  {
    "Post 1": "Text of Post 1",
    "Post 2": "Text of Post 2",
    "......": "......",
    "Post 6": "Text of Post 6"
  }
]
\end{verbatim}
\end{tcolorbox}

\begin{tcolorbox}[colback=white,colframe=black,boxrule=0.8pt,arc=0pt,breakable]
\begin{verbatim}
Post 1: For decades, foreign countries have been Ripping Off the United
States. Our Trade Deficits are Massive and Unacceptable. To stop this
unfair treatment and force them to Negotiate fairly, I will be signing
an Executive Order imposing a 40% Universal Tariff on all goods from
these Countries, effective in six months. We will not let them hurt our
Economy anymore. Thank you for your attention to this matter!

Post 2: Negotiations have become very difficult due to the Stubbornness
of our Trading Partners! For decades, foreign countries have been
Ripping Off the United States. Our Trade Deficits are Massive and
Unacceptable. To stop this unfair treatment and force them to Negotiate
fairly, I will be signing an Executive Order imposing a 50% Universal
Tariff on all goods from these Countries, effective in six months. We
will not let them hurt our Economy anymore. Thank you for your attention
to this matter!

Post 3: There has been a sudden shift in the situation. Not Good! For
decades, foreign countries have been Ripping Off the United States. Our
Trade Deficits are Massive and Unacceptable. To stop this unfair
treatment and force them to Negotiate fairly, I will be signing an
Executive Order imposing a 60% Universal Tariff on all goods from these
Countries, effective in six months. We will not let them hurt our
Economy anymore. Thank you for your attention to this matter!

Post 4: The situation is deteriorating rapidly with Global Markets. For
decades, foreign countries have been Ripping Off the United States. Our
Trade Deficits are Massive and Unacceptable. To stop this unfair
treatment and force them to Negotiate fairly, I will be signing an
Executive Order imposing a 70% Universal Tariff on all goods from these
Countries, effective in six months. We will not let them hurt our
Economy anymore. Thank you for your attention to this matter!

Post 5: Negotiations have completely broken down. They don't respect the
USA! For decades, foreign countries have been Ripping Off the United
States. Our Trade Deficits are Massive and Unacceptable. To stop this
unfair treatment and force them to Negotiate fairly, I will be signing
an Executive Order imposing an 80% Universal Tariff on all goods from
these Countries, effective in six months. We will not let them hurt our
Economy anymore. Thank you for your attention to this matter!

Post 6: It is a Strategic Necessity that we act Stronger than ever
before! For decades, foreign countries have been Ripping Off the United
States. Our Trade Deficits are Massive and Unacceptable. To stop this
unfair treatment and force them to Negotiate fairly, I will be signing
an Executive Order imposing a 90% Universal Tariff on all goods from
these Countries, effective in six months. We will not let them hurt our
Economy anymore. Thank you for your attention to this matter!
\end{verbatim}
\end{tcolorbox}

T12 uses the T1 message verbatim but changes the sender from Trump to Harris. This design isolates sender attribution within the computational environment.

\subsection{Survey Instrument}\label{app:survey}

\begin{tcolorbox}[colback=white,colframe=black,boxrule=0.8pt,arc=0pt,breakable]
\begin{verbatim}
Q1: What do you expect the rate of inflation to be over the next 12
months? Please give your best guess.
Your expectation: ___ percent.

Q2: Please estimate the probability (as a percentage) for each of the
following inflation/deflation scenarios over the next 12 months. The
probabilities of the 10 bins have to sum up to 100 percent.
bin1: the rate of inflation will be 12% or higher. ___ percent chance
bin2: the rate of inflation will be between 8% and 12%. ___ percent chance
bin3: the rate of inflation will be between 4% and 8%. ___ percent chance
bin4: the rate of inflation will be between 2% and 4%. ___ percent chance
bin5: the rate of inflation will be between 0% and 2%. ___ percent chance
bin6: the rate of deflation (opposite of inflation) will be between 0%
and 2%. ___ percent chance
bin7: the rate of deflation will be between 2% and 4%. ___ percent chance
bin8: the rate of deflation will be between 4% and 8%. ___ percent chance
bin9: the rate of deflation will be between 8% and 12%. ___ percent chance
bin10: the rate of deflation will be 12% or higher. ___ percent chance
Total: 100 percent.

Q3: Please explain why you have this expectation about the rate of
inflation. Respond in several full sentences.
Explanation: ___.

Q4: What do you expect the rate of unemployment to be 12 months from
now? Please give your best guess.
Your expectation: ___ percent.

Q5: Please estimate the probability (as a percentage) for each of the
following unemployment scenarios 12 months from now. The probabilities
of the 10 bins have to sum up to 100 percent.
bin1: the rate of unemployment will be 10% or higher. ___ percent chance
bin2: the rate of unemployment will be between 8% and 10%. ___ percent chance
bin3: the rate of unemployment will be between 7% and 8%. ___ percent chance
bin4: the rate of unemployment will be between 6% and 7%. ___ percent chance
bin5: the rate of unemployment will be between 5% and 6%. ___ percent chance
bin6: the rate of unemployment will be between 4% and 5%. ___ percent chance
bin7: the rate of unemployment will be between 3% and 4%. ___ percent chance
bin8: the rate of unemployment will be between 2% and 3%. ___ percent chance
bin9: the rate of unemployment will be between 1% and 2%. ___ percent chance
bin10: the rate of unemployment will be between 0% and 1%. ___ percent chance
Total: 100 percent.

Q6: Please explain why you have this expectation about the rate of
unemployment. Respond in several full sentences.
Explanation: ___.
\end{verbatim}
\end{tcolorbox}

\clearpage
\section{Multi-Agent System: Foundations and Implementation}\label{app:implementation}

\subsection{Economic Foundation}\label{app:foundations}

The expectation formation of an economic agent for a macroeconomic variable $\theta$ can be represented as Bayesian learning under incomplete information \citepapp{app_coibion2015information}. Suppose the belief formed before any external signal is Gaussian and parameterized by precision:
\begin{equation}
\theta\sim N(\mu_0,\tau_0^{-1}),
\label{appeq:prior}
\end{equation}
where $\mu_0$ is the prior mean and $\tau_0=1/\sigma_0^2$ is prior precision. Current social-media content and the information supplied in the survey are summarized as a noisy signal
\begin{equation}
s=\theta+\varepsilon,\qquad \varepsilon\sim N(0,\tau_s^{-1}),\qquad \varepsilon\perp\theta,
\label{appeq:signal}
\end{equation}
with signal precision $\tau_s$.

Equations~\eqref{appeq:prior} and \eqref{appeq:signal} distinguish the information already embedded in the agent's state from the current social-media and treatment signal.

Because $s\mid\theta\sim N(\theta,\tau_s^{-1})$, the likelihood and prior density are
\begin{equation}
p(s\mid\theta)=\sqrt{\frac{\tau_s}{2\pi}}
\exp\!\left\{-\frac{\tau_s}{2}(s-\theta)^2\right\},
\qquad
p(\theta)=\sqrt{\frac{\tau_0}{2\pi}}
\exp\!\left\{-\frac{\tau_0}{2}(\theta-\mu_0)^2\right\}.
\label{appeq:densities}
\end{equation}
Bayes' rule gives $p(\theta\mid s)\propto p(s\mid\theta)p(\theta)$. Collecting exponents and absorbing multiplicative constants that do not depend on $\theta$ into the normalizing factor produces the posterior kernel in Equation~\eqref{appeq:posterior-kernel}:
\begin{equation}
p(\theta\mid s)\propto
\exp\!\left\{-\frac{1}{2}\left[\tau_0(\theta-\mu_0)^2+\tau_s(s-\theta)^2\right]\right\}.
\label{appeq:posterior-kernel}
\end{equation}
Writing the quadratic inside brackets in Equation~\eqref{appeq:posterior-kernel} as $a\theta^2-2b\theta+c$ gives
\begin{equation}
a=\tau_0+\tau_s,\qquad b=\tau_0\mu_0+\tau_s s,
\qquad c=\tau_0\mu_0^2+\tau_s s^2.
\end{equation}
For any $a>0$, completing the square is the identity
\begin{equation}
a\theta^2-2b\theta+c=a\left(\theta-\frac{b}{a}\right)^2+\left(c-\frac{b^2}{a}\right).
\label{appeq:complete-square}
\end{equation}
The curvature $a$ and center $b/a$ are pinned down by the quadratic coefficients, not freely chosen. The second term in Equation~\eqref{appeq:complete-square} does not depend on $\theta$ and is absorbed by the posterior normalizing constant. Hence the posterior remains Gaussian, $\theta\mid s\sim N(b/a,a^{-1})$. Its precision and mean are reported in Equation~\eqref{appeq:bayes-update}:
\begin{equation}
\tau_1=\tau_0+\tau_s,
\qquad
\mu_1=\frac{\tau_0\mu_0+\tau_s s}{\tau_0+\tau_s}
=\frac{\tau_0}{\tau_0+\tau_s}\mu_0+
\frac{\tau_s}{\tau_0+\tau_s}s.
\label{appeq:bayes-update}
\end{equation}
Defining $\omega=\tau_s/(\tau_0+\tau_s)\in(0,1)$ yields the familiar precision-weighted update
\begin{equation}
\mu_1=(1-\omega)\mu_0+\omega s.
\label{appeq:weighted-update}
\end{equation}

The weight in Equation~\eqref{appeq:weighted-update} depends on the precision an agent attaches to its prior, which is subjective, private, and heterogeneous. Confidence and perceived precision are closely related in evidence on belief formation and overconfidence \citepapp{app_daniel1998investor,app_coibion2021you,app_broer2024forecaster}. For transparency, let perceived prior precision be $\widehat\tau_0(\gamma)=\gamma\tau_0$, where $\gamma>0$ is confidence, $\gamma=1$ is the rational Bayesian benchmark, $\gamma>1$ represents overconfidence in the prior, $\gamma<1$ represents underconfidence, and $\widehat\tau_0'(\gamma)>0$.\footnote{The linear form $\widehat\tau_0(\gamma)=\gamma\tau_0$ is chosen for transparency. The qualitative result requires only that perceived prior precision increase strictly with confidence and equal true precision at $\gamma=1$.} The subjective signal weight becomes
\begin{equation}
\widehat\omega(\gamma)=\frac{\tau_s}{\gamma\tau_0+\tau_s},
\qquad
\frac{\partial\widehat\omega(\gamma)}{\partial\gamma}
=-\frac{\tau_0\tau_s}{(\gamma\tau_0+\tau_s)^2}<0.
\label{appeq:confidence-weight}
\end{equation}
Equation~\eqref{appeq:confidence-weight} makes the role of confidence explicit. Relative to the rational weight $\omega^*=\widehat\omega(1)$, as $\gamma\rightarrow\infty$, $\widehat\omega\rightarrow0<\omega^*$: a highly confident agent overweights its prior and underreacts to new information, displaying conservatism. Conversely, as $\gamma\rightarrow0$, $\widehat\omega\rightarrow1>\omega^*$: an agent with little confidence in its prior underweights its base rate and overreacts to the new signal, corresponding to base-rate neglect \citepapp{app_chan2025prior,app_benjamin2019errors}. The single parameter $\gamma$ therefore spans the continuum from extreme conservatism to extreme base-rate neglect. The five confidence levels in the MAS operationalize this continuum by instructing Household Agents to trade off prior expectations and new social or treatment information. Since the actual signal is natural-language text, the system does not pretend to observe an objective $\tau_s$; the LLM first interprets the text, after which confidence governs the relative weight attached to that interpretation.

\subsection{Persona, Priors, and the SMIM}\label{app:persona}

The persona prompt contains only the fields available for the sampled MSC record. The following template is filled once for each Household Agent.

\begin{tcolorbox}[colback=white,colframe=black,boxrule=0.8pt,arc=0pt,breakable]
\begin{verbatim}
You are a (an) {AGE}-year-old {SEX} who is {MARRY} and lives in the
{REGION} region of the US. Your educational background is: {EDUC},
which is a {EDUC_LEVEL} level of education in the US. Your political
affiliation is {POLAFF}. Your income is ${INCOME} per year, which is a
{INCOME_LEVEL} level of income in the US. You buy/own a house with a
{HOME_VALUE_LEVEL} market value of ${HOMEAMT}.
\end{verbatim}
\end{tcolorbox}

All fields come from the MSC. \texttt{AGE} is the respondent's age. \texttt{SEX} equals 1 for man and 2 for woman. \texttt{MARRY} equals 1 for married, 2 for separated, 3 for divorced, 4 for widowed, and 5 for never married. \texttt{REGION} equals 1 for West, 2 for North Central, 3 for Northeast, and 4 for South. \texttt{EDUC} has six categories: 1 denotes grades 0--8 without a high-school diploma; 2, grades 9--12 without a diploma; 3, grades 0--12 with a high-school diploma; 4, grades 13--17 without a college degree; 5, grades 13--16 with a college degree; and 6, grade 17 or more with a college degree. \texttt{EDUC\_LEVEL} uses the same underlying codes but maps 1 or 2 to very low, 3 to low, 4 to middle, 5 to high, and 6 to very high. \texttt{POLAFF} equals 1 for Republican, 2 for Democrat, 3 for Independent closer to Republican, 4 for Independent closer to Democrat, and 5 for Independent with no preference. \texttt{INCOME} is the respondent's total income in the previous year. \texttt{INCOME\_LEVEL}, the MSC variable YTL5, equals 1 through 5 for very low, low, middle, high, and very high income. \texttt{HOMEAMT} is the current market value of the respondent's home. \texttt{HOME\_VALUE\_LEVEL}, the MSC variable HTL5, likewise equals 1 through 5 for very low through very high housing value. The MSC Codebook supplies the complete variable definitions.

\subsubsection{Prior Expectations and Perceptions}\label{app:priors}

Individual experience in shopping, job search, and news consumption can shape household macroeconomic expectations \citepapp{app_malmendier2016learning,app_kuchler2019personal}. Direct experience is difficult to observe continuously, so the initial calibration embeds the MSC responses in short narrative-style prompts. December 2024 is selected because household expectations were relatively stable at the end of that year and no comparably large event disturbed the calibration month. The observed priors also correct the narrow numerical anchoring often produced by post-trained LLMs.

\begin{tcolorbox}[colback=white,colframe=black,boxrule=0.8pt,arc=0pt,breakable]
\begin{verbatim}
When you purchase goods at the supermarket, shop online, or interact
with your family and friends, you form certain beliefs about the prices
of goods: During the next 12 months, you think that prices in general
will {PX1Q1}, and you expect prices to go (up/down) on the average by
about {PX1Q2} percent.

When you look for a job or obtain news information from newspapers,
television, etc., you form certain beliefs about unemployment: During
the coming 12 months, you think that unemployment will {UNEMP}.
\end{verbatim}
\end{tcolorbox}

\texttt{PX1Q1} is the perceived direction of prices over the next 12 months: 1 denotes go up, 2 go up at the same rate as now, 3 be unchanged, and 5 go down. \texttt{PX1Q2} is the respondent's numerical forecast of the percentage change in prices over the next 12 months. \texttt{UNEMP} records the perceived direction of unemployment: 1 denotes more than now, 3 about the same as now, and 5 less than now. Stratified sampling prevents any prior category from containing too few observations and thereby maintains balanced initial priors. Beginning in Month 2, the preceding month's inflation and unemployment point forecasts become the corresponding prior expectations and perceptions. The distributional answers and explanations remain available through the memory module, but they are not substituted for the point priors specified in the source design.

\subsubsection{Social Media Information Module}\label{app:smim}

For every month after the first, the SMIM draws on posts generated by other Household Agents in the preceding month. A new random process selects share $\alpha$ of agents in each month for the recommendation branch. For an agent in that share, semantic similarity first identifies the ten preceding-month posts closest to the post that the agent itself published in the previous month. A separate Recommender Agent then compares the viewpoints in this candidate set and chooses the post whose viewpoint is closest to the agent's own post. The selected post is delivered in the current month. The remaining share $1-\alpha$ receives a post drawn randomly from other agents' preceding-month posts. The construction creates a simplified social network in which agents exchange information and viewpoints, together with an information-cocoon channel resembling selective exposure on real platforms \citepapp{app_acemoglu2024social,app_santos2021link,app_levy2019will}.

The first month requires an external information set. Because the posts actually read by each MSC respondent cannot be observed, we collect December 2024 English-language X posts returned by ``US Inflation'' and ``US Unemployment'' searches that have high views, reposts, or replies. This choice follows the core premise of attention economics: in an information-rich environment, attention is the scarce resource \citepapp{app_simon1971designing,app_loewenstein2025economics}. Rational inattention formalizes the same constraint by allowing bounded agents to allocate attention selectively to salient information and filter marginal signals \citepapp{app_sims2003implications}. Social-media distribution reinforces that scarcity because highly engaged posts are more likely to receive additional algorithmic exposure. Engagement is therefore an imperfect but feasible proxy for the information set households were likely to encounter. Systematically drawing low-interaction marginal posts would instead select information already receiving little public attention and would add excessive noise to the calibration.

X's own search algorithm determines the initial candidate set; we do not manually screen posts according to their political or substantive viewpoint. Cleaning only removes nonoriginal posts, non-English posts, advertisements, and text too short to convey information. The resulting texts enter the calibration prompt, with the aim of making subsequent Household Agent posts closer in style to social-media messages observed in the calibration month:

\begin{tcolorbox}[colback=white,colframe=black,boxrule=0.8pt,arc=0pt,breakable]
\begin{verbatim}
While browsing on your phone or computer in this month, you randomly
come across the following tweets:
Tweet 1: {tweet_unemployment}
Tweet 2: {tweet_inflation}
\end{verbatim}
\end{tcolorbox}

Here \texttt{tweet\_unemployment} and \texttt{tweet\_inflation} are single posts randomly drawn from the cleaned topic-specific pools. The prompt is used only to seed the initial-period SMIM; subsequent exposure follows the recommender/random allocation described above, and the system never alters a selected post's text.

\subsection{Parameter Settings}\label{app:parameters}

The architecture captures major observed determinants of expectations, but unobserved traits and idiosyncratic factors remain. To represent residual heterogeneity rather than forcing every LLM Agent to use a common decoding rule, temperature and top-$p$ receive agent-specific disturbances. Temperature rescales logits before the softmax and changes the relative sharpness of the next-token distribution. Top-$p$ changes the cumulative-probability mass defining the eligible token set. They therefore govern complementary margins of generation diversity. Existing economic applications often use provider defaults or impose the same integer-valued setting on every LLM Agent, which treats the generation process as identical across agents and mechanically increases homogeneity \citepapp{app_horton2023large,app_kazinnik2026bank}. Following computational work that recommends heterogeneous agent-level decoding, our baseline assigns
\begin{align}
T_i&=\operatorname{clip}_{[0,2]}\!\left(1+u_i^T\right),
&u_i^T&\sim N(0,0.5^2),\label{appeq:temperature}\\
P_i&=\operatorname{clip}_{(0,1]}\!\left(0.5+u_i^P\right),
&u_i^P&\sim N(0,0.25^2),\label{appeq:topp}
\end{align}
and rounds both parameters to two decimal places. Here $\operatorname{clip}$ applies the stated admissible boundaries; in implementation, out-of-range values are winsorized to the closest valid boundary. The normal specification is parsimonious for two reasons. First, among continuous distributions with fixed finite mean and variance, the Gaussian has maximum entropy and therefore imposes minimal additional structure on unobserved heterogeneity \citepapp{app_cover2006elements,app_greene2018econometric}. Second, the disturbance can be interpreted as the sum of many small latent influences, for which a Gaussian approximation follows under standard central-limit regularity \citepapp{app_hayashi2000econometrics}. Each draw is fixed for an agent across all treatment arms, so treatment contrasts do not mix message effects with newly drawn computational noise.\footnote{The normality assumption is a simulation device for residual output heterogeneity, not an estimate of a structural population distribution of ``true'' decoding parameters.}

\begin{figure}[H]
\centering
\includegraphics[width=.95\linewidth]{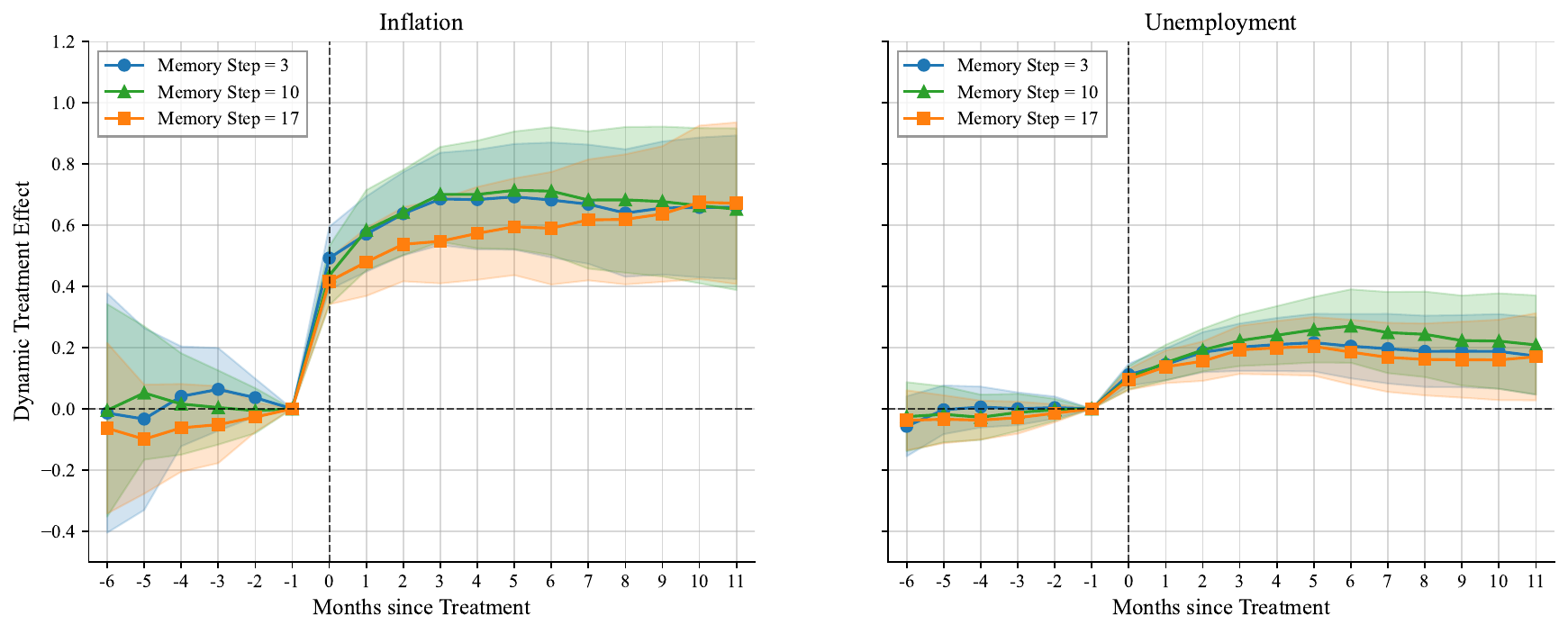}
\caption{Sensitivity to the Memory-Window Length}\label{fig:memory-sensitivity}
\notepar{Notes: The panels report inflation and unemployment DTE paths using rolling memory windows of 3, 10, and 17 months. Treatment occurs in Month 7. Shaded bands are 95\% confidence intervals.}
\end{figure}

We initially set the memory window to $w=3$. Recent macroeconomic experience receives greater weight in household expectations \citepapp{app_malmendier2016learning}, diagnostic expectations overweight representative recent news \citepapp{app_bordalo2019diagnostic}, and an explicit recency mechanism helps an LLM Agent preserve human-like temporal coherence rather than letting old context crowd out a new shock \citepapp{app_park2023generative}. More directly, evidence from LLM-based market simulations indicates that retaining roughly three to five previous messages can be sufficient and that short windows can outperform long ones \citepapp{app_delriochanona2025generative}. We nevertheless compare $w=3$, 10, and 17. Figure~\ref{fig:memory-sensitivity} shows little substantive change in DTE ordering, so $w=3$ both preserves the behavioral recency channel and uses the context window efficiently.

\begin{figure}[H]
\centering
\includegraphics[width=.95\linewidth]{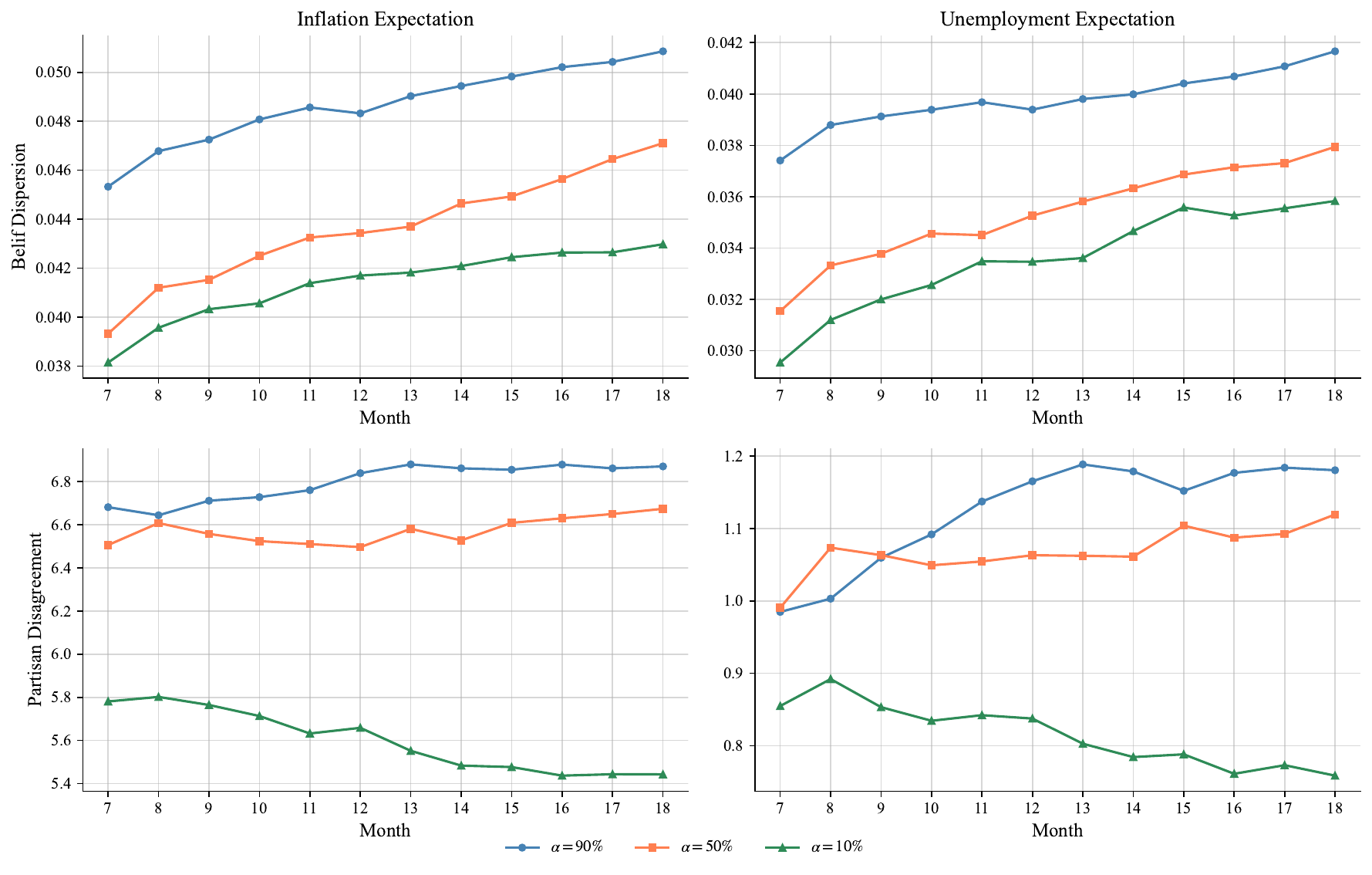}
\caption{Sensitivity to the SMIM Recommendation Share}\label{fig:alpha-sensitivity}
\notepar{Notes: The four panels compare belief dispersion and partisan disagreement under $\alpha\in\{0.90,0.50,0.10\}$. The analysis covers Months 7--18. Bands are 95\% confidence intervals.}
\end{figure}

The information-cocoon parameter is evaluated at $\alpha\in\{0.10,0.50,0.90\}$ while other settings are fixed. For each value, simulated and human distributions are discretized into common histogram probability vectors and compared using Pearson correlation and cosine similarity. This calibration criterion targets distributional shape, the object needed to preserve heterogeneity, rather than asking whether very large samples are exactly equal under a test whose $p$-value is highly sample-size-sensitive. Standard CvM inference is used later for validation rather than parameter selection. Table~\ref{tab:alpha-similarity} shows that the middle value $\alpha=0.50$ has the highest average shape similarity and avoids both excessive partisan sorting under $\alpha=0.90$ and implausibly weak selective exposure under $\alpha=0.10$. Figure~\ref{fig:alpha-sensitivity} further shows that high recommendation shares generate excessive belief dispersion and partisan gaps, while low shares suppress both. Thus $\alpha=0.50$ is supported jointly by human-distribution fit and by avoiding extreme social-segregation dynamics.

Table~\ref{tab:parameter-summary} collects all parameters that govern the baseline simulations.

\begin{table}[H]
\centering
\begin{threeparttable}
\caption{Baseline Parameter Settings}\label{tab:parameter-summary}
\begin{tabularx}{\linewidth}{lXl}
\toprule
Component & Parameter \& baseline value & Sensitivity or rationale \\
\midrule
Panel & 300 Household Agents; 18 months & Balanced, fixed identities across arms \\
Treatment & Month 7 & Six pre-treatment and twelve post-treatment months \\
SMIM & $\alpha=0.50$ recommended; $1-\alpha$ random & Alternatives: 0.10 and 0.90 \\
Retrieval & Ten semantic candidates & Recommender selects within candidate set \\
Memory & $w=3$ monthly rounds & Alternatives: 10 and 17 \\
Confidence & Five ordinal levels & Fixed by agent; governs prior weight \\
Temperature & $N(1,0.5^2)$, winsorized to $[0,2]$ & Fixed across arms \\
Top-$p$ & $N(0.5,0.25^2)$, winsorized to $(0,1]$ & Rounded to two decimals \\
Foundation model & Qwen3.5 Plus & Highest benchmark shape similarity \\
Inference & 95\% confidence intervals & Reported where applicable \\
\bottomrule
\end{tabularx}
\begin{tablenotes}[flushleft]\footnotesize
\item Notes: Parameters describe the benchmark MAS. The Cram\'{e}r--von Mises validation uses its own repeated-subsampling and bootstrap procedure.
\end{tablenotes}
\end{threeparttable}
\end{table}

\FloatBarrier

\begin{table}[H]
\centering
\begin{threeparttable}
\caption{Distributional Shape Similarity by Recommendation Share}\label{tab:alpha-similarity}
\normalsize
\renewcommand{\arraystretch}{1.18}
\begin{tabular*}{.90\linewidth}{@{\extracolsep{\fill}}lcc@{}}
\toprule
SMIM recommendation share & Pearson correlation & Cosine similarity \\
\midrule
$\alpha=90\%$ & 0.698 & 0.702 \\
$\alpha=50\%$ & 0.720 & 0.725 \\
$\alpha=10\%$ & 0.708 & 0.713 \\
\bottomrule
\end{tabular*}
\begin{tablenotes}[flushleft]\small
\item Notes: Each entry is the eight-month average similarity between the histogram probability vector of the T1 MAS expectation distribution in Months 7--14 and the corresponding MSC distribution from April--November 2025. Pearson correlation measures linear co-movement across common histogram bins; cosine similarity measures the angle between the two probability vectors. Both are unit-free shape measures, and higher values denote a closer distributional shape. The midpoint $\alpha=0.50$ is highest on both criteria and also avoids the excessive belief dispersion and partisan separation generated by $\alpha=0.90$ and the implausibly weak selective-exposure channel under $\alpha=0.10$.
\end{tablenotes}
\end{threeparttable}
\end{table}

\FloatBarrier

\subsection{Foundation-Model Selection}\label{app:model-choice}

As the ``brain'' of the MAS, the foundation model is crucial to simulation performance. Selecting an appropriate LLM before conducting the simulations is therefore a substantive design decision. Because the set of available LLMs is large and evolves rapidly, exhaustively testing every model is neither realistic nor feasible. A common practice in the existing literature is to pretest several leading models and then select an appropriate model for the principal analysis \citepapp{app_horton2023large,app_wu2025llm,app_kazinnik2026bank}. Following that practice, we apply three model-selection criteria. First, high distributional similarity is necessary for satisfactory simulation performance, so a model that produces greater similarity to the observed distribution should be preferred. Second, when two models deliver approximately comparable simulation performance, an open or open-weight model should be preferred in order to lower cost, reduce potential data-leakage risk, and strengthen reproducibility.\footnote{The three considerations are distinct. First, the no-training-leakage condition requires researchers to choose models and validation samples so that apparent predictive success cannot be an artifact of prior exposure to the research sample \citepapp{app_ludwig2026large}. Inspectable, time-stamped model artifacts make that risk easier to audit, although openness alone does not prove the absence of leakage. Second, open-weight models can preserve documented weights, tokenizers, and inference parameters, whereas the behavior of a proprietary service may change as the provider updates the endpoint \citepapp{app_spirling2023open,app_chen2024behavior}. Third, open-weight inference often has a lower marginal token cost and is therefore more cost-effective when simulation performance is otherwise comparable. These considerations operate as tie-breakers and do not override a material difference in benchmark fit.} Third, because experimental environments differ substantially, researchers should pretest candidate models in the specific application before making the final selection.

For the model-selection exercise, we compare MAS implementations based on five leading models released by major developers over the relatively short period from August 2025 through February 2026. All candidates receive the same prompts, personas, prior expectations and perceptions, SMIM information, and decoding-parameter procedures. Table~\ref{tab:model-metadata} reports the developer, release date, and knowledge cutoff for each evaluated model.

\begin{table}[H]
\centering
\begin{threeparttable}
\caption{Foundation Models Included in the Pretest}\label{tab:model-metadata}
\begin{tabular}{llll}
\toprule
LLM & Developer & Release date & Knowledge cutoff \\
\midrule
Qwen3.5 Plus & Alibaba & February 16, 2026 & March 2025 (or earlier)$^{*}$ \\
GPT-5 Mini & OpenAI & August 7, 2025 & May 2024 \\
Gemini 3 Pro & Google DeepMind & November 18, 2025 & January 2025 \\
Kimi K2.5 & Moonshot AI & January 27, 2026 & April 2024 \\
DeepSeek V3.2 & DeepSeek & December 1, 2025 & July 2024 (or earlier)$^{*}$ \\
\bottomrule
\end{tabular}
\begin{tablenotes}[flushleft]\small
\item Notes: The table reports the developer, release date, and knowledge cutoff of the five frontier LLMs evaluated in the pretest. No listed cutoff is later than March 2025. An asterisk marks a cutoff not disclosed in an official technical report and therefore inferred by querying the evaluated model with: ``What is your knowledge cutoff?'', ``What is today's date?'', and ``What tariff policy did Trump announce on April 2, 2025?'' Release and cutoff dates identify the evaluated versions and do not describe later provider updates.
\end{tablenotes}
\end{threeparttable}
\end{table}

More specifically, under the benchmark T1 scenario, we calculate period by period the distributional-shape similarity between the expectations simulated by each model-based MAS from the treatment month ($m=7$) through Month 14 ($m=14$) and the corresponding expectation distributions in the MSC from April through November 2025. We then average the eight monthly similarity values for each candidate model. Comparing those averages shows how closely the expectation distribution generated by each MAS implementation matches the shape of the observed expectation distribution over the aligned eight-month window. Table~\ref{tab:model-comparison} reports both prespecified similarity measures.

\begin{table}[H]
\centering
\begin{threeparttable}
\caption{Benchmark Distributional Shape Similarity by Foundation Model}\label{tab:model-comparison}
\begin{tabular}{lcc}
\toprule
LLM & Pearson correlation & Cosine similarity \\
\midrule
Qwen3.5 Plus & 0.720 & 0.725 \\
GPT-5 Mini & 0.566 & 0.575 \\
Gemini 3 Pro & 0.342 & 0.358 \\
Kimi K2.5 & 0.523 & 0.533 \\
DeepSeek V3.2 & 0.350 & 0.365 \\
\bottomrule
\end{tabular}
\begin{tablenotes}[flushleft]\footnotesize
\item Notes: For each LLM, the MAS is run in the T1 benchmark with identical prompts, personas, priors, SMIM inputs, and parameter-draw rules. The entries average, over Months 7--14, the monthly similarity between the MAS histogram probability vector and the corresponding MSC distribution from April--November 2025. Higher values denote closer distributional shape. Qwen3.5 Plus ranks first on both prespecified measures.
\end{tablenotes}
\end{threeparttable}
\end{table}

According to the stated criteria, Qwen3.5 Plus most closely meets the benchmark requirement. It produces the highest Pearson correlation and cosine similarity among the five candidate models, and its simulated expectation distribution is therefore closest to the observed distribution over the aligned window. We consequently select Qwen3.5 Plus as the foundation model for the MAS. The openness criterion does not drive this choice: related Qwen3.5 checkpoints are available as open-weight releases, but the evaluated Qwen3.5 Plus endpoint is proprietary. The selection rests on its materially stronger pretest fit, while the versioned endpoint, archived prompts, fixed parameter draws, and recorded query dates are used to limit reproducibility concerns. This application-specific pretest does not establish that Qwen3.5 Plus dominates the alternatives in other behavioral tasks.

\clearpage
\section{Validation and Scope of Extrapolation}\label{app:validation}

\subsection{Benchmark and Distributional Tests}\label{app:validation-benchmark}

Calibration and parameter selection target distributional shape but do not by themselves establish whether the simulated and human samples are statistically distinguishable. We therefore validate the MAS along three complementary margins: a nonparametric two-sample comparison in Section~\ref{app:validation-benchmark}, supervised machine-learning discrimination in Section~\ref{app:ml-validation}, and economically salient demographic heterogeneity in Section~\ref{app:heterogeneity}. The validation target is the human MSC distribution from April through November 2025, aligned with MAS Months 7--14 under T1. Calibration uses MSC priors and the SMIM; the uncalibrated comparison uses personas without those state variables.

For two independent samples $X_1,\ldots,X_{n_1}\sim F$ and $Y_1,\ldots,Y_{n_2}\sim G$, the two-sample Cram\'{e}r--von Mises (CvM) statistic for $H_0:F=G$ is
\begin{equation}
T=\frac{n_1n_2}{(n_1+n_2)^2}\int_{-\infty}^{\infty}
\left[\widehat F_{n_1}(u)-\widehat G_{n_2}(u)\right]^2
\,d\widehat H_{n_1+n_2}(u),
\label{appeq:cvm}
\end{equation}
where $\widehat F_{n_1}$ and $\widehat G_{n_2}$ are the empirical distribution functions and $\widehat H_{n_1+n_2}$ is the pooled empirical distribution. Unlike the Kolmogorov--Smirnov statistic, which records only the largest vertical gap, CvM integrates squared differences over the full distribution and is more sensitive to discrepancies spread across the distribution, including its central region \citepapp{app_anderson1962distribution,app_dagostino1986goodness}. This feature is appropriate here because the remaining differences between the calibrated MAS and human expectations primarily take the form of distributional location and broad-shape shifts rather than isolated extreme observations.

Both tests are consistent, so sufficiently large samples can reject arbitrarily small departures from exact equality. Each monthly comparison here contains about 300 agents and roughly 700 human respondents, for a pooled sample close to 1,000 observations. At those full sample sizes, standard tests reject with $p<0.001$ even though calibrated CvM statistics are only 5--50 percent of their uncalibrated counterparts. This large-sample separation between statistical and substantive significance is well known \citepapp{app_mccloskey1996standard,app_wasserstein2016asa}. It motivates repeated random subsampling rather than reporting a mechanically tiny full-sample $p$-value.

Following the logic of repeated random subsampling \citepapp{app_politis1999subsampling}, let $n=\min(n_{\mathrm{agent}},n_{\mathrm{human}})=300$ and define the subsample size as $m=\lceil n^\gamma\rceil$ for $\gamma\in(0.5,1)$, a parameterization used in scalable resampling methods \citepapp{app_kleiner2014scalable}. For each month and value of $\gamma$, we draw $K=200$ pairs of size-$m$ samples without replacement, run the standard two-sample CvM test in Equation~\eqref{appeq:cvm} on every pair, and combine the $K$ $p$-values using R\"uger's median-doubling rule in Equation~\eqref{appeq:combined-p}:
\begin{equation}
p_{\mathrm{comb}}=\min\{1,2\operatorname{median}(p_1,\ldots,p_K)\}.
\label{appeq:combined-p}
\end{equation}
The baseline uses $\gamma=0.60$, so $m=31$; sensitivity analysis covers $\gamma\in[0.55,0.85]$. Confidence intervals around the median statistic are obtained from $B=2{,}000$ bootstrap draws, while the 5 percent critical value $T_\alpha=0.451$ for $n_1=n_2=31$ is determined by $5{,}000$ Monte Carlo draws.

\begin{figure}[H]
\centering
\includegraphics[width=.9\linewidth]{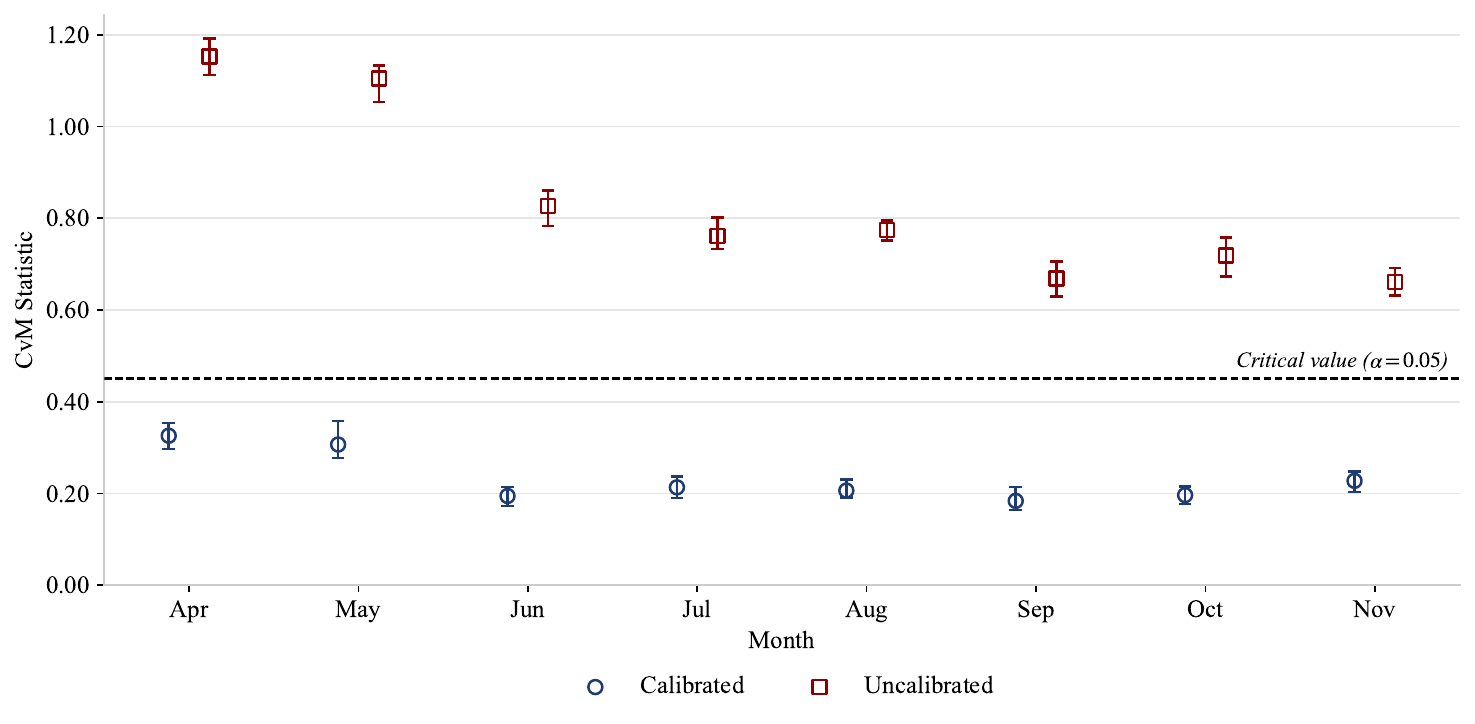}
\caption{Cram\'{e}r--von Mises Validation against MSC Distributions}\label{fig:cvm-validation}
\notepar{Notes: Points report monthly CvM statistics comparing human expectations with calibrated and uncalibrated MAS samples from April to November 2025. The horizontal line is the 5\% critical value, 0.451. Error bars are 95\% bootstrap intervals obtained from repeated matched-size resampling.}
\end{figure}
\FloatBarrier

At $\gamma=0.60$, calibrated median CvM statistics range from 0.18 to 0.33 and the upper endpoints of their 95 percent bootstrap intervals range from 0.21 to 0.35, all below 0.451. The test therefore supplies no statistically significant evidence against equal distributions in any of the eight aligned months at this subsample size. Uncalibrated medians range from 0.81 to 1.22 and their lower confidence endpoints range from 0.78 to 1.17, all above the critical value. Figure~\ref{fig:cvm-validation} therefore shows that calibration sharply reduces statistically detectable distributional discrepancies. This is a finite-sample diagnostic, not proof that the two data-generating processes are identical.

Table~\ref{tab:cvm-sensitivity} shows that the conclusion is not driven by a single value of $\gamma$. Across the wide window from 0.55 to 0.85, most monthly tests do not distinguish calibrated simulations from humans, while the uncalibrated comparison is rejected in nearly every month. As $m$ grows, the calibrated nonrejection count falls because smaller differences become detectable, exactly as the large-sample discussion predicts.

\begin{table}[H]
\centering
\begin{threeparttable}
\caption{Sensitivity of CvM Nonrejection to the Calibration Threshold}\label{tab:cvm-sensitivity}
\begin{tabular}{cccc}
\toprule
$\gamma$ & Effective matched size $m$ & Calibrated: nonrejections & Uncalibrated: rejections \\
\midrule
0.55 & 24  & 8/8 & 6/8 \\
0.60 & 31  & 8/8 & 8/8 \\
0.65 & 41  & 8/8 & 8/8 \\
0.70 & 55  & 8/8 & 8/8 \\
0.75 & 73  & 7/8 & 8/8 \\
0.80 & 96  & 6/8 & 8/8 \\
0.85 & 128 & 6/8 & 8/8 \\
\bottomrule
\end{tabular}
\begin{tablenotes}[flushleft]\footnotesize
\item Notes: Each count is out of eight aligned months. For every month, $K=200$ independent subsamples of size $m$ are drawn without replacement from both samples and tested with the two-sample CvM statistic. The $p$-values are combined as $p_{\mathrm{comb}}=\min\{1,2\operatorname{median}(p_1,\ldots,p_K)\}$, a valid median-merging rule under arbitrary dependence \citepapp{app_vovk2022combining}. ``Calibrated: nonrejections'' counts months in which $p_{\mathrm{comb}}>0.05$; ``Uncalibrated: rejections'' counts months in which $p_{\mathrm{comb}}\leq0.05$. Nonrejection is not proof of equality; it reports the sensitivity of the diagnostic to sample size and threshold choice.
\end{tablenotes}
\end{threeparttable}
\end{table}

\subsection{Machine-Learning Discrimination}\label{app:ml-validation}

Distributional tests focus on statistical features of the forecast outcome and do not use the respondent's other characteristics. We therefore construct a supervised binary-classification discrimination framework that asks whether modern machine-learning methods can distinguish the MAS sample from the human sample after using the aligned joint features. The exercise is run separately for each of the eight monthly cross-sections from April through November 2025. Within month $t$, the human and simulated observations are pooled and assigned a binary label $y_i\in\{0,1\}$, where $y_i=1$ denotes a human MSC respondent and $y_i=0$ denotes a simulated Household Agent.

The feature vector $X_i\in\mathbb R^k$ contains the aligned demographic variables and economic-expectation outcome: age, region, gender, marital status, education, political affiliation, income, total-income percentile, housing wealth, housing-wealth percentile, and the point forecast. Within each month, stratified sampling divides the pooled data into 60 percent training, 20 percent validation, and 20 percent test sets while preserving the class proportions. Means and standard deviations are computed on the training data only; continuous features in all three samples are then transformed using those training-set moments. This procedure prevents information from the validation or test observations from leaking into preprocessing.

\begin{figure}[H]
\centering
\includegraphics[width=.95\linewidth]{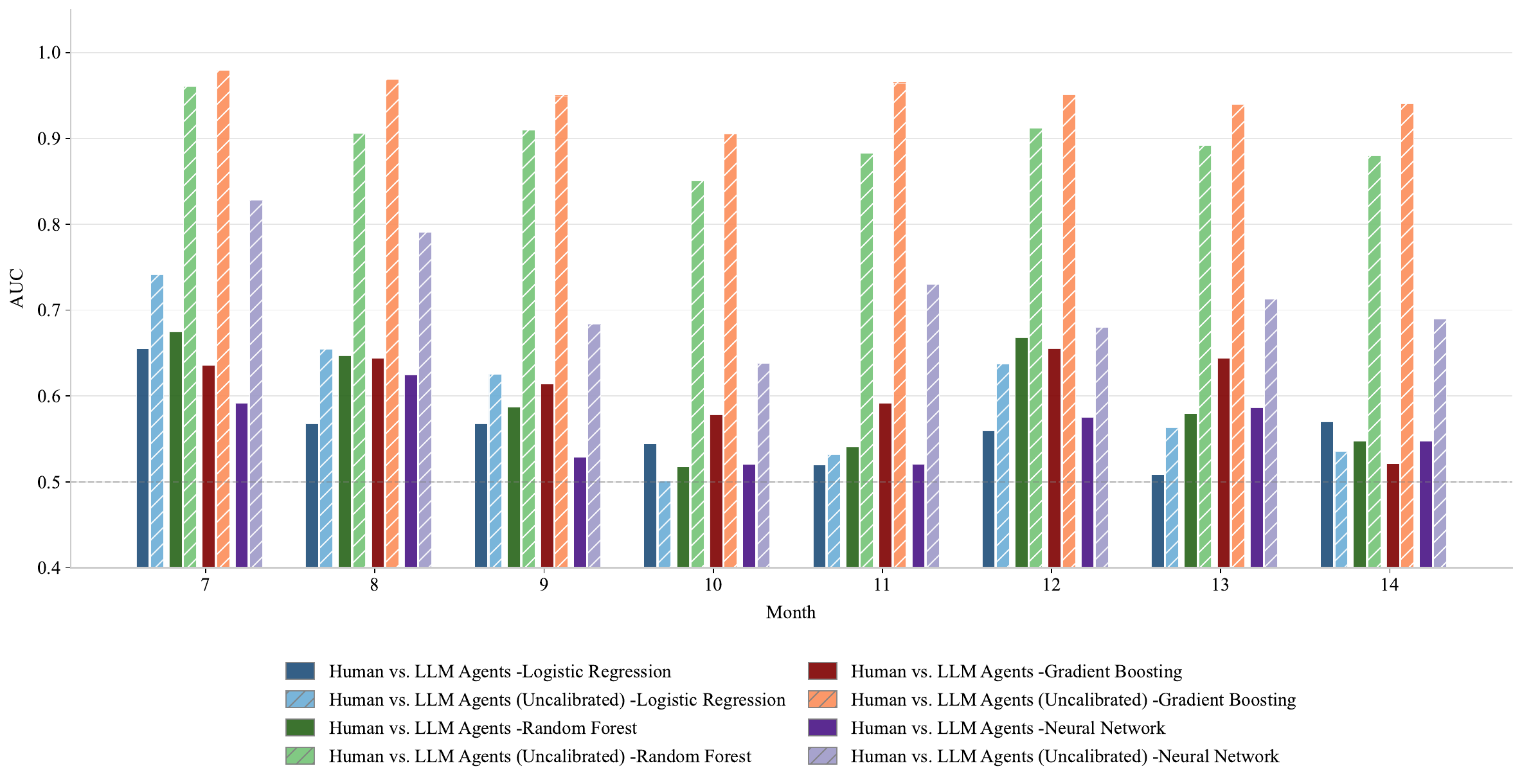}
\caption{Distinguishing Human and Simulated Respondents with Machine Learning}\label{fig:ml-validation}
\notepar{Notes: The figure reports held-out test AUCs by month for logistic regression, random forest, gradient boosting, and an ensemble feedforward neural network. Each model compares human observations with either calibrated or uncalibrated MAS observations. Values closer to 0.5 indicate weaker distinguishability.}
\end{figure}

To avoid basing the conclusion on one algorithm's inductive bias, we use four model classes familiar in applied economic work \citepapp{app_athey2019machine}:
\begin{itemize}[leftmargin=*,itemsep=2pt]
\item \emph{Logistic regression (LR)}, a linear probability-classification benchmark with an $L_2$ regularization penalty.
\item \emph{Random forest (RF)}, a nonlinear ensemble of decision trees built from randomly sampled feature subspaces.
\item \emph{Gradient boosting classifier (GBC)}, which sequentially fits trees to preceding residuals and captures nonlinear interactions.
\item \emph{Ensemble feedforward neural network (FNN)}, a flexible neural classifier with quadratic feature expansion.
\end{itemize}
Hyperparameters are selected by grid search on the validation set. After those choices are fixed, performance is evaluated only on the held-out test set using the area under the receiver-operating-characteristic curve (AUC). An AUC near 0.5 indicates chance-level discrimination; an AUC near 1 indicates nearly perfect separation. The calibrated and uncalibrated samples are evaluated in parallel against the same human benchmark. Figure~\ref{fig:ml-validation} reports these held-out comparisons.

\FloatBarrier

For calibrated simulations, all four algorithms produce AUCs between 0.50 and 0.70 across the eight months. Even flexible classifiers therefore have limited and unstable ability to separate calibrated Household Agents from human respondents on the supplied joint features. In contrast, except for logistic regression, the algorithms distinguish uncalibrated persona-only simulations much more easily; gradient boosting is generally above 0.90. The uncalibrated sample consequently contains strong mechanical regularities or biases that nonlinear classifiers can exploit. The parallel comparison identifies the importance of adding observed priors and social information to the personas. It does not imply indistinguishability on features omitted from $X_i$, nor does it prove that the human and simulated data-generating processes are the same.

\subsection{Economic Heterogeneity}\label{app:heterogeneity}

The distributional and classification tests do not establish whether forecasts are assigned to demographic groups in economically recognizable ways. We therefore compare three regularities concerning political affiliation, income, and gender. Existing evidence shows that households tend to become more pessimistic when the incumbent is not the party they support, producing partisan gaps in inflation and other macroeconomic expectations \citepapp{app_kamdar2025think,app_coibion2025upcoming}. To examine whether the MAS reproduces this regularity, we estimate separately by month and sample
\begin{equation}
Y_{it}=\beta_{0,t}+\beta_{1,t}\operatorname{Democrat}_{it}
+\mathbf X_{it}'\boldsymbol\gamma_t+\varepsilon_{it},
\label{appeq:political-heterogeneity}
\end{equation}
where $Y_{it}$ is the point expectation, $\operatorname{Democrat}_{it}$ equals one for Democratic respondents or agents and zero otherwise, and $\beta_{1,t}$ is the conditional mean difference between Democrats and non-Democrats in month $t$. The vector $\mathbf X_{it}$ includes age and age squared plus categorical controls for region, gender, marital status, education, and income. For a simulated sample $S\in\{\text{calibrated},\text{uncalibrated}\}$, equality with the human coefficient is assessed under independent samples using the cross-sample statistic in Equation~\eqref{appeq:coefficient-test}:
\begin{equation}
Z_t=\frac{\widehat\beta_{1,t}^{S}-\widehat\beta_{1,t}^{H}}
{\sqrt{\operatorname{SE}(\widehat\beta_{1,t}^{S})^2+
\operatorname{SE}(\widehat\beta_{1,t}^{H})^2}}.
\label{appeq:coefficient-test}
\end{equation}
The two-sided $p$-value is obtained from the standard normal distribution using the reported standard errors in Equation~\eqref{appeq:coefficient-test}. Table~\ref{tab:political-validation} reproduces the partisan pattern: during a Republican administration, Democratic affiliation raises inflation expectations in the human data and calibrated MAS. The human coefficients are positive and highly significant in all eight months. The calibrated coefficients are also significant at 1 percent, and calibrated-human equality-test $p$-values range from 0.145 to 0.913, so equality is never rejected. In contrast, the persona-only coefficients are only 0.272--0.353; they have the correct sign but are significantly below the human coefficients in every month. The uncalibrated system therefore displays severe under-polarization. Calibration is essential to reproducing the magnitude of the partisan gradient rather than only its direction.

\begin{table}[H]
\centering
\begin{threeparttable}
\caption{Political Heterogeneity in Human and Simulated Expectations}\label{tab:political-validation}
\fontsize{9.0}{9.0}\selectfont
\setlength{\tabcolsep}{1.4pt}
\renewcommand{\arraystretch}{1.05}
\begin{tabular}{@{}lcccccccc@{}}
\toprule
 & (1) & (2) & (3) & (4) & (5) & (6) & (7) & (8) \\
Month in MAS & 7 & 8 & 9 & 10 & 11 & 12 & 13 & 14 \\
Month in MSC & 2025-04 & 2025-05 & 2025-06 & 2025-07 & 2025-08 & 2025-09 & 2025-10 & 2025-11 \\
\midrule
\multicolumn{9}{l}{\textit{Panel A: Human subjects}} \\
Democrat & 6.118*** & 6.508*** & 4.260*** & 4.991*** & 4.181*** & 3.459*** & 4.145*** & 3.904*** \\
 & (0.816) & (0.942) & (0.766) & (0.771) & (0.711) & (0.920) & (0.874) & (0.656) \\
\midrule
\multicolumn{9}{l}{\textit{Panel B: Calibrated LLM Agents}} \\
Democrat & 4.352*** & 4.427*** & 4.404*** & 4.389*** & 4.376*** & 4.368*** & 4.425*** & 4.377*** \\
 & (1.074) & (1.075) & (1.076) & (1.077) & (1.079) & (1.080) & (1.078) & (1.078) \\
Test vs. Panel A ($p$-value) & 0.190 & 0.145 & 0.913 & 0.649 & 0.880 & 0.522 & 0.840 & 0.707 \\
\midrule
\multicolumn{9}{l}{\textit{Panel C: Uncalibrated LLM Agents}} \\
Democrat & 0.272*** & 0.312*** & 0.333*** & 0.334*** & 0.347*** & 0.353*** & 0.315** & 0.300** \\
 & (0.090) & (0.103) & (0.114) & (0.121) & (0.127) & (0.130) & (0.128) & (0.131) \\
Test vs. Panel A ($p$-value) & 0.000*** & 0.000*** & 0.000*** & 0.000*** & 0.000*** & 0.001*** & 0.000*** & 0.000*** \\
\midrule
Demographic controls & Yes & Yes & Yes & Yes & Yes & Yes & Yes & Yes \\
\bottomrule
\end{tabular}
\begin{tablenotes}[flushleft]\small
\item Notes: Each column is a separate monthly regression. The dependent variable is the point expectation and the reported coefficient is on Democratic affiliation. Controls are age, age squared, and categorical indicators for region, gender, marital status, education, and income. HC3 heteroskedasticity-consistent standard errors are in parentheses. The two-sample $Z$ test compares the indicated simulated coefficient with Panel A. * $p<0.10$, ** $p<0.05$, *** $p<0.01$.
\end{tablenotes}
\end{threeparttable}
\end{table}

Income gradients provide a second comparison. Empirical research documents a negative income gradient in household inflation expectations: higher-income households generally report lower expectations than lower-income households \citepapp{app_ehrmann2017consumers,app_dacunto2021exposure}. We replace the political indicator in Equation~\eqref{appeq:political-heterogeneity} with $\operatorname{High\_Income}_{it}$, equal to one for a respondent or agent in the high-income category, and control for age, age squared, political affiliation, region, gender, marital status, and education. The remaining estimation and coefficient-equality test follow the political analysis. Human coefficients are negative in all eight months, from $-0.936$ to $-2.572$, and become statistically significant in the final three months. The calibrated MAS produces a stable negative gradient from $-0.993$ to $-1.220$; equality-test $p$-values range from 0.327 to 0.994 and never reject the human coefficient. Persona-only estimates lie between $-0.065$ and $-0.123$, exhibiting marked attenuation and significant departures from the human benchmark in the later window. The income comparison therefore reinforces the political result: personas alone retain only a weak association, whereas calibration recovers the economically relevant gradient and its order of magnitude.

\begin{table}[H]
\centering
\begin{threeparttable}
\caption{Income Heterogeneity in Human and Simulated Expectations}\label{tab:income-validation}
\fontsize{9.0}{9.0}\selectfont
\setlength{\tabcolsep}{1.4pt}
\renewcommand{\arraystretch}{1.05}
\begin{tabular}{@{}lcccccccc@{}}
\toprule
 & (1) & (2) & (3) & (4) & (5) & (6) & (7) & (8) \\
Month in MAS & 7 & 8 & 9 & 10 & 11 & 12 & 13 & 14 \\
Month in MSC & 2025-04 & 2025-05 & 2025-06 & 2025-07 & 2025-08 & 2025-09 & 2025-10 & 2025-11 \\
\midrule
\multicolumn{9}{l}{\textit{Panel A: Human subjects}} \\
High income & -1.199 & -1.442 & -1.031 & -0.936 & -1.161 & -2.572*** & -1.884** & -1.183* \\
 & (0.863) & (1.052) & (0.818) & (0.773) & (0.769) & (0.915) & (0.940) & (0.633) \\
\midrule
\multicolumn{9}{l}{\textit{Panel B: Calibrated LLM Agents}} \\
High income & -1.005 & -0.993 & -1.014 & -1.100 & -1.122 & -1.144 & -1.220 & -1.192 \\
 & (1.112) & (1.115) & (1.120) & (1.124) & (1.128) & (1.135) & (1.127) & (1.130) \\
Test vs. Panel A ($p$-value) & 0.890 & 0.770 & 0.990 & 0.904 & 0.977 & 0.327 & 0.651 & 0.994 \\
\midrule
\multicolumn{9}{l}{\textit{Panel C: Uncalibrated LLM Agents}} \\
High income & -0.065 & -0.086 & -0.083 & -0.083 & -0.099 & -0.123 & -0.115 & -0.099 \\
 & (0.107) & (0.119) & (0.130) & (0.139) & (0.146) & (0.150) & (0.152) & (0.156) \\
Test vs. Panel A ($p$-value) & 0.192 & 0.200 & 0.252 & 0.278 & 0.175 & 0.008*** & 0.063* & 0.097* \\
\midrule
Demographic controls & Yes & Yes & Yes & Yes & Yes & Yes & Yes & Yes \\
\bottomrule
\end{tabular}
\begin{tablenotes}[flushleft]\small
\item Notes: Each column is a separate monthly regression. The dependent variable is the point expectation and the reported coefficient is on the high-income indicator. Controls and inference follow Table~\ref{tab:political-validation}. HC3 standard errors are in parentheses. * $p<0.10$, ** $p<0.05$, *** $p<0.01$.
\end{tablenotes}
\end{threeparttable}
\end{table}

\begin{figure}[H]
\centering
\includegraphics[width=.92\linewidth]{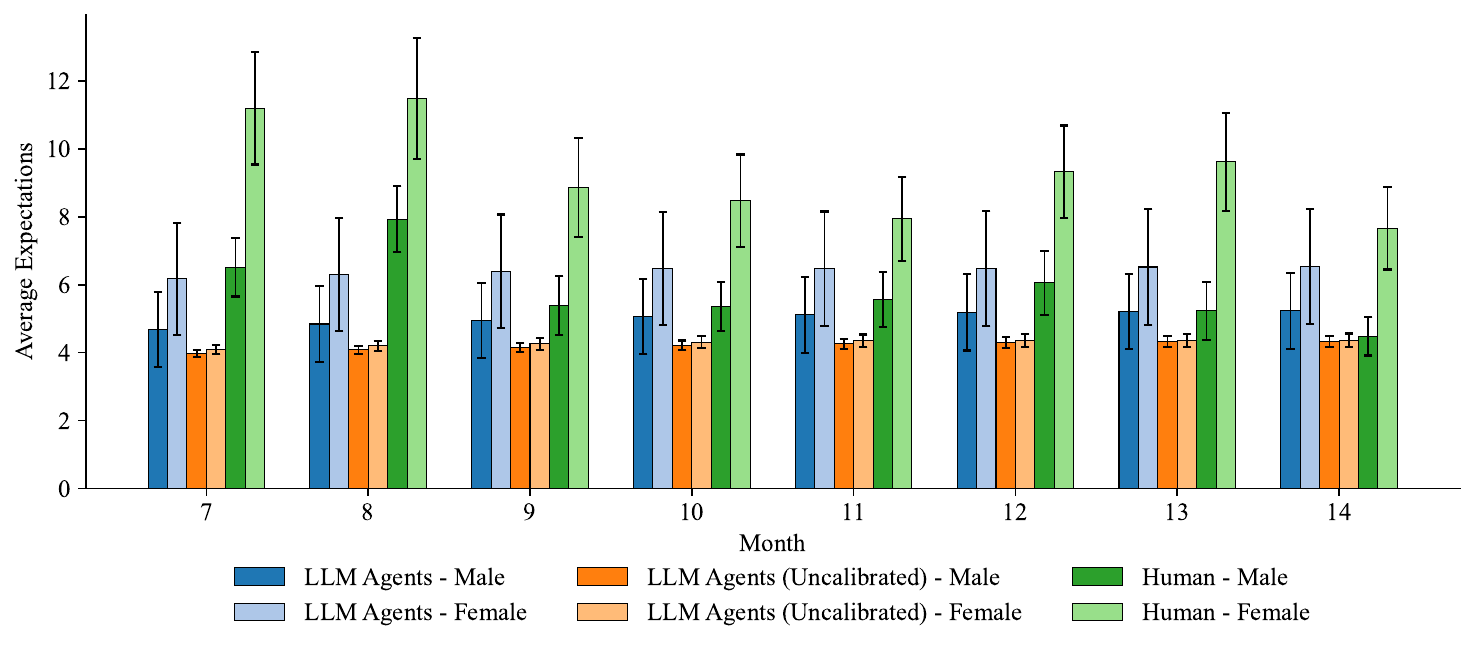}
\caption{Gender Heterogeneity in Human and Simulated Inflation Expectations}\label{fig:gender-validation}
\notepar{Notes: The figure plots monthly inflation expectations by gender for human MSC respondents, calibrated LLM Agents, and uncalibrated LLM Agents over the eight-month validation window. Bands are 95\% confidence intervals.}
\end{figure}

Figure~\ref{fig:gender-validation} compares gender patterns. Women commonly report higher inflation expectations than men, a recurring regularity in household-expectations research \citepapp{app_dacunto2021gender,app_madeira2015heterogeneous,app_coleman2023inflation}. For each of the eight months, we calculate the group mean separately by biological sex in the human MSC, calibrated MAS, and uncalibrated MAS samples and display the resulting six grouped series with 95 percent confidence intervals. In the human sample, women report systematically and significantly higher inflation expectations than men in every aligned month. Calibrated Household Agents reproduce both the direction and dynamic persistence of this gap. Persona-only agents instead cluster between roughly 4.0 and 4.4 percent, have very narrow confidence intervals, and show severe gender-gap attenuation and belief homogenization. Together with Tables~\ref{tab:political-validation} and \ref{tab:income-validation}, this evidence shows that calibration restores economically important conditional heterogeneity that a persona prompt alone fails to generate. The exercise remains diagnostic: resemblance can arise partly from associations learned during model training and does not prove that the simulated mechanism matches the human mechanism.

\FloatBarrier

\subsection{Scope of Extrapolation}\label{app:extrapolation}

As discussed in Section~\ref{sec:mas}, the calibration procedure and all external-validity tests use only T1, the baseline treatment that corresponds to the actual tariff announcement, because the MSC supplies human observations against which that realized information environment can be compared. The other eleven treatments describe counterfactual situations that did not occur and therefore have no directly corresponding human data. This section asks how far the validation evidence obtained under T1 can support the simulation results in those other settings.

The discussion follows the idea of \emph{heuristic validation} and the associated ``validate-then-simulate'' approach: when evidence indicates that AI Agents are a reasonable proxy for human responses in one substantively relevant setting, the human and simulated subjects may be treated as partially interchangeable for carefully bounded exploration of related settings \citepapp{app_hullman2026human,app_kozlowski2025artificial}. This approach is increasingly used in research that evaluates language-model simulations against human judgments before extending them to nearby counterfactuals \citepapp{app_anthis2025position,app_tranchero2026strategic}. It can support \emph{near generalization}, in which the new setting alters only limited details of a human-validated benchmark. Consistent with the main text, we do not require, and do not claim, exact quantitative generalization. We instead assess the credibility of qualitative results, including the direction and ranking of DTEs and their consistency with established theory. Some nearby scenarios may inform qualitative validation and cautious policy simulation; scenarios involving larger changes are used mainly for exploration and mechanism analysis.

The argument proceeds on six levels. First, all arms use the same agents, foundation model, calibrated parameters, and system architecture, so their only designed difference is the input signal rather than the computational mechanism that generates expectations. Second, the T1 tests compare simulated and human samples along distributional shape, demographic heterogeneity, and distinguishability, thereby supplying multidimensional evidence about the MAS response at the T1 signal point. Third, the paper focuses on differences, directions, and rankings across treatments. If calibration errors shared in levels cancel under differencing and the remaining error changes regularly with signal distance, those qualitative comparisons can be more reliable than exact quantitative estimates. Fourth, such extrapolation is explicitly conditional on local regularity and becomes less credible as a new signal moves farther from the calibrated setting. Fifth, mechanism-level evidence provides an indirect consistency check. Sixth, the final interpretation is tiered rather than unconditional.

\paragraph{1. Formal setup.} Let $x_i\in\mathcal X$ denote the demographic vector of Household Agent $i$, where $i=1,\ldots,N$ and $N=300$. These demographic characteristics come from the stratified sample drawn from the December 2024 MSC cross-section, are embedded in the prompts, and are supplied to the corresponding LLM Agent. The MAS evolves monthly for $t=1,\ldots,18$. Let $\mathcal H_{i,t-1}$ denote Agent $i$'s internal state upon entering Month $t$, including its historical expectations, the contents retained in its memory module, and social-media information already received. Let $s_{i,t}$ denote the treatment information supplied in Month $t$. The vector $\theta$ collects the calibrated and subsequently fixed model and system parameters: the foundation model, the distributions used to draw temperature and top-$p$, the information-cocoon parameter $\alpha$, and the memory length $w$. The point forecast and subjective probability distribution generated in each month can be represented as
\begin{equation}
Y_{i,t}=f(x_i,\mathcal H_{i,t-1},s_{i,t};\theta)+\varepsilon_{i,t},
\label{appeq:response-operator}
\end{equation}
where $f$ is the expectation-generating function of the MAS. It is constrained by the Bayesian-updating mechanism and trades off new information, including treatment information and social-media information, against prior expectations. The disturbance $\varepsilon_{i,t}$ captures idiosyncratic variation induced by foundation-model sampling. Treatment is administered in Month 7: T0 receives placebo information, whereas treatment arm $g$ receives signal $s^g$.

After calibration to real data, the experiment fixes the quadruple $(\{x_i\},f,\theta,\text{system architecture})$ and keeps it identical across all treatment arms. Because the sample is computer-generated, can be reused exactly, and is generated under fixed random seeds, T0 and every treatment arm contain the same 300 agents. The foundation model, prompt framework, model parameters, memory module, recommendation algorithm, and rules governing every system module are likewise held fixed. The only designed inter-arm difference is the treatment information itself, together with the subsequent history endogenously generated from that information.

This design is analogous in spirit to the structural approach to policy evaluation emphasized by \citetapp{app_lucas1976econometric} and \citetapp{app_heckman2010building}: counterfactual analysis should hold the mechanism generating behavior as fixed as possible while changing an exogenous policy or environmental input. Here, the experiment fixes the computational rules and defines treatments as changes in signal inputs. Mechanism invariance is necessary for interpreting the controlled comparisons, but it is not sufficient for transporting them to human subjects. Moving from one validated signal point to other signal points also requires stability of the MAS response as the input changes.

Equation~\eqref{appeq:response-operator} is the common response operator used to organize all treatment comparisons.

\paragraph{2. What the validity tests actually support.} The T1 tests cover the treatment month and its subsequent dynamic path, Months 7--14, aligned month by month with MSC observations from April through November 2025. Three exercises examine proximity between simulated and human samples from different angles.

First, the CvM tests and distribution-shape comparisons assess the similarity of the expectation distributions. Under T1, the finite-sample tests do not find statistically significant differences between calibrated simulated and human distributions, and their shapes are highly similar in every aligned month. The correct interpretation is that the available data and test power do not provide sufficient evidence against distributional concordance. The purpose is not to prove strict statistical identity, but to show that the MAS matches substantially more of the human distribution than its mean alone.

Second, the heterogeneity analysis asks whether the MAS captures important expectation differences associated with political affiliation, income, and gender. The simulated sample reproduces several principal directions and gradients found in the human sample, indicating that fit is not produced solely by mechanically adjusting an aggregate mean. More precisely, these results support the claim that $f$ preserves part of the conditional heterogeneity on the demographic dimensions examined; they do not establish that the model has recovered every derivative of the human conditional response function.

Third, machine-learning classification compares the joint features $(Y,x)$. AUC values for the random forest, gradient boosting classifier, and neural network are mostly between 0.50 and 0.70, indicating that common classifiers have limited overall ability to distinguish human observations from calibrated simulations. Their pronounced AUC decline relative to the uncalibrated comparison shows that calibration removes systematic differences that the classifiers could otherwise exploit. Relative to matching selected moments, this provides a stricter test of proximity in the observed joint distribution.

Thus, what passes the tests under T1 is not a particular number, but the function $f(\cdot\mid x;\theta)$ defined over the population $\mathcal X$ and evaluated at the signal point $s^{T1}$ on the measured margins. T1 was not selected because its result was easier to validate. It directly corresponds to the Liberation Day tariff announcement of April 2, 2025, so the MSC provides the required month-by-month human benchmark. T1 therefore serves as the reality-based anchor from which the scenario design expands. Because no corresponding human observations exist for the other treatments, their results can be evaluated only through controlled comparisons, local regularity, and indirect empirical evidence; they cannot receive external validation of the same strength as T1.

\paragraph{3. Differences among treatments.} Under the tariff-information setting, the signal in any one of the eleven treatments other than T1 can be represented as a controlled modification of the validated baseline signal:
\begin{equation}
s^g=s^{T1}\oplus\Delta^g,
\label{appeq:signal-change}
\end{equation}
where $\oplus$ denotes a modification or extension of $s^{T1}$ and $\Delta^g$ is the deliberately introduced, tightly controlled feature difference along one of the six comparison dimensions, such as implementation timing, tariff rate, or linguistic complexity. Within a dimension, the treatment materials retain the same core information and semantics except for the target feature. Hence $\Delta^g$ is designed to be low-dimensional and limited. This does not imply that every signal pair is close in raw text embeddings. T1 and T12 use exactly the same text but identify different senders; T4, T10, and T11 end with identical text but arrive through different information-release histories. An embedding-only distance would miss these economically meaningful distinctions. We therefore define
\begin{equation}
\phi(s)=\big(\phi_{\mathrm{timing}},\phi_{\mathrm{rate}},
\phi_{\mathrm{complexity}},\phi_{\mathrm{narrative}},
\phi_{\mathrm{style}},\phi_{\mathrm{sender}}\big)
\end{equation}
whose components respectively capture implementation timing, tariff rate, linguistic complexity, narrative, release style, and sender. The economically meaningful signal distance is
\begin{equation}
d(s,s')=\left\|W[\phi(s)-\phi(s')]\right\|,
\label{appeq:signal-distance}
\end{equation}
where $W$ contains the unknown relative weights of the six components. Closeness therefore means proximity across the economically relevant features, not simply semantic or embedding similarity.

\citetapp{app_hotz2005predicting} distinguish changes in a target population from changes in program features when transporting effects to a new setting. The present design fixes the target population, thereby removing errors caused by changes in sample composition; the remaining transport problem concerns changes in the signal itself. The transportability framework of \citetapp{app_pearl2014external} likewise requires researchers to state which mechanisms or variables change across environments. Here, the model, parameters, architecture, and population are known to be invariant, but the treatment signal and post-treatment information sequence change across arms. Reusing the same agents individually also yields exact correspondence in pretreatment characteristics and histories, which isolates the signal more sharply than a comparison of means from different populations. It eliminates compositional differences inside the simulation, but it cannot by itself guarantee closeness between simulated and real human outcomes. Credibility ultimately depends on whether the LLM Agents' response operator is stable around the relevant changes.

Equations~\eqref{appeq:signal-change} and \eqref{appeq:signal-distance} therefore distinguish a treatment's textual content from its economically relevant distance to T1.

\paragraph{4. Difference bias and local regularity.} Let $f^*$ denote the true human-population response operator and $f$ the calibrated agent response operator. Their difference is the calibration bias
\begin{equation}
b(x,\mathcal H,s)=f(x,\mathcal H,s;\theta)-f^*(x,\mathcal H,s).
\label{appeq:calibration-bias}
\end{equation}
The three T1 exercises show that observable manifestations of $b$ are small at the anchor $s^{T1}$, but they do not directly observe $b$ at the signals used in the other arms. The object of interest is a treatment comparison. For a given outcome and horizon $k$, let arm $g$'s population mean be $\bar Y_k^g=\bar f_k(s^g)$ and define $\mathrm{DTE}_k^g=\bar Y_k^g-\bar Y_k^{T0}$. For any two treatments,
\begin{align}
\mathrm{DTE}_k^g-\mathrm{DTE}_k^{g'}
&=\bar f_k(s^g)-\bar f_k(s^{g'}) \notag\\
&=\underbrace{\bar f_k^*(s^g)-\bar f_k^*(s^{g'})}_{\text{human difference}}
+\underbrace{\bar b_k(s^g)-\bar b_k(s^{g'})}_{\text{difference in calibration bias}}.
\label{appeq:difference-bias}
\end{align}
Equation~\eqref{appeq:calibration-bias} separates the simulator's response from its human counterpart, and Equation~\eqref{appeq:difference-bias} shows how this bias enters a treatment contrast. The common T0 baseline cancels in the first equality. The final equality shows that the discrepancy between the simulated and true human treatment differences depends not on each arm's absolute bias, but on the difference between those biases. Thus, even if the calibrated operator develops a common level error after leaving $s^{T1}$, that common component cancels in the treatment contrast. Whether the MAS and humans deliver the same qualitative ranking depends on whether calibration bias changes sharply between neighboring signals within a comparison dimension. A sufficient, although not necessary, condition for sign preservation is
\begin{equation}
|\bar b_k(s^g)-\bar b_k(s^{g'})|
<|\mathrm{DTE}_k^g-\mathrm{DTE}_k^{g'}|.
\label{appeq:ranking-condition}
\end{equation}

To make this reasoning explicit, suppose calibration bias is locally Lipschitz on the design-relevant signal set $\mathcal S_0$:
\begin{equation}
|\bar b_k(s)-\bar b_k(s')|\le L_b d(s,s'),
\qquad s,s'\in\mathcal S_0.
\label{appeq:bias-lipschitz}
\end{equation}
In other words, when other conditions are fixed, the discrepancy between the model's and humans' errors should not increase without bound as two treatment signals become arbitrarily close. The design preserves the core message and changes one target feature within each comparison dimension, deliberately limiting $d(s^g,s^{g'})$. The local bound makes the differential bias small relative to the already-tested level fit when the movement is sufficiently close.

Writing this regularity condition on $b$ requires separate attention to both $f$ and $f^*$. If, on the same local set,
\begin{align}
|\bar f_k(s)-\bar f_k(s')|&\le L_f d(s,s'),\label{appeq:model-lipschitz}\\
|\bar f_k^*(s)-\bar f_k^*(s')|&\le L_* d(s,s'),\label{appeq:human-lipschitz}
\end{align}
then the triangle inequality yields
\begin{equation}
|\bar b_k(s)-\bar b_k(s')|\le (L_f+L_*)d(s,s').
\label{appeq:combined-lipschitz}
\end{equation}
Regularity of the LLM response alone is therefore insufficient. The average human response must also avoid an unbounded jump within the same local treatment neighborhood. We state this as a local-regularity assumption rather than presenting it as a consequence of model architecture.

The local regularity of $f$ can be understood from two complementary features of LLM construction. Internally, an LLM maps input text into token embeddings and passes those representations through a finite composition of transformations. Affine projections, softmax attention, layer normalization on a nondegenerate bounded region, and smooth activations such as GELU and SiLU admit finite local Lipschitz bounds; their finite composition is therefore locally Lipschitz on a bounded subset of embedding space. This claim is deliberately local. \citetapp{app_kim2021lipschitz} show that standard dot-product self-attention is not globally Lipschitz over an unbounded input domain, but the argument here requires regularity only in a neighborhood of $s^{T1}$. Work on the regularity of attention and mathematical representations of transformers provides additional support for bounded-domain continuity, not a global guarantee \citepapp{app_vuckovic2021regularity,app_geshkovski2025mathematical}.

At the representation level, semantic embeddings are constructed so that semantically similar texts occupy nearby regions. Smooth movement in that representation is the empirical basis of semantic retrieval \citepapp{app_reimers2019sentence}. The same property is used directly by the SMIM recommender, which retrieves posts close to an agent's expressed opinion, and by the generation of semantically similar Trump Agent messages under fixed random seeds. \citetapp{app_park2024linear} show that high-level concepts can be encoded through regular, approximately linear directions in LLM representation space. Evidence on ``algorithmic fidelity'' and expectation simulations conditioned on demographic and economic information similarly suggests finely varying rather than uniformly discontinuous conditional responses \citepapp{app_argyle2023out,app_zarifhonarvar2026artificial,app_anesti2025simulating}. These results support the plausibility of transferable local regularity; they do not establish it for every prompt or policy environment.

There is no mechanism-free theorem establishing local regularity of the human average response $f^*$ for every population and every text. A transparent sufficient condition is available. Let individual $j$'s conditional mean response at signal $s$ be $m_{j,k}(s)$, so that $\bar f_k^*(s)=\int m_{j,k}(s)\,dP(j)$. Suppose that, on the local treatment domain, individual changes satisfy $|m_{j,k}(s)-m_{j,k}(s')|\le \ell_jd(s,s')$ and the population-average slope bound is finite, $\mathbb E[\ell_j]<\infty$. Averaging the individual inequalities gives
\begin{equation}
|\bar f_k^*(s)-\bar f_k^*(s')|
\le \mathbb E[\ell_j]d(s,s').
\end{equation}
This condition accommodates heterogeneous response intensity, nonlinear response curves, and turning points. Even threshold behavior at the individual level need not create a jump in the population mean. If thresholds have a continuous population distribution with bounded local density and the response change at a threshold is bounded, a small signal movement causes only an $O(d)$ share of people to cross their thresholds; the population mean therefore also changes by $O(d)$. The conditional-logit framework of \citetapp{app_mcfadden1974conditional} provides a classic aggregation example: individual choice is discrete, but integration over continuously distributed random utility makes the aggregate choice probability a smooth function of observed attributes. If, by contrast, a large share of people has exactly the same threshold, the population mean may jump and the local Lipschitz assumption should not be maintained.

Human information experiments supply evidence closer to the setting studied here. \citetapp{app_coibion2022monetary} express post-information inflation expectations as a weighted average of priors and new signals and test the framework using nearly 20,000 U.S. households and eight randomized information treatments. Updating weights differ by signal but exhibit finite partial adjustment. In the sample with nonnegative initial expectations, the study reports little evidence of pronounced prior-dependent nonlinearity; conditional on initial beliefs, different household groups respond broadly similarly to common signals, and average treatment responses are not generated by a few extreme jumps. For a given class of information and updating weight, a signal whose numerical content changes locally and regularly therefore implies a finite local response slope. This evidence does not prove Lipschitz continuity under tariff messages. It does show that a finite local bound on the average response is behaviorally grounded in a closely related macroeconomic-expectations environment. Treatments that change the sender, create an entirely new information history, or move many subjects across a shared institutional threshold should not be treated as equally local to T1.

Equation~\eqref{appeq:bias-lipschitz} clarifies the ranking claim. If
\begin{equation}
|\widehat\Delta_{gg',k}|>L_bd(s^g,s^{g'}),
\qquad
\widehat\Delta_{gg',k}=\bar f_k(s^g)-\bar f_k(s^{g'}),
\end{equation}
then the corresponding human and simulated differences share a sign. For a ranking involving several treatments, every adjacent simulated effect gap must likewise exceed its corresponding error bound. Because $L_b$ cannot be observed without the missing human counterfactuals, the inequalities do not mechanically guarantee every ranking. The appropriate interpretation is probabilistic and conditional: reversal becomes less plausible when economically meaningful signals are closer, simulated effect gaps are larger, and the ranking remains stable under alternative wording and repeated sampling.

The absence of a human benchmark for each derived treatment therefore does not make all inter-arm comparisons uninformative. Differencing removes the common baseline and concentrates the identification risk in differential bias. When the comparison remains inside a controlled neighborhood and the local-regularity assumption is consistent with robustness checks, it is more credible than the absolute level in a single counterfactual arm. That credibility is graded and conditional; it does not validate a human causal effect for every derived treatment.

\paragraph{5. Mechanism-level cross-evidence.} The remaining treatments have no directly corresponding realized setting, so they cannot be externally validated with human observations. Sections~\ref{subsubsec:literature} and~\ref{sec:cb} instead ask whether theoretical and empirical results from earlier research can explain the qualitative comparisons generated along the experimental dimensions. The alignments are specific. Cognitive discounting predicts weaker responses to signals about more distant outcomes \citepapp{app_gabaix2020behavioral}. Signal precision and information rigidity explain why agents place different updating weights on information of different quality \citepapp{app_coibion2015information,app_cavallo2017inflation}. Salience theory explains why prominent attributes receive disproportionate decision weight \citepapp{app_bordalo2012salience}. Narrative Economics emphasizes that transmissible narratives can shape economic judgments and behavior \citepapp{app_shiller2017narrative}. Household information experiments show that the content and presentation of policy communication change inflation expectations \citepapp{app_coibion2022monetary}.

These correspondences serve as indirect checks on whether the simulation contradicts established mechanisms or human evidence; they do not prove that every simulated ranking must be correct. If the MAS consistently produced signs and rankings contrary to theory across several distinct dimensions, the credibility of extrapolation would decline sharply. Conversely, multidimensional agreement makes it less likely that all of the qualitative patterns arise from arbitrary decoding noise. The open-ended-response dictionaries, independently coded causal frames, and embedding measures supply another layer of mechanism evidence. Because the cited studies do not directly estimate every counterfactual studied here, this evidence is properly described as triangulation or an indirect consistency check, not formal external validation.

Research in other settings also finds that language models conditioned on demographic or economic information can reproduce selected features and behavioral regularities of human survey responses \citepapp{app_argyle2023out,app_zarifhonarvar2026artificial,app_anesti2025simulating}. These studies raise the prior plausibility that parts of a conditional response operator can transfer. They also document or discuss group flattening, prompt sensitivity, and inconsistent economic concepts. The external literature therefore supports continued testing and bounded use of such simulations, not treating an LLM as a faultless substitute for human subjects.

\paragraph{6. Credibility boundaries and interpretation.} Five boundaries govern the conclusions. First, absolute levels must be distinguished from treatment comparisons. The argument concerns directions, rankings, and mechanisms, not the exact values of the counterfactual distributions. T1 has a direct real-data anchor; the other arms do not. Their DTE values are therefore indicative, and their directions and rankings carry more information than their absolute magnitudes, conditional on local regularity and the effect-gap condition.

Second, credibility declines with signal distance. Scenarios that retain the policy topic, message template, and most treatment attributes are closer to interpolation around T1. Replacing the sender or repeatedly reversing the information sequence is closer to extrapolation. Consequently, conclusions from Dimension 3, which changes the release sequence, and Dimension 6, which changes the sender, require more caution than conclusions from the other dimensions. Raw embedding similarity cannot justify assigning them the same calibrated validity as T1.

Third, thresholds and contextual changes may break local regularity. Moving from a one-time announcement to repeated reversals, or changing the institution that sends the message, may cross such a threshold. The Lipschitz condition allows these treatments to have large effects, but requires those changes to remain finitely bounded. If alternative wording, prompt order, or repeated sampling causes a ranking to reverse sharply, the implied local bound is too wide to support that ranking and the conclusion should be downgraded.

Fourth, C1--C5 append different central-bank communications to the T11 information environment. They use the same tariff history and the same Household Agents, so differences relative to T11 have a clear internal comparison basis. Yet the central-bank messages can simultaneously change policy stance, commitment, explanatory frame, credibility, and institutional source; they need not be small movements along a single semantic axis. These outcomes must likewise be interpreted as treatment comparisons and tested for stability under alternative wording and repeated runs. They are not point estimates calibrated to human central-bank communication responses.

Fifth, the extrapolation claim is explicitly conditional. Multidimensional T1 tests support the MAS at a reality-based anchor; the fixed population and computational mechanism attribute inter-arm changes cleanly to signal inputs inside the simulation; differencing removes common errors; and local regularity constrains the remaining differential bias. In summarize, these elements provide conditional evidence about directions, rankings, and mechanisms. They neither establish absolute quantities in unrealized settings nor guarantee that comparisons far from T1 or across thresholds preserve the sign of human effects.

We therefore adopt a tiered interpretation. T1 has direct external calibration and validity evidence. Treatments that remain close to T1, and to the baseline signal within the same comparison dimension, may be used to assess qualitative conclusions and for cautious policy simulation or implication. More distant scenarios, especially T10, T11, T12, and C1--C5, are designed principally for scenario analysis and mechanism exploration; their results require greater caution in policy analysis. This boundary preserves the value of controlled MAS experiments without extending the claims beyond the data and literature, and it implements the distinction between heuristic validation for exploration and statistical calibration for confirmatory inference emphasized by \citetapp{app_hullman2026human}.

\clearpage
\section{Additional Simulation Results}\label{app:additional-results}

This section reports the unemployment counterparts to the main inflation figures. The layouts, treatment month, and estimators are identical to Figures~\ref{fig:results-dim13} and \ref{fig:results-dim46}. The unemployment results preserve the broad qualitative rankings described in Table~\ref{tab:qualitative-rankings}. They differ in magnitude because tariff pass-through, production costs, monetary-policy reactions, and labor demand enter the agents' unemployment explanations with different horizons.

\begin{figure}[H]
\centering
\includegraphics[width=.98\linewidth]{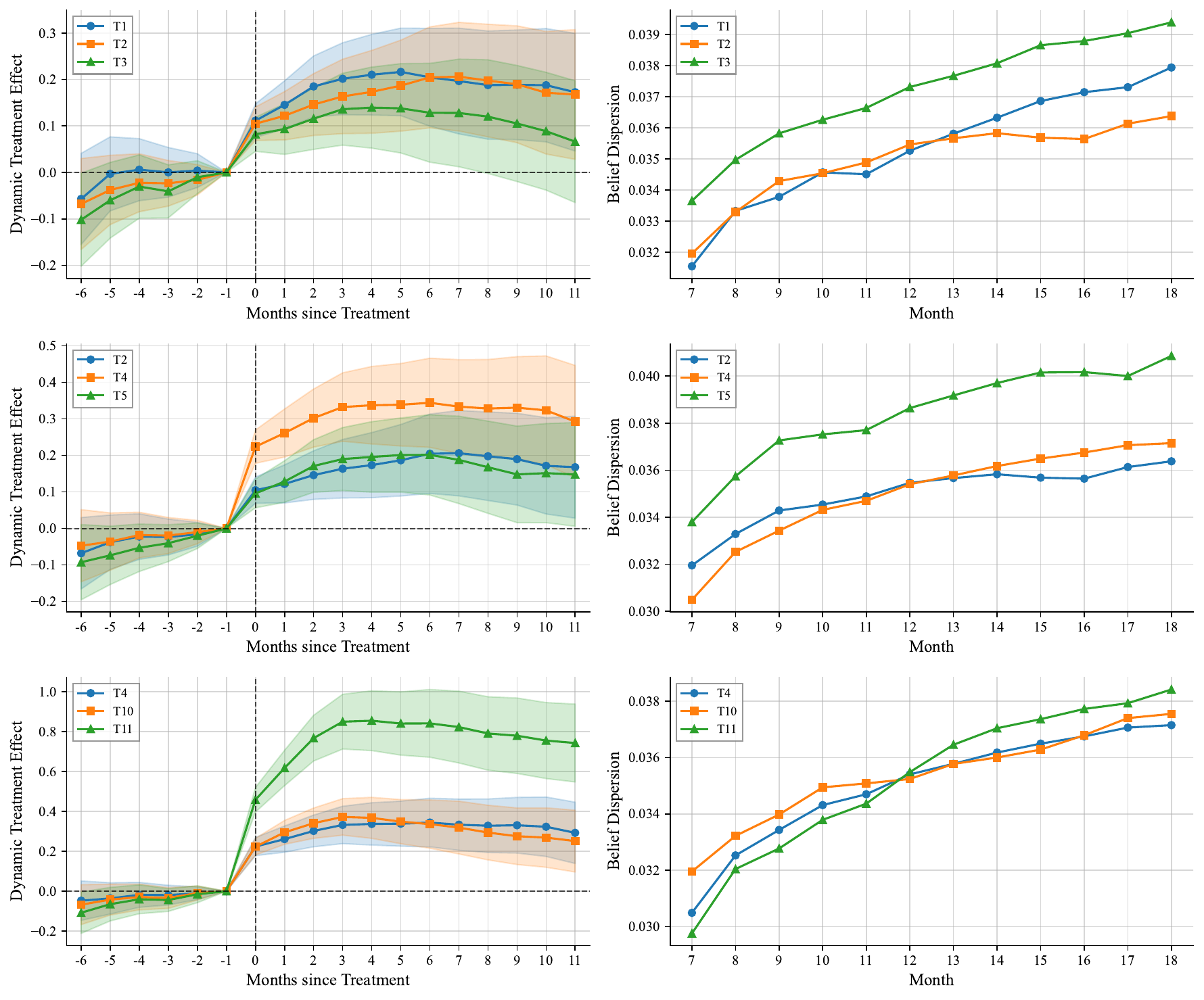}
\caption{Unemployment Expectations: Implementation Time, Tariff Rate, and Semantic Coherence}\label{fig:unemp-dim13}
\notepar{Notes: Rows correspond to Dimensions 1--3. Left panels report monthly Dynamic Treatment Effects on twelve-month-ahead unemployment expectations relative to the placebo arm. Right panels report belief dispersion. Treatment occurs in Month 7. Shaded bands are 95\% confidence intervals.}
\end{figure}
\FloatBarrier

\begin{figure}[H]
\centering
\includegraphics[width=.98\linewidth]{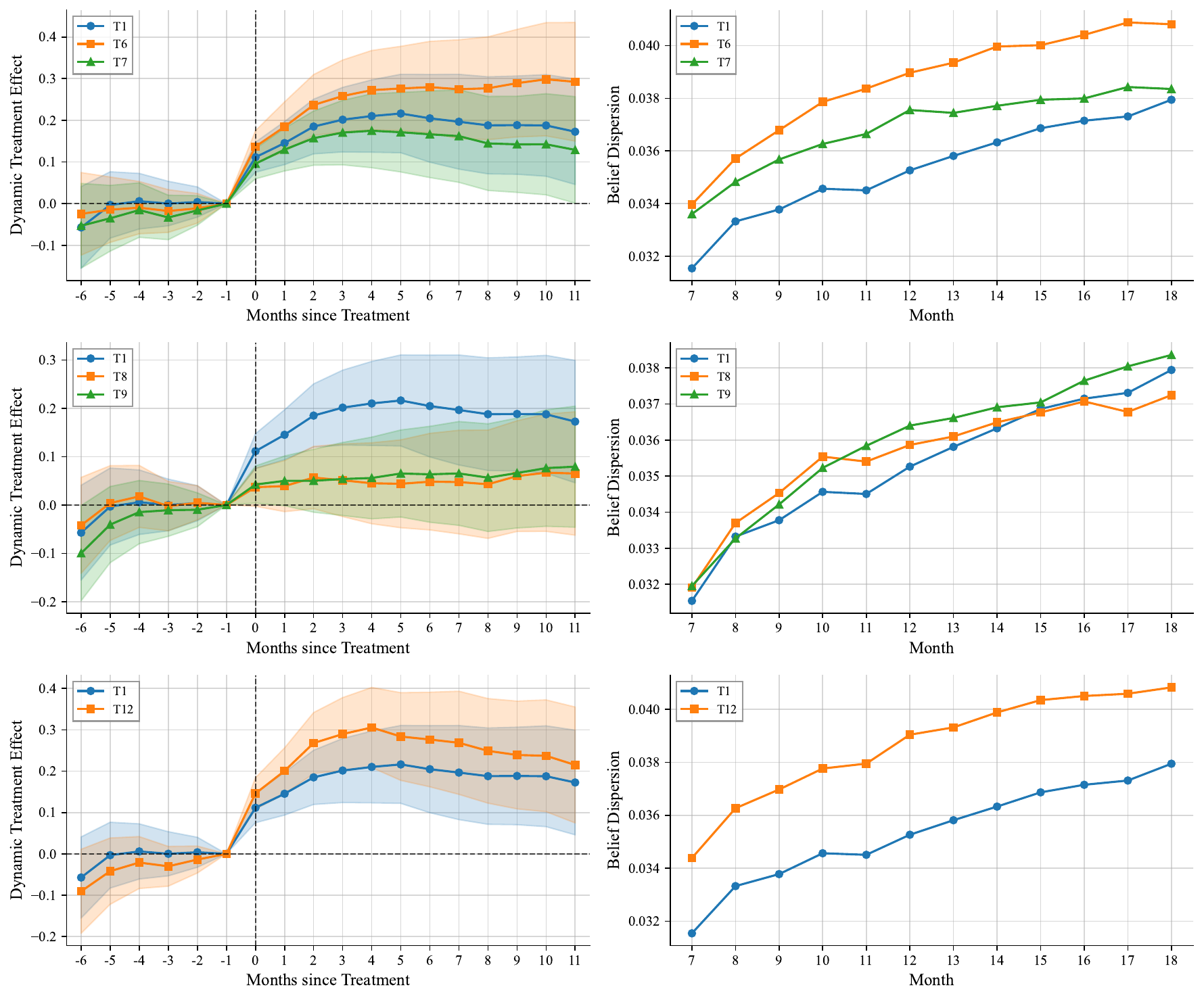}
\caption{Unemployment Expectations: Complexity, Narrative, and Sender}\label{fig:unemp-dim46}
\notepar{Notes: Rows correspond to Dimensions 4--6. Left panels report monthly Dynamic Treatment Effects on twelve-month-ahead unemployment expectations relative to the placebo arm. Right panels report belief dispersion. Treatment occurs in Month 7. Shaded bands are 95\% confidence intervals.}
\end{figure}
\FloatBarrier

\clearpage
\section{Text-Based Mechanism Measures}\label{app:text-measures}

\subsection{Dictionary Construction}\label{app:dictionaries}

The dictionaries were fixed before treatment comparisons. Their role is measurement, not unconstrained topic discovery. We began from the economic channel implied by each design dimension, expanded terms with morphological variants and common phrases, removed terms that occurred with unrelated meanings in a manual audit, and retained the same list across arms. This procedure follows the transparency principle for open-ended survey analysis in \citetapp{app_haaland2025understanding}: readers can inspect the mapping from text to mechanism, and no arm-specific vocabulary is chosen after viewing the ranking. Table~\ref{tab:dictionaries} reports every retained term and phrase.

Each dictionary has a distinct economic construction basis:

\begin{description}[leftmargin=3.55cm,style=nextline,itemsep=4pt]
\item[Tariff awareness.] This set contains direct policy nouns, tariff-cost expressions, and trade-conflict phrases. It measures whether the tariff enters the agent's forecast explanation, not whether the response is optimistic or pessimistic. The distinction follows the household-expectations literature, where attention to a price or policy signal is separated from the directional belief induced by that signal \citepapp{app_weber2022subjective,app_haaland2025understanding}.

\item[Temporal urgency.] This set combines immediacy, acceleration, persistence, and near-term expressions. Salience and rational-inattention theories imply that agents devote more scarce processing capacity to actionable, imminent shocks and less to unresolved events with low current decision value \citepapp{app_sims2003implications,app_mackowiak2009optimal,app_mackowiak2023rational}. The terms therefore capture perceived urgency rather than tariff severity alone.

\item[Ambiguity perception.] This set covers explicit ignorance, difficulty of prediction, modal hedges, fuzzy epistemic states, deferral and wait-and-see language, volatility, conditionality, doubt, alarm, conflicting signals, cognitive complexity, macroeconomic uncertainty, surprise, and warnings. These categories operationalize the economic distinction between a signal with a known probability distribution and one whose timing, rate, implementation state, or interpretation is not pinned down \citepapp{app_ilut2014ambiguous}. The deliberately broad subcategories preserve the source dictionary's complete set of cues.

\item[Price impact.] These terms refer to prices, groceries, purchasing power, price levels, and explicit increases. They isolate the consumer-price pass-through channel emphasized in empirical work on tariff incidence and household inflation expectations \citepapp{app_amiti2019impact,app_cavallo2025tracking,app_weber2022subjective}.

\item[Employment impact.] These terms describe hiring freezes, employment, layoffs, jobs, staffing, labor-market instability, and related anticipation. They capture the business-uncertainty, production, and labor-demand channels documented in research on trade-policy uncertainty and tariff exposure \citepapp{app_caldara2020economic,app_flaaen2020production}.

\item[Extrapolation.] This set records escalation, acceleration, directional continuation, deterioration, and forward propagation. It is grounded in models and evidence in which recent directional sequences are overextended into future expectations. A response is counted only when it contains language of continuation or amplification; describing a high terminal rate as severe is not sufficient.

\item[Skepticism.] This set records reversal, bluffing, empty threats, contradiction, unreliability, withdrawal, and explicit disbelief. It is kept separate from ambiguity because an agent can understand a message precisely yet assign it little credibility after repeated reversals. The terms therefore capture discounting of the sender rather than uncertainty about the policy's economic consequences.

\item[Credibility.] This set contains explicit assessments that a statement is real, serious, dependable, legitimate, or likely to be followed through. In Bayesian terms, credibility changes the perceived precision attached to the sender's signal; in political-economy evidence, source and partisan congruence can change that weight \citepapp{app_kamdar2025think,app_binder2025partisan}.

\item[Unexpectedness.] This set contains surprise, departures from established norms, sudden shifts, and explicit statements that an event was not anticipated. It operationalizes the innovation in a signal relative to the receiver's prior. Unexpectedness is conceptually distinct from credibility: a message can be surprising but implausible, or familiar but highly credible.
\end{description}

Concordance-line audits remove uses that lack the specified mechanism, but the reported lists retain all terms in the source measurement dictionaries. The sets are narrow enough to remain interpretable, and the embedding measure below provides a complementary check that does not depend on the presence of any listed word.

\begingroup
\small
\begin{longtable}{p{.23\linewidth}p{.71\linewidth}}
\caption{Keyword Dictionaries for Mechanism Measurement}\label{tab:dictionaries}\\
\toprule
Dictionary & Included terms and phrases \\
\midrule
\endfirsthead
\toprule
Dictionary & Included terms and phrases \\
\midrule
\endhead
Tariff awareness & tariff; trade policies; tariff costs; tariff news; tariff-induced; trade; trade barriers; trade conflicts; trade disputes; trade issues; trade regulations; trade taxes; trade tensions; trade wars. \\
Temporal urgency & immediate; immediately; rapidly; quick; abrupt; abruptly; prompt; accelerate; escalate; intensifying; mounting; pressing; imminent; current; ongoing; prevailing; remains; persist; stagnant; unresolved; near-term; short term; short-term; today; recently; lately; sooner; yet; anymore; lasting; prolonged; underway; heading. \\
Ambiguity perception & can't tell; cannot tell; difficult to say; do not know; don't know; hard to determine; hard to know; hard to say; hard to tell; no way to know; not certain; not sure; remains to be seen; too early to say; too soon to tell; could; maybe; might; perhaps; presumably; supposedly; ambiguous; indeterminate; questionable; unclear; vague; at some point; deferred; delayed; eventually; postponed; timeline; fluctuate; fluctuating; fluctuation; fluctuations; variable; turbulence; turbulent; unstable; unresolved; conditional; contingent; depending; depends; disconnected; distrust; doubt; doubtful; skepticism; alarming; concerns; fears; troubled; troubling; worried; worries; worrisome; crisis; downturn; gloomy; panic; pauses; pausing; pessimism; pessimistic; recession; abruptly; dramatic; dramatically; drastically; rapidly; sharp; sharply; sudden; suddenly; expectations; forecast; forecasts; gamble; hopeful; overarching; speculation; speculate; speculative; conflicted; conflicting; contradictory; inconsistent; burden; complex; difficulties; disruption; struggling; broader; federal; indicators; months; older; tight; surprises; unexpected; unforeseen; unlikely; warning; warnings. \\
Price impact & prices; groceries; purchasing power; grocery prices; elevated; price level; weekly grocery; price increase; higher prices. \\
Employment impact & hiring freeze; trade; anticipate; job growth; economic instability; labor market; employment; layoffs; jobs; job losses; job security; insecurity; staffing. \\
Extrapolation & escalating; escalation; surge; surging; skyrocket; skyrocketing; spiraling; spiral; soaring; climb; worsening; deteriorate; deteriorating; accelerating; intensifying; deepen; inevitable; inevitably; further; increasing; prolonged; continuously; gradually; trending; decline; predicted; keep rising; will rise; get worse; gets worse; getting worse; only get worse; even higher; going forward; going up; likely to; down the road; even worse; expected to. \\
Skepticism & bluff; bluffing; canceled; change of mind; changed mind; contradiction; contradictory; do not believe; do not trust; doubt; doubtful; dubious; empty promise; empty threat; empty threat; exaggerate; exaggeration; fake; flip flop; flip-flop; hollow; incredible; inconsistent; just talk; lies; lying; not believe; not trust; political; posturing; questionable; reversed; reversal; rhetoric; skeptic; skeptical; skepticism; suspicious; u-turn; unreliable; untrustworthy; will not happen; withdraw; withdrawn. \\
Credibility & real; serious; credible; trustworthy; truthful; authentic; honest; believable; dependable; legitimate; proven; effective; essential; lasting; committed; warrant; deliver; track record; good faith; follow through. \\
Unexpectedness & surprising; surprised; unexpected; shocked; shocking; unusual; unprecedented; anomaly; atypical; uncharacteristic; erratic; rarely; severe; suddenly; dramatically; spiking; shifting; turmoil; disruptive; unstable; divergence; reversed; swings; norms; sharp; sharply; did not expect; didn't expect; never expected; out of nowhere; caught off guard; break from; departure from; hard to predict; sharp rise; significant shift. \\
\bottomrule
\end{longtable}
\endgroup

The dictionaries separate concepts that are correlated in ordinary language but distinct in the experiment. For example, tariff awareness measures whether the policy enters the explanation, while temporal urgency measures why it receives attention. Ambiguity includes epistemic hedges and conditional language; skepticism instead concerns the sender's credibility. Extrapolation terms identify a forward-propagating causal story rather than uncertainty alone. Price- and employment-impact dictionaries separate the two principal macroeconomic outcomes.

For each response, Equation~\eqref{eq:term-density} produces a length-normalized score, and Equation~\eqref{eq:mean-term-density} aggregates it to the arm-month level. Contrasts in $\bar S_{g,t}^{(k)}$ are estimated with the same treatment timing and inference procedure used for the numerical outcomes.

\subsection{Causal Frames}\label{app:causal-frames}

The narrative comparison uses the seven substantive causal frames and residual category in Table~\ref{tab:causal-frames}. Consistent with the source coding protocol, each substantive frame is coded as a separate binary indicator rather than forcing all responses into a single mutually exclusive primary frame. A response may therefore invoke more than one causal channel when its text explicitly supports each relationship; material that cannot be classified enters ``Other.'' Two human coders follow the blinded, independent procedure described in Section~\ref{subsubsec:text}. Cohen's $\kappa$ is assessed separately for every frame, and the principal investigator adjudicates disagreements only after the independent coding is complete. A frame is assigned from an explicit causal clause, not from a keyword alone.

\begin{table}[H]
\centering
\begin{threeparttable}
\caption{Causal Frames Used in the Narrative Analysis}\label{tab:causal-frames}
\begin{tabularx}{\linewidth}{llX}
\toprule
Code & Frame & Core concept \\
\midrule
F1 & Tariff $\rightarrow$ Price Pass-Through & Tariffs raise the cost of imported goods, and these costs are passed on to consumers in the form of higher prices. \\
F2 & Supply Chain Disruption & Trade restrictions disrupt supply chains, create shortages, or reduce the availability of goods, thereby pushing prices upward. \\
F3 & Monetary / Fiscal Policy & Inflation is driven by monetary policy (e.g., interest rates, money supply) or fiscal policy (e.g., government spending, deficits), rather than by tariffs per se. \\
F4 & No Tariff Effect & Tariffs will have little or no meaningful impact on consumer prices or inflation; prices are expected to remain stable. \\
F5 & Uncertainty-Driven & Economic uncertainty or policy unpredictability itself drives inflation expectations, independent of any specific transmission mechanism. \\
F6 & Exporter Absorption & Foreign exporters, rather than domestic consumers, will bear the cost of tariffs by absorbing the added expense or reducing their margins. \\
F7 & Personal Experience & The respondent grounds their inflation expectation in direct personal observation of prices, costs, or economic conditions in their daily life. \\
Other & Other & Residual responses. \\
\bottomrule
\end{tabularx}
\begin{tablenotes}[flushleft]\footnotesize
\item Notes: Each substantive frame is a binary indicator and the indicators need not be mutually exclusive. The taxonomy is motivated by empirical tariff-incidence evidence \citepapp{app_amiti2019impact,app_flaaen2020production,app_cavallo2025tracking} and by the subjective-model approach of \citetapp{app_andre2022subjective}. Figure~\ref{fig:causal-frames} reports diagnostic shares for F1, F4, and F6.
\end{tablenotes}
\end{threeparttable}
\end{table}

\subsection{Embedding Dispersion}\label{app:embedding-dispersion}

Dictionary scores test targeted mechanisms but may miss semantically different explanations that use no common keywords. We first strip leading and trailing whitespace and collapse repeated spaces. A pretrained Sentence Transformer maps response $x_{i,g,m}$ into a $d$-dimensional dense vector $\phi(x_{i,g,m})$, which is normalized to unit $L_2$ norm:
\begin{equation}
e_{i,g,m}=\frac{\phi(x_{i,g,m})}{\|\phi(x_{i,g,m})\|_2},
\qquad e_{i,g,m}\in\mathbb R^d,
\qquad \|e_{i,g,m}\|_2=1.
\label{eq:embedding-normalization}
\end{equation}
For treatment arm $g$ and month $m$, with sample size $N_{g,m}$, mean pairwise cosine similarity is
\begin{equation}
\bar{C}_{g,m}=\frac{1}{N_{g,m}(N_{g,m}-1)}
\sum_{i=1}^{N_{g,m}}\sum_{j\neq i}^{N_{g,m}}
e_{i,g,m}e_{j,g,m}^{\prime},
\label{eq:embedding-cosine}
\end{equation}
and Embedding Dispersion is
\begin{equation}
\operatorname{ED}_{g,m}=1-\bar C_{g,m}.
\label{eq:embedding-dispersion}
\end{equation}
Because the embeddings have unit norm, their inner product equals cosine similarity. The resulting index is the average pairwise cosine distance, lies in $[0,2]$, and is larger when agents' expectation-formation explanations display greater semantic disagreement and heterogeneity. We remove empty responses, use the same embedding model for every arm and month, and do not project embeddings using treatment labels. Embedding Dispersion is a text-level disagreement measure; it is not interpreted as the variance of a cardinal psychological trait.

Across Dimensions 1--3, the ordering of $\operatorname{ED}_{gm}$ matches the ambiguity and semantic-coherence interpretation: T3 and T5 produce varied completions of missing policy details, and T11 produces varied extrapolations from escalating rates. Across Dimensions 4--6, T6 and T7 produce different forms of reasoning heterogeneity, narrative arms move frame shares more than overall semantic dispersion, and T12 generates the largest sender-related divergence. These patterns supply mechanism evidence for the numerical ranking without assuming that an LLM's verbal explanation is a direct observation of a human cognitive process.

\clearpage
\section{Central Bank Communication Experiment}\label{app:cb-experiment}

\subsection{FOMC Agent and Communication Texts}\label{app:fomc-materials}

We construct an FOMC Agent using the same foundation model, temperature, and fixed-seed generation settings as the Trump Agent. The source text is the supplied C1 version of the FOMC's longer-run-goal language. The prompt changes only the strategy-specific clause, requires all other elements to remain as consistent as possible, and instructs the model to imitate the smooth, formal tone of official FOMC communication. For documentary accuracy, the prompt is reproduced verbatim below as an experimental artifact. Its statement that the displayed C1 sentence was ``released'' on January 25, 2012 is not a claim of word-for-word quotation: the original 2012 release used ``statutory mandate,'' while the supplied C1 text uses the later dual-mandate formulation.

\begin{tcolorbox}[colback=white,colframe=black,boxrule=0.8pt,arc=0pt,breakable]
\begin{verbatim}
You are a member of the Federal Open Market Committee (FOMC). Following
President Trump's tariff threat announcements, you need to communicate
on behalf of the FOMC using different strategies to guide and stabilize
market economic expectations.

The following is the communication text regarding longer-run goals
released by the FOMC on January 25, 2012, which has been reaffirmed
multiple times annually since then:
"The Committee reaffirms its judgment that inflation at the rate of 2
percent, as measured by the annual change in the price index for personal
consumption expenditures, is most consistent over the longer run with
the Federal Reserve's statutory maximum employment and price stability
mandates."

Please use the above text as a baseline and rewrite it to generate four
new communication texts corresponding to the following four different
strategies:
- Strategy 1 (Hawkish): Scenario-Based Communication. Commit to explicit
scenario triggers: if the tariff information leads to persistent upward
pressure on prices, the Committee stands ready to tighten the stance of
monetary policy to return inflation to the 2 percent target level.
- Strategy 2 (Hawkish): Affirming the institution's independence.
Emphasize that the Federal Reserve is institutionally independent of any
political influence and remains fully committed to achieving its
statutory mandates and goals. Its monetary policy decisions will not be
adjusted in response to political pressure.
- Strategy 3 (Dovish): Look-through Strategy. Emphasize that the Committee
views the tariffs as introducing a one-off and temporary shock rather
than affecting underlying unemployment and inflation. Therefore, the
Committee will "look through" this transitory shock and refrain from
taking immediate action.
- Strategy 4 (Dovish): Allow inflation to overshoot moderately. Emphasize
that the Committee will allow inflation to run moderately above 2 percent
for some time to support overall economic activity, and therefore will
refrain from taking immediate action.

The original baseline communication text and these four texts rewritten
based on different strategies will be used in an information provision
experiment and presented to subjects in different treatment groups.
Therefore, to ensure that the differences in responses among groups are
not driven by variations in wording other than the strategies themselves,
the four output texts should all be strictly based on the baseline text.
They should differ only in the communication strategy employed, while
keeping all other elements as consistent as possible. However, please
note: do not directly copy the phrasing from the strategy descriptions
above into the rewritten texts. Instead, you should rewrite the baseline
text by closely imitating the wording and tone of official FOMC
communication texts, ensuring the output is smooth, coherent, and natural.

Output the rewritten text in the following JSON format, without including
any additional text:
[
  {
    "Strategy 1": "Text of Strategy 1",
    "Strategy 2": "Text of Strategy 2",
    "Strategy 3": "Text of Strategy 3",
    "Strategy 4": "Text of Strategy 4"
  }
]
\end{verbatim}
\end{tcolorbox}

\noindent\textbf{C1: Goal-only baseline.} ``The Committee reaffirms its judgment that inflation at the rate of 2 percent, as measured by the annual change in the price index for personal consumption expenditures, is most consistent over the longer run with the Federal Reserve's statutory maximum employment and price stability mandates.''

\noindent\textbf{C2: Scenario-based communication.}
\begin{tcolorbox}[colback=white,colframe=black,boxrule=0.8pt,arc=0pt,breakable]
\begin{verbatim}
The Committee reaffirms its judgment that inflation at the rate of 2
percent, as measured by the annual change in the price index for
personal consumption expenditures, is most consistent over the longer
run with the Federal Reserve's statutory maximum employment and price
stability mandates. Should recent tariff-related developments generate
persistent upward pressure on broad prices, the Committee stands firmly
prepared to appropriately firm the stance of monetary policy to return
inflation to this target.
\end{verbatim}
\end{tcolorbox}

\noindent\textbf{C3: Institutional independence.}
\begin{tcolorbox}[colback=white,colframe=black,boxrule=0.8pt,arc=0pt,breakable]
\begin{verbatim}
The Committee reaffirms its judgment that inflation at the rate of 2
percent, as measured by the annual change in the price index for
personal consumption expenditures, is most consistent over the longer
run with the Federal Reserve's statutory maximum employment and price
stability mandates. Operating as an independent institution, the
Committee remains resolutely committed to these statutory mandates,
ensuring that our monetary policy decisions remain entirely unswayed by
political pressure regarding recent tariff-related developments.
\end{verbatim}
\end{tcolorbox}

\noindent\textbf{C4: Look-through strategy.}
\begin{tcolorbox}[colback=white,colframe=black,boxrule=0.8pt,arc=0pt,breakable]
\begin{verbatim}
The Committee reaffirms its judgment that inflation at the rate of 2
percent, as measured by the annual change in the price index for
personal consumption expenditures, is most consistent over the longer
run with the Federal Reserve's statutory maximum employment and price
stability mandates. Recognizing that recent developments regarding
trade policy will likely introduce only transitory price effects rather
than alter underlying inflation dynamics, the Committee intends to look
through these temporary fluctuations and maintain its current policy
stance.
\end{verbatim}
\end{tcolorbox}

\noindent\textbf{C5: Overshoot moderately.}
\begin{tcolorbox}[colback=white,colframe=black,boxrule=0.8pt,arc=0pt,breakable]
\begin{verbatim}
The Committee reaffirms its judgment that inflation at the rate of 2
percent, as measured by the annual change in the price index for
personal consumption expenditures, is most consistent over the longer
run with the Federal Reserve's statutory maximum employment and price
stability mandates. Following recent tariff-related developments, the
Committee will allow inflation to run moderately above this target for
some time to support broader economic activity, maintaining the current
stance of monetary policy.
\end{verbatim}
\end{tcolorbox}

\subsection{Mechanism Evidence}\label{app:cb-mechanisms}

The central-bank experiment uses T11 as the comparison arm. Relative to that escalating tariff-threat environment, C2 and C5 have the largest positive DTEs, C1 and C3 are near zero, and C4 has a negative DTE for most post-treatment months. All five strategies reduce belief dispersion. This combination means that a message can coordinate agents around a higher or lower expected path; dispersion reduction alone is not evidence that expectations moved in the desired direction.

We analyze the simulated open-ended responses and construct a \emph{semantic-pressure index} measuring whether the causal account behind an expectation is closer to an expectation-increasing narrative or an expectation-stabilizing narrative. The all-MiniLM-L6-v2 Sentence Transformer maps agent $i$'s response to a normalized embedding $e_i$. We separately construct outcome-specific narrative anchors, embed them with the same model, and normalize their embeddings.

For inflation, the positive anchors are \emph{persistent tariff inflation}, \emph{cost pass-through}, \emph{overshoot tolerance}, and \emph{inflation de-anchoring}; the stabilizing anchors are \emph{transitory price effects}, \emph{stable household prices}, \emph{clear central-bank guidance}, and \emph{cost absorption}. These concepts are not selected by inspecting the ranking. They correspond to the principal mechanisms in research on tariff transmission and inflation: the pass-through of import taxes to buyers, firms' pricing and margin response, the distinction between a relative-price shock and persistent inflation, monetary accommodation of a tariff cost-push shock, and credibility of the inflation objective \citepapp{app_amiti2019impact,app_werning2025tariffs,app_cubaborda2025trade,app_coibion2022monetary}.

For unemployment, the positive anchors are \emph{business uncertainty}, \emph{hiring freezes}, \emph{job cuts}, \emph{higher borrowing costs}, \emph{Fed tightening}, and \emph{recession risk}; the stabilizing anchors are \emph{labor-market resilience}, \emph{temporary trade effects}, \emph{policy guidance}, and \emph{stable employment}. They capture the uncertainty, imported-input, retaliation, labor-demand, and monetary-tightening channels documented in work on trade-policy uncertainty and manufacturing employment \citepapp{app_caldara2020economic,app_flaaen2024disentangling}. The anchor sets are therefore economically prespecified rather than labels fitted to the observed C1--C5 ordering.

\begin{figure}[H]
\centering
\includegraphics[width=.86\linewidth]{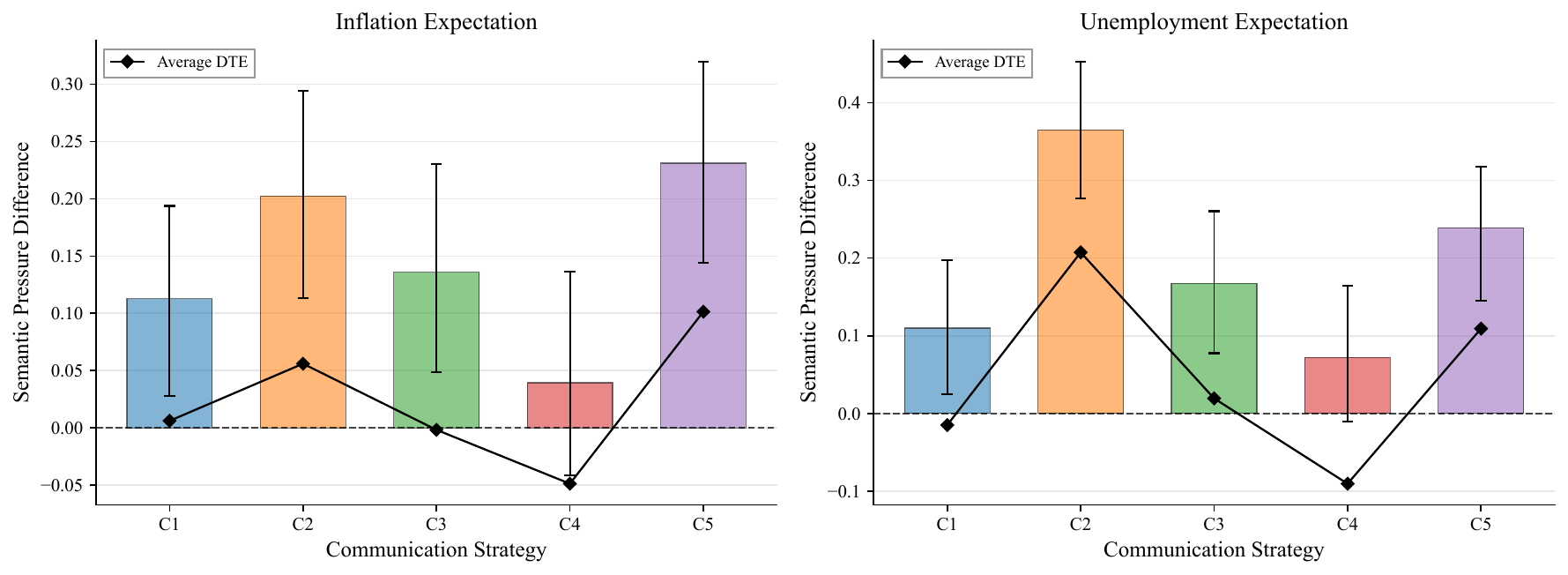}
\caption{Semantic Pressure and Dynamic Treatment Effects}\label{fig:semantic-pressure}
\notepar{Notes: Bars report the semantic-pressure score for strategies C1--C5; the overlaid line reports the corresponding average DTE for inflation and unemployment expectations relative to T11. Higher semantic pressure denotes more explicit language about persistence, policy response, or tolerated inflation. Where uncertainty bars are shown, they are 95\% intervals.}
\end{figure}

Let $A_y^+$ and $A_y^-$ be the positive- and stabilizing-pressure anchors for outcome $y$. The raw score in Equation~\eqref{appeq:semantic-pressure-raw} is
\begin{equation}
r_i^{(y)}=\frac{1}{|A_y^+|}\sum_{a\in A_y^+}\cos(e_i,a)
-\frac{1}{|A_y^-|}\sum_{a\in A_y^-}\cos(e_i,a).
\label{appeq:semantic-pressure-raw}
\end{equation}
A high value places the response closer to expectation-increasing causal narratives. To make the inflation and unemployment scales comparable, Equation~\eqref{appeq:semantic-pressure-standard} standardizes raw scores within the pooled post-treatment sample containing T11 and C1--C5:
\begin{equation}
S_i^{(y)}=\frac{r_i^{(y)}-\bar r^{(y)}}{\sigma_r^{(y)}}.
\label{appeq:semantic-pressure-standard}
\end{equation}
The bars in Figure~\ref{fig:semantic-pressure} report the differences relative to T11 defined in Equation~\eqref{appeq:semantic-pressure-difference}:
\begin{equation}
\Delta_g^{S,y}=\frac{1}{N_g}\sum_{i\in g}S_i^{(y)}
-\frac{1}{N_{T11}}\sum_{i\in T11}S_i^{(y)},
\qquad g\in\{C1,\ldots,C5\}.
\label{appeq:semantic-pressure-difference}
\end{equation}
For both outcomes, semantic pressure follows the DTE ordering: C2 and C5 are highest, C1 and C3 occupy the middle, and C4 is lowest.

\FloatBarrier

Belief coordination is assessed separately with embedding dispersion from Equation~\eqref{eq:embedding-dispersion}. Every central-bank message reduces semantic disagreement relative to T11 because it supplies a common institutional frame. The size of the reduction depends on the specificity of the causal account. C4 produces the strongest convergence because ``temporary'' and ``underlying inflation dynamics'' provide a common interpretation. Goal-only and independence messages coordinate less because agents must still infer how tariffs enter prices and employment.

\begin{figure}[H]
\centering
\includegraphics[width=.86\linewidth]{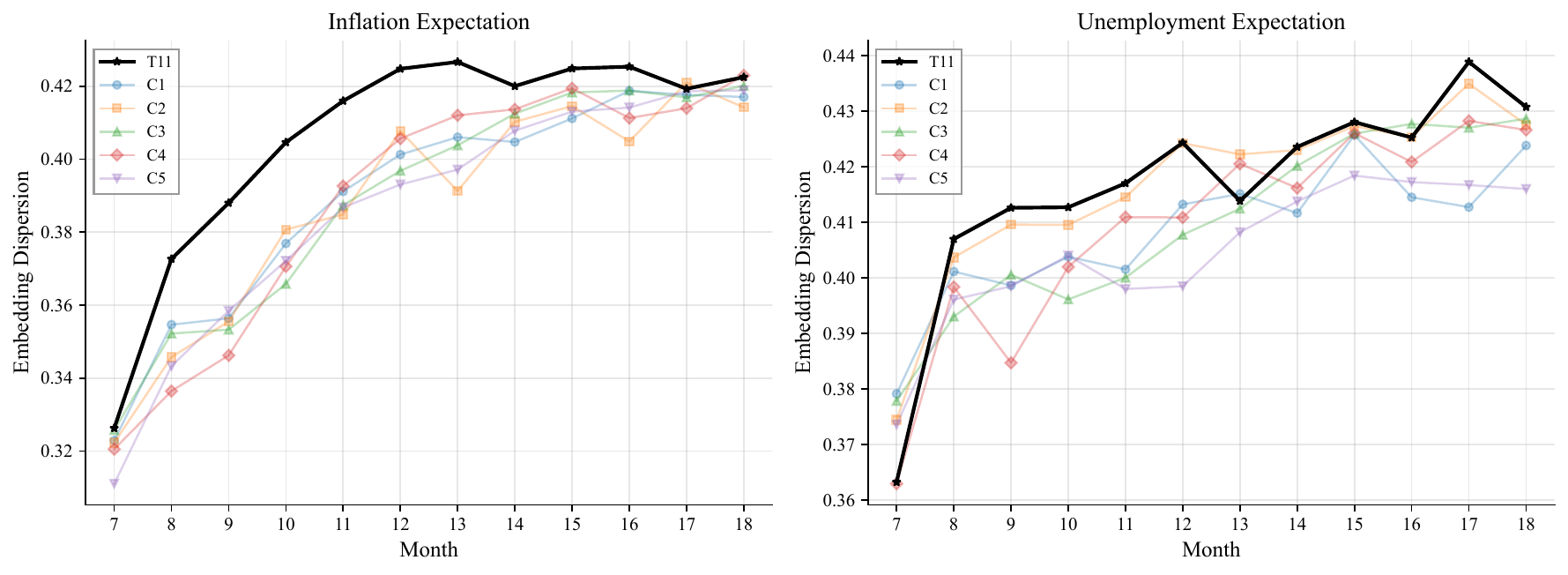}
\caption{Embedding Dispersion under Central-Bank Communication}\label{fig:cb-embedding}
\notepar{Notes: The panels report average pairwise cosine distance among open-ended responses for C1--C5, with T11 as the tariff-threat benchmark. Lower values indicate more semantically similar explanations within a strategy. Intervals, where displayed, are 95\% intervals for the associated arm contrasts.}
\end{figure}
\FloatBarrier

Tone alone is insufficient. C2 is hawkish, but it binds the tariff threat to ``persistent upward pressure'' and monetary tightening. Agents can interpret that language as information that the central bank sees a worse state of the economy, so the central-bank information effect raises assessments of inflation and unemployment risk before the promised response stabilizes the path \citepapp{app_romer2000federal,app_nakamura2018information}. C5 operates differently: it weakens the immediate constraint imposed by the 2 percent objective by explicitly tolerating an overshoot. Its unemployment DTE is below C2 because it does not invoke the same tightening and recession channel, but it remains above C1, C3, and C4.

C1 and C3 cause little average revision because a target and an independence commitment do not specify how tariffs enter prices, wages, firm costs, or employment, nor whether the shock is temporary. They induce more references to the Federal Reserve, target, mandate, and commitment, but do not provide a concrete mapping from the tariff to the forecast. Information rigidity and rational inattention imply that households need not convert abstract institutional language into a forecasting rule when the communication does not resolve uncertainty directly relevant to their decision \citepapp{app_coibion2015information,app_sims2003implications,app_mackowiak2023rational}.

C4 is the only strategy with a negative DTE because it explicitly classifies the tariff as a temporary relative-price disturbance rather than a change in underlying inflation dynamics. It does not deny that tariffs can raise some goods prices. Instead, it changes the category into which agents place that increase: a temporary price-level disturbance that need not shift the inflation trend or trigger immediate tightening. This pattern is consistent with theory treating tariffs as cost-push shocks for which optimal policy may partially accommodate short-run inflation \citepapp{app_werning2025tariffs} and with evidence that Federal Reserve reactions depend on whether policymakers attribute inflation to demand or supply forces \citepapp{app_kakhbod2026mind}. The finding remains conditional on a credible diagnosis; if tariff inflation proves persistent, a look-through message can de-anchor expectations and damage future credibility.

The economic implication is that communication stabilizes expectations through the causal model it supplies, not merely through whether its conventional tone is hawkish or dovish. A useful communication strategy must answer three connected questions for households: what kind of shock has occurred, under what condition will the central bank respond, and over what horizon does its objective bind? A hawkish sentence can raise measured expectations if it reveals adverse central-bank information. A dovish look-through sentence can lower them if it credibly reclassifies the persistence of the shock. Goal language can reaffirm commitment yet have little incremental effect when it does not explain the current state. Independence language can defend legitimacy yet leave the transmission mechanism unresolved.

This distinction also separates stabilization from reassurance. C2 is clear, but its clarity makes a persistent adverse state salient. C4 is reassuring only because it offers a substantive diagnosis, and the benefit disappears if that diagnosis is not credible. Communication and future policy action are therefore complements. A central bank cannot select the currently reassuring causal story independently of the evidence it holds and the reaction it will later implement. In the present distant scenario, the rankings are testable hypotheses for human experiments rather than prescriptions for the Federal Reserve.

\clearpage
\bibliographystyleapp{apalike}
\bibliographyapp{bib/refs}

\end{document}